\documentclass[twocolumn,trackchanges]{aastex701}

\usepackage{float}
\newcommand{\mycomment}[1]{}
\usepackage{algpseudocode}
\usepackage{amsmath}

\begin{document}

\title{Year-timescale changes in AGN radio luminosity as seen by the ASKAP Variables and Slow Transients Survey}

\correspondingauthor{Chloe Klare}
\author[0000-0002-8899-3769]{Chloe Klare}
\affiliation{Department of Astronomy and Astrophysics, 525 Davey Lab, 251 Pollock Road, The Pennsylvania State University, University Park, PA 16802, USA}
\email[show]{chloe.klare@psu.edu}  

\author[0000-0002-4557-6682]{Charlotte Ward} 
\affiliation{Department of Astronomy and Astrophysics, 525 Davey Lab, 251 Pollock Road, The Pennsylvania State University, University Park, PA 16802, USA}
\email{cvw5890@psu.edu}

\author[0000-0003-0699-7019]{Dougal Dobie}
\affiliation{Sydney Institute for Astronomy, School of Physics, The University of Sydney, Sydney, New South Wales, Australia}
\affiliation{ARC Centre of Excellence for Gravitational Wave Discovery (OzGrav), Australia}
\email{d.dobie@sydney.edu.au}
\author[0000-0001-6295-2881]{David L. Kaplan}
\affiliation{Department of Physics and Astronomy, University of Wisconsin-Milwaukee, P.O. Box 413, Milwaukee, WI 53201, USA}
\email{kaplan@uwm.edu}

\author[0000-0002-8935-9882]{Akash Anumarlapudi}
\affiliation{Department of Physics and Astronomy, University of North Carolina at Chapel Hill, 120 E. Cameron Ave., Chapel Hill, NC 27514, USA}
\email{akasha@unc.edu} 

\author[0000-0002-2686-438X]{Tara Murphy}
\affiliation{Sydney Institute for Astronomy, School of Physics, The University of Sydney, Sydney, New South Wales, Australia}
\affiliation{ARC Centre of Excellence for Gravitational Wave Discovery (OzGrav), Australia}
\email{tara.murphy@sydney.edu.au}

\author[0000-0003-1575-5249]{Joshua Pritchard}
\affiliation{Australia Telescope National Facility, CSIRO, Space \& Astronomy, PO Box 76, Epping, 1710 NSW, Australia}
\email{joshua.pritchard@csiro.au}

\author[0000-0002-4405-3273]{Laura N. Driessen}
\affiliation{Sydney Institute for Astronomy, School of Physics, The University of Sydney, Sydney, New South Wales, Australia}
\email{laura.driessen@sydney.edu.au}
\author[0000-0002-8977-1498]{Igor Andreoni}
\affiliation{Department of Physics and Astronomy, University of North Carolina at Chapel Hill, 120 E. Cameron Ave., Chapel Hill, NC 27514, USA}
\email{Igor.Andreoni@unc.edu}

\begin{abstract}
A few dozen previously radio-quiet active galactic nuclei (AGN) have been observed to transition to radio-loud at $1$--$3$~GHz frequencies over timescales of more than a decade, and this has has been interpreted to be due to newly launched jets. We identified 101 compact radio sources out of a sample of 64\,972 non-blazar AGN which increased in flux density by $80-1800\%$ over $1$--$6$ years in the $887.5$~MHz Australian SKA Pathfinder Variables and Slow Transients survey. We obtained optical spectra and radio SEDs using new observations and archival survey data. We determined 60 sources were consistent with extrinsic variability due to refractive interstellar scintillation and 41 were variable due to intrinsic causes, with 26 continuously brightening and two transitioning from radio-quiet to radio-loud. We concluded that young radio jets launched by either tidal disruption events or changes in the accretion properties were responsible for the continuously brightening AGN. These sources were non-variable at higher frequencies over the same time period, as expected for an expanding emission region. Fourteen sources had inverted or peaked SEDs initially which either flattened below the turnover or evolved into steep SEDs, consistent with young, expanding jets. Twelve sources had non-variable steep or gigahertz-peaked SEDs, which suggested these hosted more slowly evolving jets. We investigated the previously discovered AGN with newly launched jets, and found several have faded at multiple frequencies, which suggested the observed radio-loudness was temporary, rather than the onset of a sustained period of radio activity.
\end{abstract}


\section{Introduction}
While the existence of radio emission from synchrotron jets in some active galactic nuclei (AGN) has been well established for decades \citep[see, for example][]{Kellerman1969,Kellermann1983,Readhead1994}, only recently has evidence emerged that these jets may be able to form or undergo major luminosity changes on decade-long timescales \citep{Nyland2020,wolowska2021,Zhang2022}. A few dozen previously radio-quiet AGN have been observed to brighten considerably across 2--3 epochs spread over 20 years \citep{Nyland2020, wolowska2021, Zhang2022}, and the AGN which continued to be monitored exhibited stable emission on day--month timescales \citep{wolowska2021}.  Multi-frequency radio follow-up observations of some of these AGN revealed they had peaked radio spectral energy distributions (SEDs), which are associated with jet activity in compact, non-blazar AGN \citep{Nyland2020,wolowska2021}. At higher frequencies, an AGN with an off-axis jet has a characteristic synchrotron power-law SED, $S_\nu\propto\nu^\alpha$ and, at lower frequencies (below the turnover frequency, $\nu_p$), the jet becomes optically thick due to synchrotron self-absorption (SSA), or, in especially dense environments, free-free absorption \citep[FFA;][]{Kellerman1969, Ballieux2024, Odea2021}. The newly radio-loud peaked-spectrum AGN have been interpreted as either recently launched jets, with the low-frequency absorption dominated by SSA, or newly visible jets which were previously suppressed by the local interstellar medium, with the absorption caused by FFA \citep{Nyland2020,wolowska2021}. 

These results challenge our current understanding of AGN evolution, but further investigation is needed to determine the underlying physics of the changing jet activity. The newly radio-loud AGN were not detected between 1993--2009 in the Faint Images of the Radio Sky at Twenty-Centimeters \citep[FIRST;][]{Becker1995} survey, which had a sensitivity of $S_{1.4\text{GHz}}=1$~mJy~beam\textsuperscript{$-1$}. They were detected in the late 2010s with $S_{3\text{GHz}}>5$~mJy~beam\textsuperscript{$-1$} in the first $2$ epochs of the Very Large Array (VLA) Sky Survey \citep[VLASS;][]{Lacy2020} \citep{Nyland2020,Zhang2022} or the Caltech-NRAO Stripe 82 Survey \citep{wolowska2021}. However, with sparse, $3$-epoch light curves, it is challenging to differentiate jet activity from other periodic variability, transient events, or propagation effects which occur over much shorter timescales. Additionally, these AGN were observed at different frequencies in the early and late epochs, which creates an ambiguity between spectral and time variability.

The search for and characterization of rare events in AGN, such as the birth of a jet, has historically been challenging due to the landscape of existing radio surveys. Most time-domain radio studies of AGN have been extremely limited either in sensitivity, cadence, or field of view. However, this is changing with modern and upcoming radio surveys \citep{Murphy2026}, such as the $887.5$~MHz Australia SKA Pathfinder \citep[ASKAP;][]{Hotan2021} survey for Variables and Slow Transients \citep[VAST;][]{Murphy2013}. VAST extragalactic observations \citep{deruiter2026} cover $10,000$~deg\textsuperscript{$2$} of the sky with a two month cadence to a depth of $\sim0.24$~mJy~beam\textsuperscript{$-1$}, which makes it an ideal survey for investigating radio variability in AGN. The large footprint increases the likelihood of capturing rare events, such as jet launching in AGN \citep{Murphy2013}. Identifying a sample of variable AGN distributed broadly across the sky minimizes the risk that the variability is due to a single distorting plasma structure along our line of sight \citep{Murphy2013,Marchili2025}. The relatively high cadence and depth enable us to detect and monitor variability in even very faint or distant AGN. While the spatial resolution of VAST ($10\arcsec$) is limited compared to other wide field surveys, this is less important for isolated, extremely compact radio AGN which are unlikely to be resolved even in higher resolution surveys. The duration of the VAST survey is ideal for understanding long-term variability in AGN. VAST observations commenced in December 2022 and are slated to finish in 2027. VAST was preceded by a smaller pilot survey \citep{Murphy2021} which ran from 2019--2020, yielding light curves that are up to 7 years long.

The goal of this work was to use ASKAP VAST to identify AGN with new radio jets, trace their evolution, and study their radio SEDs to understand the physical reason for the variability.

We assumed a standard $\Lambda$CDM cosmology, with $H_0=67.4$km~s\textsuperscript{$-1$}Mpc\textsuperscript{$-1$}, $\Omega_{m0}=0.315$, and $\Omega_{\lambda 0}=0.685$ throughout \citep{cosmology}.
\section{Candidate Selection}
Our goal was to identify the most promising candidates for newly launched radio jets in AGN based on their VAST light curves. While we expected the radio emission in our candidates to be confined to a small emitting region and therefore unresolved at the $10\arcsec$ VAST resolution, we lacked constraints on the expected rate and magnitude of the increase. Our strategy was to broadly select AGN with compact radio emission which had increased in flux density over the survey lifetime. 
\subsection{Initial AGN Catalog}
We first constructed a catalog of AGN visible to ASKAP ($\delta\leq+30^\circ$). Our sample included 390\,413 spectroscopically identified AGN from the Sloan Digital Sky Survey $17$\textsuperscript{th} Data Release \citep{Abdurrouf2022} and 3\,555\,715 Wide Infrared Sky Explorer (WISE) W2--W1 color selected AGN candidates in the reliability-optimized R90 WISE AGN catalog \citep{Assef2018}. This initial catalog included blazars, which are highly radio-variable due to relativistic beaming of their synchrotron jets \citep{Readhead1994,Lister2001}. To avoid mistaking a flaring blazar for a young, off-axis jet, we removed all sources within $1.5\arcsec$ of any blazar candidates presented in the RomaBZ catalog \citep{Massaro2015} or the WISE Candidate $\gamma$-ray Blazar catalog \citep{dabrusco2014}. These candidates were selected based on their optical spectra, X-ray luminosity, and spectral index between $0.8$--$1.4$~GHz. After removing blazar contaminants, our catalog contained 389\,441 optically-selected AGN and 3\,551\,261 infrared color-selected AGN. To avoid double counting any AGN identified in both the optical and infrared, given the broad WISE point-spread function, we removed any infrared AGN candidates within $1.5\arcsec$ of an optical AGN. This yielded a final catalog of 3\,793\,162 unique AGN.
\subsection{Initial VAST catalog}
For our initial radio source list, we used the VAST pilot survey \citep{Murphy2021} and the VAST extragalactic first data release \citep{deruiter2026}. These catalogs were produced with the VAST pipeline \citep{Pintaldi2021,Stewart2024}, which ingests images, along with auxillary maps and source lists generated with the ASKAP source finder \citep[{\fontfamily{qcr}\selectfont
Selavy};][]{Whiting2012}. The pipeline carries out source association to link sources across epochs and generate light curves. The VAST pipeline also carries out forced photometry for any images containing the position of a source in which the source was not detected by {\fontfamily{qcr}\selectfont
Selavy}. We combined the VAST pilot and VAST extragalactic source catalogs, which contained 944\,091 and 2\,279\,118 objects, respectively.

For complex galaxies with multiple resolved components, sources with transient, extended radio flares, or especially bright objects, {\fontfamily{qcr}\selectfont Selavy} can sometimes split a single, continuous emitting region into multiple sources. In these cases, the source association between images breaks down and results in light curves with artificial variability. We conservatively selected only sources with surjective mappings across epochs (2\,079\,853 VAST extragalactic sources and 836\,305 VAST pilot sources) and zero epochs with two or more sources fit to the same emitting region (1\,299\,626 VAST extragalactic sources and 487\,886 VAST pilot sources).

Since we were interested solely in AGN with compact, unresolved radio emission, we removed any sources with an average compactness $\langle S_{\rm int}/S_{\rm peak}\rangle\geq 1.4$, where $S_{\rm int}$ is the flux density integrated across the entire emitting region and $S_{\rm peak}$ is the flux density integrated over the primary beam. For a perfect point source within the survey resolution, this ratio would be unity. The VAST pipeline extracts both $S_{\rm peak}$ and $S_{\rm int}$ only for {\fontfamily{qcr}\selectfont
Selavy} detections, so we did not include epochs with forced photometry measurements (for which $S_{\rm peak}=S_{\rm int}$) in calculating the average compactness.
This cut left 1\,068\,139 sources in the VAST extragalactic catalog and 390\,568 sources in the VAST pilot catalog.
\subsection{VAST-AGN Cross-match}
We cross-matched our AGN catalog with both VAST source catalogs, using a $2.5\arcsec$ radius. There were 57\,631 AGN with counterparts in the VAST extragalactic catalog and 21\,417 with counterparts in the VAST pilot catalog. A few dozen AGN were cross-matched to multiple VAST sources in the same catalog, and we removed these entirely to prevent source confusion. This left us with 21\,405 and 57\,583 sources in the VAST pilot and VAST extragalactic catalogs, respectively. We merged the catalogs by assuming any source in the VAST pilot catalog matched to the same AGN as a VAST extragalactic source were the same astrophysical object. This left us with a catalog of 64\,792 unique radio AGN.
\subsection{Light curve morphology selection}
We wanted to select the most promising newly launched jet candidates for follow-up investigation based on their VAST light curves. However, we were probing a timescale and frequency of AGN variability which has been relatively unexplored, so we lacked strong constraints on the light curve morphology we expected to observe for a young jet. The AGN hosting young radio jets discovered by \citet{Nyland2020}, \citet{wolowska2021}, and \citet{Zhang2022} were identified at much higher frequencies ($>1$~GHz), and were found to be gigahertz-peaked spectrum sources. These AGN should be optically thick at the VAST frequency, so we expected to find a different population of radio variable AGN. 
As a further complication, our light curves varied in length from less than a month to six years, and in number of epochs from $1$--$100$. Given these challenges, we chose our light curve morphology criteria to broadly select the sources with the most drastically increasing flux density, while avoiding any with false variability introduced by pipeline or observational artifacts. A summary of our criteria is given in Table~\ref{tab:lightcurvecuts}.
For our light curve selection and throughout this work, we used $S_\nu=S_{\text{peak}}$, rather than the integrated flux density. These values should be roughly equal for compact sources, but the integrated flux density measurements may have been compromised by the VAST pipeline mistaking artifacts as extended structure. 
Before performing any morphology-based cuts, we removed all VAST epochs with greater than $15\%$ error in their flux density measurements. This was to avoid selecting sources appearing to be variable due to epochs with poor data quality or those with variability that can be explained by large uncertainties in their flux densities.

We next removed sources with fewer than three available epochs, including forced photometry measurements, which reduced our sample size to 41\,607. To select the AGN with the most dramatic flux density changes, we excluded sources with $S_{\nu,\text{max}}<1.8S_{\nu,\text{min}}$, which left 3\,100 objects. Since we expected to see sustained radio emission from a jet, rather than a transient flare, we removed sources with $S_{\nu,\text{final}}\leq1.5S_{\nu,\text{initial}}$ (276 remaining) and $S_{\nu,\text{final}}\leq0.7S_{\nu,\text{max}}$ (260 remaining). Next, we removed sources for which the brightest epoch occurred earlier than the faintest epoch, which left 242 sources. Our next two cuts applied only to sources detected in both VAST catalogs: we excluded sources which reached their maximum flux density during the pilot survey (2019--2020) or their minimum flux density during the extragalactic survey (2023--2026), yielding a sample size of 234 and 226, respectively.

\setcounter{figure}{0}
\begin{figure*}[t!]
    \centering
    \includegraphics[width=0.95\linewidth]{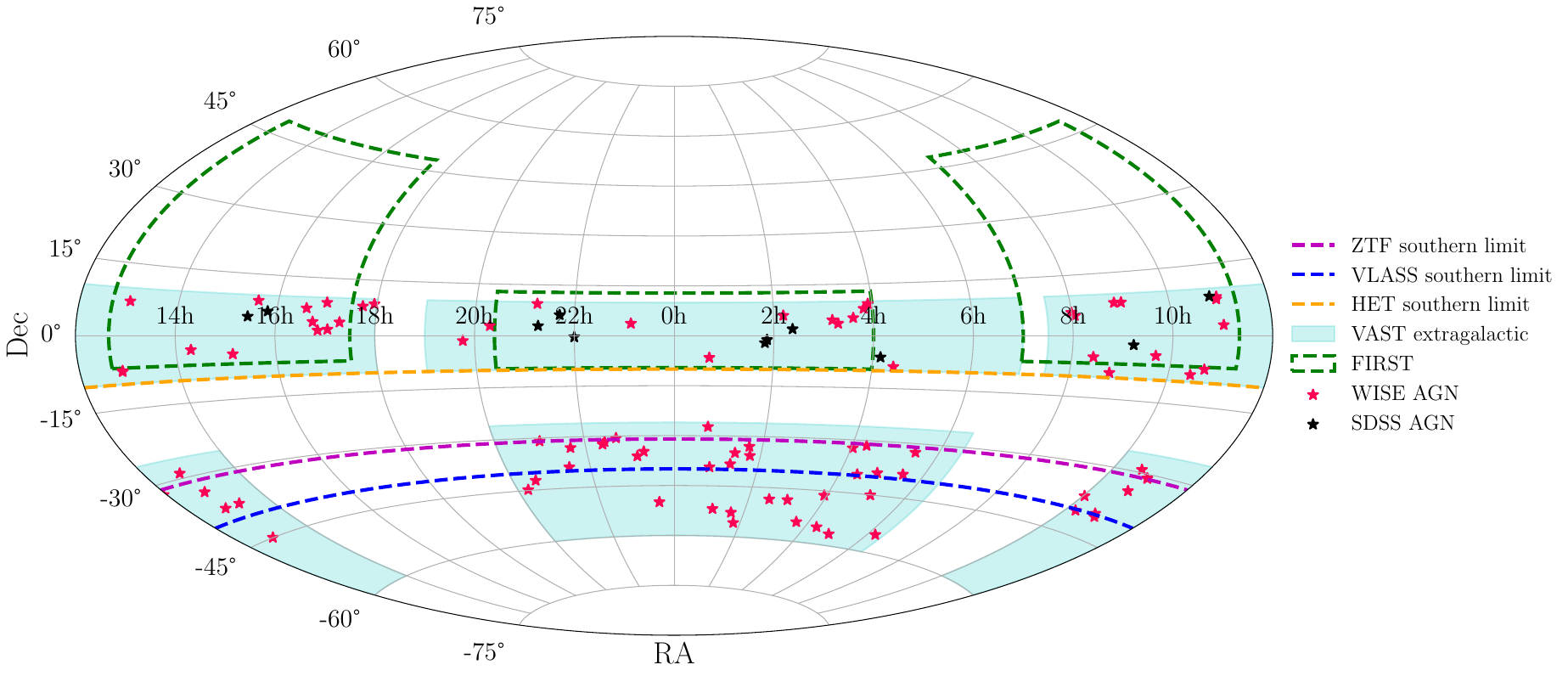}
    \caption{Aitoff projection of the 101 AGN in our sample. Black stars represent the optically selected AGN and red stars represent the infrared selected AGN. The VAST extragalactic fields are shown in cyan. The FIRST footprint is indicated by green dashed lines, and the Southern declination limits of ZTF, VLASS, and the HET are denoted by purple, blue, and orange dashed lines, respectively. \label{fig:candidate-map}}
\end{figure*}
\begin{deluxetable*}{lrr}[hbpt!]
    \tablecaption{Summary of our light curve morphology selection cuts. Note that since we visually inspected each epoch during the final sample cuts, we did not need to remove epochs with $>15\%$ error as a data quality check.}
    \label{tab:lightcurvecuts}
    \tablehead{\colhead{Criteria} & \colhead{Initial Sample} & \colhead{Final Sample}}
\startdata
    Total radio AGN & 64\,792 & 143 \\ 
    After removing all epochs with $\sigma_{S_{\nu}} >0.15S_{\nu}$  & 53\,385 & ... \\ 
    There are $\geq3$ epochs & 41\,607 & 141 \\
    $S_{\nu,\text{max}}\geq 1.8S_{\nu,\text{min}}$ & 3100 & 135\\
    $S_{\nu,\text{final}}>1.5S_{\nu,\text{initial}}$ & 276 & 107\\
    $S_{\nu,\text{final}}\geq 0.7S_{\nu,\text{max}}$ & 260 & 107\\
    $S_{\nu,\text{min}}$ occurs before $S_{\nu,\text{max}}$ & 242 & 107\\
    If the light curve spans both VAST surveys, the peak occurs in the full survey & 234 & 107\\
    If the light curve spans both VAST surveys, the minimum occurs in the pilot survey & 226 & 107\\
    $S_{\nu,\text{final}}\geq S_{\nu,\text{mean}}$ & 221 & 106 \\
    After $S_{\nu,\text{max}}$, $S_{\nu}>0.5S_{\nu,\text{max}}$ & 218 & 106\\
    Before $S_{\nu,\text{max}}$, $S_{\nu}$ does not decrease by more than $10\%$ & 207 & 104\\
    After $S_{\nu,\text{max}}$, $S_{\nu}$ does not decrease by more than $0.3(S_{\nu,\text{max}}-S_{\nu,\text{min}})$ & 194 & 103\\
    After $S_{\nu,\text{max}}$, $S_{\nu}$ does not decrease by more than $10\%$ & 188 & 103\\
    Before $S_{\nu,\text{peak}}$, $S_{\nu}$ does not decrease by more than $0.3(S_{\nu,\text{max}}-S_{\nu,\text{min}})$ & 178 & 102 \\
    If there are two measurements less than two days apart, they are in agreement & 143 & 101 \\
\enddata
\end{deluxetable*}

To avoid transient sources which may have faded, we required $S_{\nu,\text{final}}>S_{\nu,\text{average}}$ (221) and $S_{\nu,\text{initial}}>0.5S_{\nu,\text{max}}$ for all epochs occurring after the brightest epoch (218). Next, we removed sources with variability more likely to be stochastic or periodic; prior to the brightest epoch, if any local maxima in the light curve were followed by a decrease in flux density of more than 90$\%$ (207) or within 30$\%$ of the difference between the global extrema across the light curve (194), we removed the source. We then repeated this cut for the section of the light curve following the brightest epoch, which left sample sizes of 188 and 178, respectively. Lastly, we removed any sources with two flux density measurements taken within the same 24 hour period which were not within the error bounds of one another, since intrinsic AGN variability does not occur over such small timescales \citep{Kankkunen2025II,Nyland2020}. This left 143 sources in our sample.
\begin{figure*}[hptb!]
    \gridline{\fig{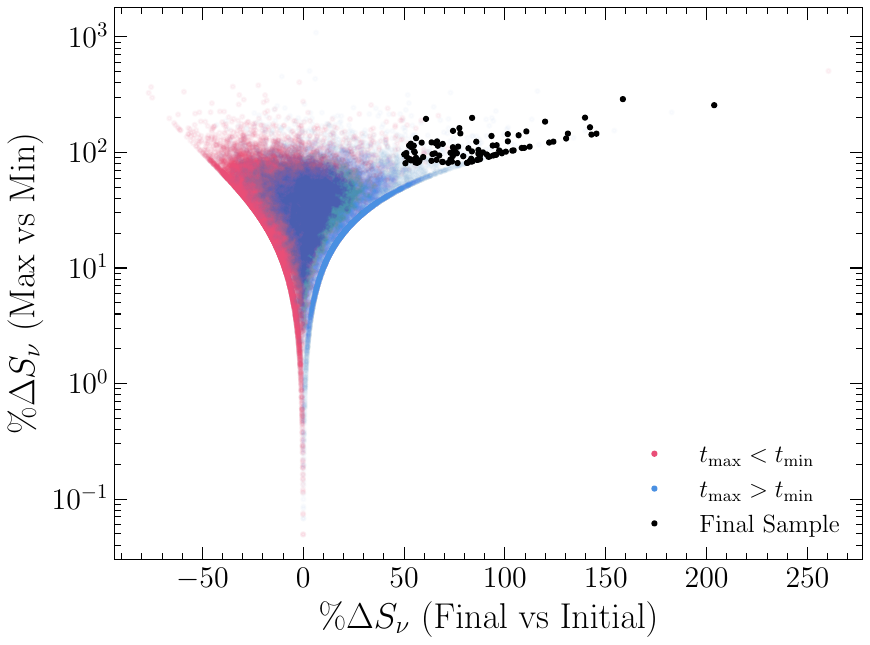}{0.48\textwidth}{}\fig{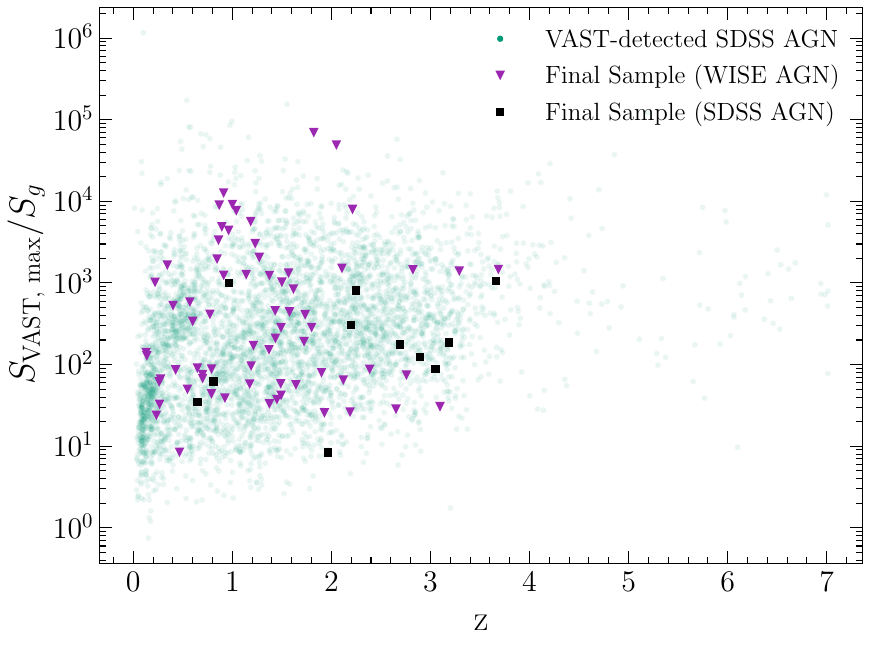}{0.48\textwidth}{}}
    \caption{Comparison of the initial catalog of VAST-detected AGN and our final sample. Left: difference in flux density between the brightest and faintest epochs compared to the difference between the first and last epoch, two parameters used in our light curve morphology cuts. Pink markers represent sources in the initial sample which reached $S_{\nu,\text{max}}$ before $S_{\nu,\text{min}}$, and blue markers represent sources in the initial sample which reached $S_{\nu,\text{max}}$ after $S_{\nu,\text{min}}$. Black markers represent the sources in our final sample. Right: radio-loudness compared to redshift. Cyan circles represent sources in the initial VAST-detected SDSS AGN catalog. The infrared-color detected AGN in the initial VAST catalog are excluded here since redshift and optical magnitudes were not available. Black squares represent the optically selected AGN in our final sample, and purple triangles represent the infrared selected AGN in our final sample for which we obtained redshifts and optical magnitudes.}
    \label{fig:agn_dist}
\end{figure*}
\subsection{Final Sample Selection}
For the 143 radio AGN meeting our light curve morphology criteria, we visually inspected all images generated by the VAST pipeline. We found 28 sources to have at least one unreliable epoch: in five sources, the VAST pipeline had identified an additional source in some epochs and failed to flag it; one source was centered on the edge of a field; in the other 22, the source was too far from the center of the primary beam for the flux density measurement to be considered reliable. Examples of epochs we removed are shown in Figure~\ref{fig:bad_epochs} in Appendix~\ref{appx:bad_epochs}. In total, we removed 53 epochs.

After removing the unreliable epochs, we re-ran our light curve morphology cuts on our source sample, including the epochs with $>15\%$ error in their flux density measurements, which we had excluded initially. This yielded a final sample size of 101 sources. The number of sources meeting each criteria is shown in the right-most column in Table~\ref{tab:lightcurvecuts}. Figure~\ref{fig:candidate-map} shows the distribution of our sources in the VAST footprint. Figure~\ref{fig:agn_dist} shows how our final sample compared to the original radio-AGN catalog.
\section{Results}
\begin{figure}[htpb!]
    \gridline{\fig{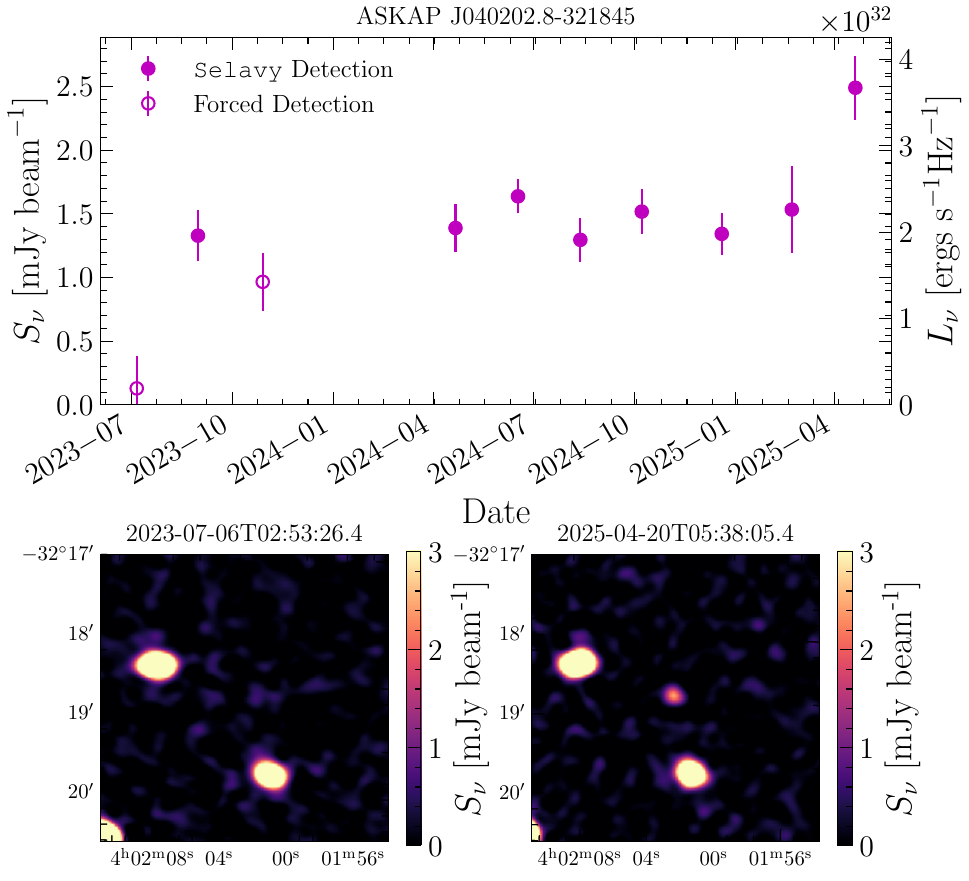}{0.47\textwidth}{}}
    \caption{ASKAP J040202.8$-$321845, one source which met our criteria for variable AGN in VAST. The left panel shows the first VAST epoch, the middle shows the final VAST epoch, and the right panel shows the corresponding light curve. Solid markers denote {\fontfamily{qcr}\selectfont Selavy} detections and unfilled markers denote forced photometry measurements. \label{fig:radio_variable_agn}}
\end{figure}

Table~\ref{tab:candidates} shows an overview of the AGN in our sample, including their flux density increases, minimum and maximum radio-loudness, redshifts, and maximum luminosities. Our sources exhibited flux density increases between $81-1830\%$ between the faintest and brightest epochs. The light curve for one source is shown with its first and last epoch images in Figure~\ref{fig:radio_variable_agn}. The VAST light curves of all sources meeting our selection criteria for variable AGN are shown in Appendix~\ref{appx:vast_light_curves}.
\subsection{Archival Optical Observations}
We obtained archival photometric and spectroscopic observations for our sources, which are outlined in the following sections.
\subsubsection{Optical Magnitudes}
To determine the radio-loudness of our sources, we acquired g-band magnitudes for 92 sources from the following catalogs: the 16\textsuperscript{th} SDSS-IV data release \citep[SDSS IV;][]{Ahumada2020}, the Dark Energy Spectroscopic Instrument \citep[DESI;][]{Desi2026} Legacy Imaging surveys \citep{Dey2019}, and the SkyMapper Southern Sky survey \citep[SMSS;][]{Onken2024}. ASKAP J201813.9$+$025307 was not included in these catalogs, so we obtained its G-band magnitude from the 3\textsuperscript{rd} Gaia data release \citep[Gaia;][]{Vallenari2023}. Three sources, ASKAP J112810.8$-$305103, ASKAP J210007.1$-$305617, and ASKAP J222325.2$-$322403 were too faint to be detected in the DESI Legacy Imaging surveys, so we assumed a lower limit $g=24.0$, which was the minimum depth for the surveys \citep{Dey2019}. ASKAP J115600.0$-$315504 was not detected in the Legacy Imaging surveys in the g-band, so we used the r-band magnitude instead. ASKAP J171659.5$+$033419, ASKAP J174307.9$+$075614, and ASKAP J194551.5$-$012544 were non-detections in the g-band Zwicky Transient Facility \citep[ZTF;][]{Bellm2019} Forced Photometry Service \citep{masci2023}, so we assumed a lower limit $g=21.5$. We were unable to find optical magnitudes or limits for two sources, ASKAP J101432.5$-$401536 and ASKAP J125050.8$-$445757. For the remaining 99, we determined the maximum radio-loudness, $S_{\nu,\text{max}}/S_{g}$, which ranged from $8$--$71,000$. The optical measurements were not necessarily contemporaneous with our VAST measurements.
\subsubsection{Cataloged Redshifts}
We obtained spectroscopic redshifts for 44 sources from the following catalogs: SDSS, DESI, the Gaia extragalactic catalog \citep{BailerJones2023}, and the Calán-Tololo survey Version VI   \citep{Maza1995}. For 28 sources with no archival spectroscopic redshift data, we acquired photometric redshifts from SDSS IV, the DESI photometric redshifts catalog \citep{duncan2022}, Version 2 of the Millions of Optical-Radio/X-ray Associations catalog \citep[MORX;][]{Flesch2024}, and the Kilo Degree survey catalog \citep[KiDS;][]{wright2024}. For each source with an available redshift, we determined the maximum radio luminosity, assuming non-beamed, isotropic emission using the maximum flux density. 
\subsubsection{Optical Light Curves}
\begin{figure*}[t!]
    \centering
    \gridline{\fig{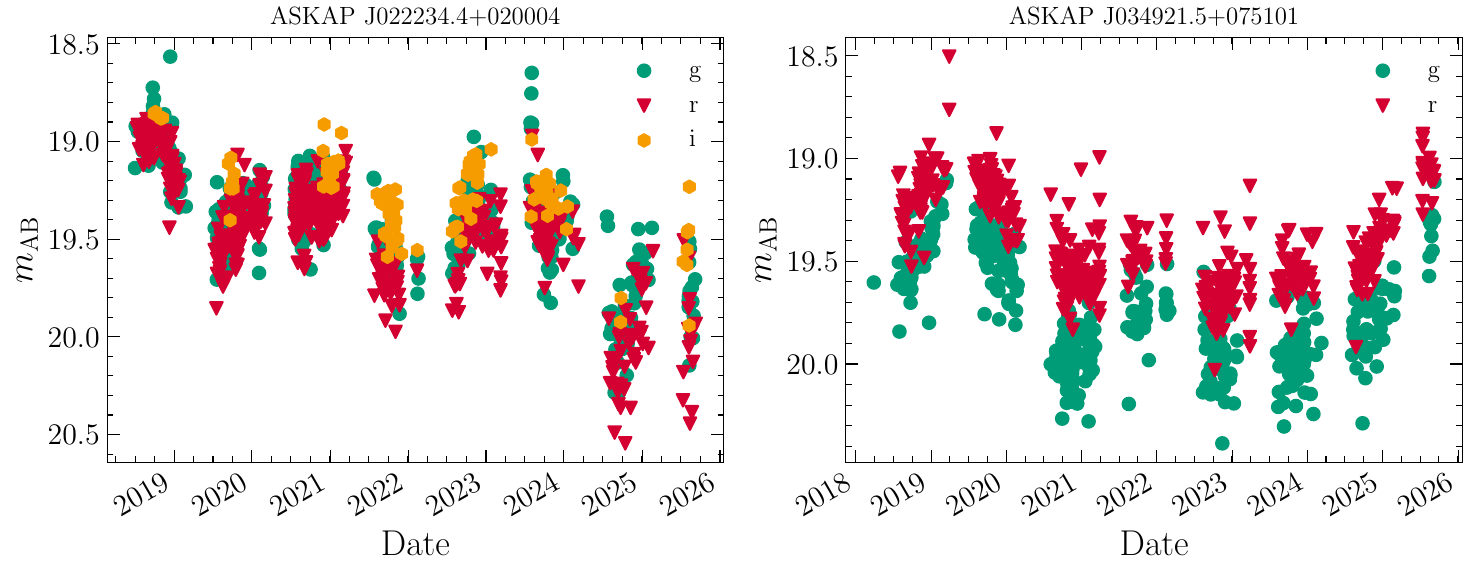}{0.98\textwidth}{}}
    \caption{ZTF forced-photometry light curves showing typical AGN-like variability. Orange hexagons indicate $i$-band measurements, red triangles indicate $r$-band measurements, and green circles indicate $g$-band measurements. No TDE or SNe signatures were present in any optical light curves.}
    \label{fig:ztf}
\end{figure*}

To investigate whether the increases in radio emission were accompanied by an optical counterpart, we obtained ZTF $g$, $r$, and $i$-band light curves, which were contemporaneous with VAST, using the Forced Photometry Service. Forced photometry was available for 52 sources ($\delta>-30^\circ$), but 17 of these were too faint ($g>22$) to be detected by ZTF. We obtained light curves for the remaining 35.  For two sources with no cataloged $g$-band magnitudes, ASKAP J171659.5$+$033419 and ASKAP J194551.5$-$012544, there was no reference magnitude available, indicating that these sources are below the ZTF detection threshold. There were no forced difference flux density measurements available for ASKAP J015150.5$-$011219 ($g=18.9$), for unknown reasons. In the remaining light curves, we discarded measurements not meeting the recommendations outlined in \citet{masci2023}: {\fontfamily{qcr}\selectfont
infobitssci} $=0$, {\fontfamily{qcr}\selectfont
scisigpix} $<25$, {\fontfamily{qcr}\selectfont
sciinpseeing} $<4$, {\fontfamily{qcr}\selectfont
forceddiffimchisq} $<2 $. We removed any epochs with SNR$<3$. 
Five of the resulting light curves were too sparse to derive any meaningful information. In one case, ASKAP J174307.9$+$075614, which had no cataloged g-band magnitude, the light curve consisted of only a few $r\sim22$ measurements with SNR$<5$, indicating that this source was likely too faint to be detected with ZTF. All 27 remaining light curves showed slow variability consistent with a damped random walk, as expected for AGN \citep{He2026}. No clear optical flares, such as those observed in tidal disruption events \citep[TDEs;][]{Hammerstein2023}  and supernovae \citep[SNe;][]{He2026}, were present in any light curve. Two optical light curves are shown in Figure~\ref{fig:ztf}. 
\subsection{X-ray Associations}
To help differentiate between types of radio transients and confirm the AGN nature of our infrared selected sources, we cross-matched our sample with the MORX, the eROSITA All-Sky survey \citep{Merloni2024}, and the eROSITA Final Equatorial Depth survey \citep{Aydar2025} X-ray source catalogs. Eight of our source had an X-ray counterpart in at least one catalog: ASKAP J012404.4$-$350117, ASKAP J041916.2$-$312707, ASKAP J042550.4$-$084716, ASKAP J080329.2$+$045904, ASKAP J091307.4 $-$020936, ASKAP J141806.4$-$031153, ASKAP J150801.4$-$041936, and ASKAP J215954.4$-$002149.
\subsection{Archival radio observations}
We obtained multi-frequency radio observations from archival surveys for each source in our sample.
\subsubsection{Rapid ASKAP Continuum Survey}
The Rapid ASKAP Continuum Survey \citep[RACS;][]{McConnell2020} covered the entire sky visible to ASKAP with a sensitivity of $\sim0.2$~mJy~beam\textsuperscript{$-1$} and resolution of $\sim15\arcsec$ in three bands: $887.5$~MHz \citep[RACS-low;][]{Hale2021}, $1367.5$~MHz \citep[RACS-mid;][]{Duchesne2024}, and $1655.5$~MHz \citep[RACS-high;][]{Duchense2025}. RACS-low observations were carried out in 2019, RACS-mid in 2021, and RACS-high in 2021--2022. We cross-matched our sample with the RACS-high (Duchesne 2025, private communication), RACS-mid, and RACS-low time-domain source catalogs. These catalogs used the same source finder as VAST, {\fontfamily{qcr}\selectfont
Selavy}. We removed any detections flagged as spurious by the RACS pipeline. We found 69 of our sources had counterparts in the RACS-low catalog, 84 had counterparts in the RACS-mid catalog, and 91 had counterparts in the RACS-high catalog.
\subsubsection{FIRST}
FIRST was a $1.4$~GHz VLA survey with a resolution of $5\arcsec$ and an average detection threshold of $1$~mJy~beam\textsuperscript{$-1$} and was carried out between 1993--2011. We cross-matched the 30 sources in our sample within the FIRST footprint with the FIRST source catalog \citep{Becker1995}. Twenty-six sources were detected in FIRST, and four were not.
\subsubsection{VLASS}
VLASS is a $3$~GHz survey covering the entire sky above declination $-40^\circ$ with a $2.5\arcsec$ angular resolution and average detection threshold of $1$~mJy~beam\textsuperscript{$-1$}. We obtained 3--4 epoch light curves from the VLASS Cutout Service\footnote{\url{https://dillon-z-dong.github.io/vlass-image-browser/}} for the 83 sources in our sample in the VLASS footprint. We found 80 were detected with SNR~$>3$ in at least one epoch.
\subsubsection{LOFAR Two-meter Sky Survey}
The International Low Frequency Array (LOFAR) Two-Centimeter Sky Survey \citep[LoTSS;][]{Shimwell2017} is an ongoing $144$~MHz survey with a detection limit of $\sim0.45$~mJy~beam\textsuperscript{$-1$}. We cross-matched our sample with the $3$\textsuperscript{rd} LoTSS data release source catalog \citep{Shimwell2026}, which does not include observation dates. Twenty-two of our sources in the LoTSS footprint had counterparts in the LoTSS catalog, and seven did not.
\subsubsection{Non-detections in Archival Radio Surveys}
There were 34 sources which did not have a counterpart in the RACS-low catalog, 17 with no counterpart in RACS-mid, 10 with no counterpart in the RACS-high catalog, four which were in the FIRST footprint and had no counterpart in the FIRST source catalog, and seven which were in the LoTSS footprint and not listed in the LoTSS source catalog. To determine whether they were missed in the source catalogs or were non-detections, we obtained image cutouts centered on each source's position from the CSIRO ASKAP Science Data Archive (RACS), the FIRST Cutout Server\footnote{\url{https://sundog.stsci.edu/cgi-bin/firstcutout}} and the LOFAR Data Archive (LoTSS). We used the {\fontfamily{qcr}\selectfont
cgcurs} task with {\fontfamily{qcr}\selectfont options$=$imstat} in {\fontfamily{qcr}\selectfont
miriad} \citep{Sault1995} to determine the flux density at each source position and the RMS of each image. We used a forced detection cutoff of $S_{\nu}>3$RMS and found ten sources were non-detections in RACS-low, one was a non-detection in RACS-mid, three were non-detections in RACS-high, one was a non-detection in FIRST, and one was a non-detection in LoTSS. Images of non-detections in RACS and LoTSS are shown in Figure~\ref{fig:forced_photometry} in Appendix~\ref{appx:forced_photometry}.
\subsection{Radio SEDs and Multi-Frequency Light Curves}
\begin{figure*}[thpb!]
    \centering
    \gridline{\fig{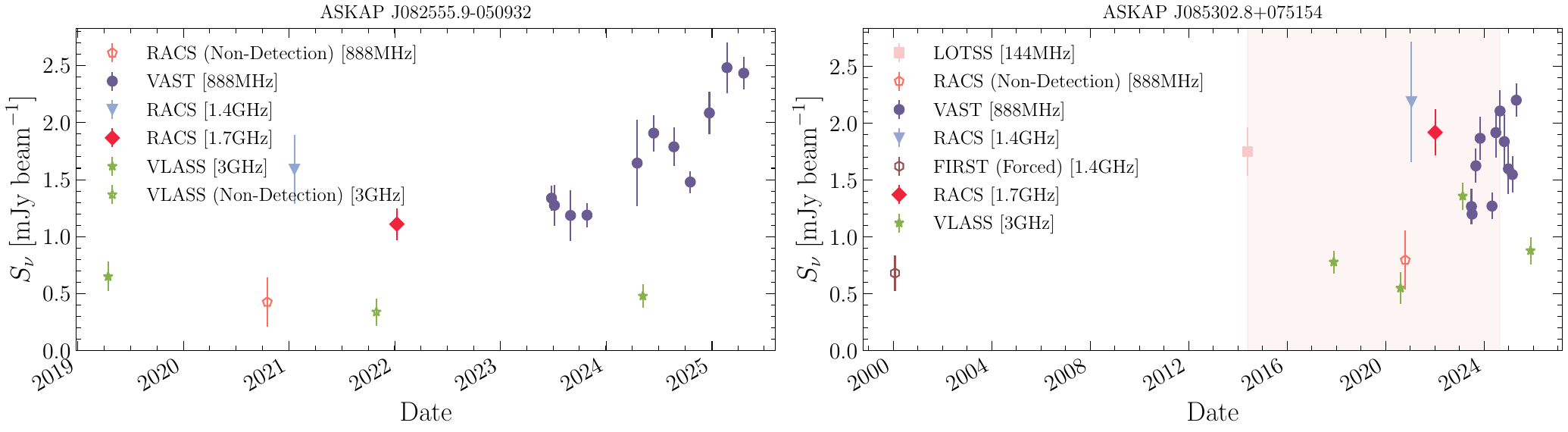}{0.95\textwidth}{}}
    \caption{Multi-frequency radio light curves of two VAST sources constructed using archival survey data. Brown hexagons indicate FIRST measurements, green stars indicate VLASS measurements, purple circles indicate VAST measurements, orange pentagons indicate RACS-low measurements, blue triangles indicate RACS-mid measurements, red diamonds indicate RACS-high measurements, and pink squares indicate LoTSS measurements. Solid markers indicate catalog detections and unfilled markers indicate non-detections or forced detections. The observation dates of LoTSS were unknown, so the survey time frame is represented by the pink region. \label{fig:long_lightcurves}} 
\end{figure*}

Using the archival radio data, we created multi-frequency light curves. Two are shown in Figure~\ref{fig:long_lightcurves}. 
\begin{figure*}[htbp!]
    \gridline{\fig{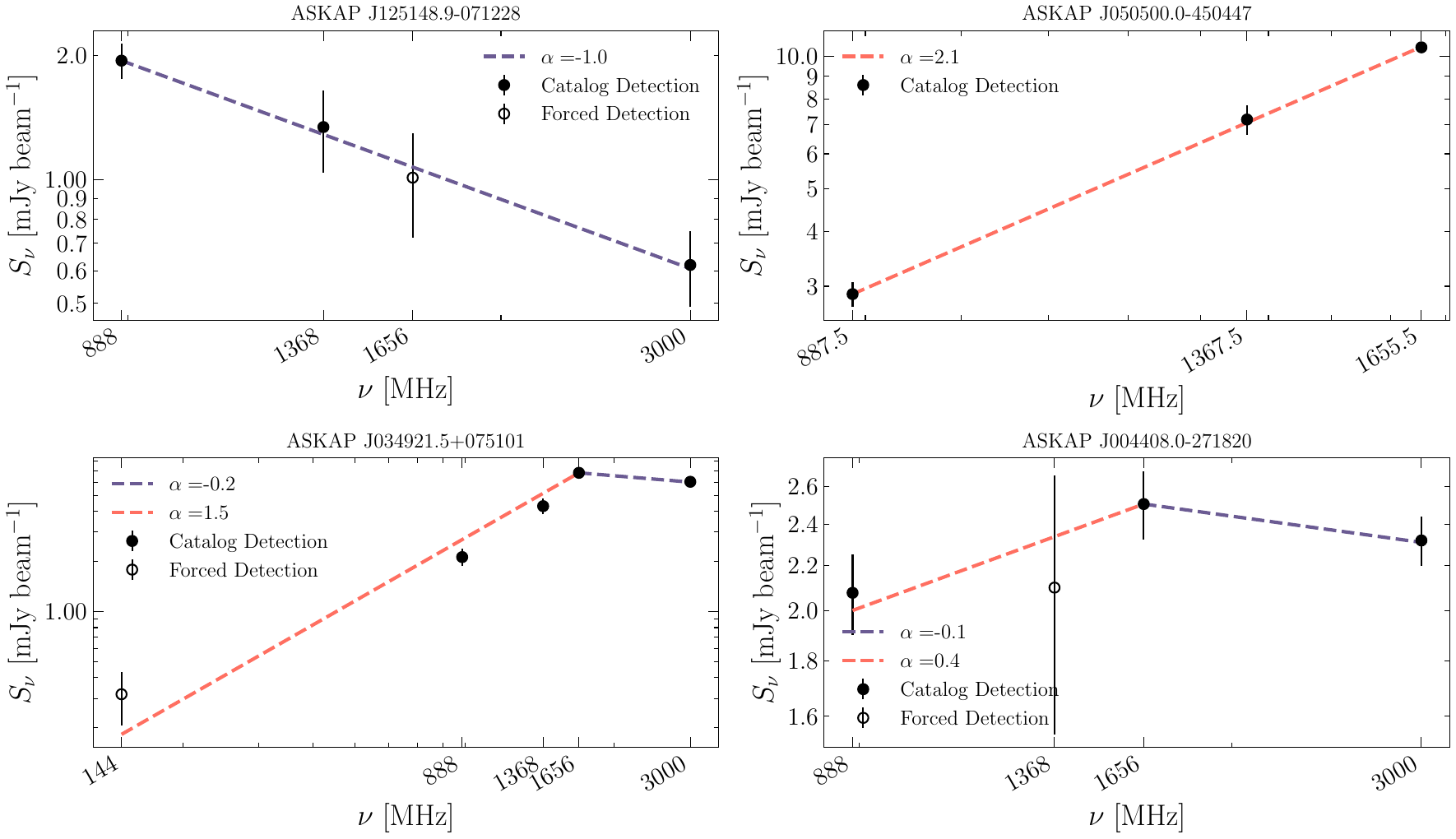}{0.98\textwidth}{}}
    \caption{Non-contemporaneous radio SEDs and power-law spectral fits for the four spectral classifications. In clockwise order from the top left: steep, inverted, peaked, and (peaked) flat SEDs. Solid black markers indicate catalog detections and unfilled black markers indicate forced photometry measurements. Purple dashed lines indicate steep fits ($\alpha<0$) and orange dashed lines indicate inverted fits ($\alpha>0$). Steep and inverted spectra were classified as flat if $|\alpha|<0.5$, and peaked spectra were classified as flat if $|\alpha|<0.5$ on both sides of the peak. \label{fig:seds}}
\end{figure*}
We created radio SEDs with all available survey data, using the median VLASS and median VAST flux densities. We classified each SED as steep ($S_{\nu,\text{max}}=S_{\nu_{\text{low}}}$), inverted ($S_{\nu,\text{max}}=S_{\nu_{\text{high}}}$), or peaked ($\nu_{\text{low}}<\nu_p< \nu_{\text{high}}$). We fit a power law SED ($S_\nu\propto\nu^{\alpha}$), with $-4\leq\alpha\leq4$, using $\chi^2$ minimization. For peaked SEDs, we fit a power law to each side of the turnover. We classified any steep or inverted SEDs with $|\alpha|\leq0.5$ and any peaked SEDs with $|\alpha|\leq0.5$ on both sides as flat. An example of a steep, inverted, peaked, and flat SED are shown in Figure~\ref{fig:seds}. We determined 10 SEDs were inverted, with $\alpha_{\text{inv}}$ ranging from $0.5$ to $2.1$; 25 were steep, with $\alpha_{\text{steep}}$ ranging from $-1.9$ to $-0.5$; 15 were flat, with $\alpha$ ranging from $-0.4$ to $0.3$; and 51 were peaked, with $\alpha_{\text{inv}}$ ranging from $0.1$ to $2.1$ and $\alpha_{\text{steep}}$ ranging from $-4.0$ to $0.0$. Inverted and peaked SEDs in radio AGN have been interpreted as the early phases of jet activity, while steep SEDs are thought to be more evolved jets \citep{Odea2021, Ballieux2024}. In peaked spectrum sources, $\nu_p$ is correlated with the linear size of the emitting region: high-frequency peaked sources ($\nu_p\geq5$~GHz) are the most compact, with linear sizes $\leq500$~pc; gigahertz-peaked sources $\nu_p\sim1$--$5$~GHz have linear sizes $<1$~kpc; and sources with $\nu_p<1$~GHz, compact steep-spectrum sources and megahertz-peaked sources, have the largest linear sizes ($1-20$~kpc) \citep{Ballieux2024}. Flat SEDs are associated with highly beamed jets in radio AGN \citep{degouveia2005}. We used our spectral index fits to extrapolate the flux density at $5$~GHz, which is the canonical radio frequency used to define radio-loudness in AGN \citep[$S_{5\text{GHz}}/S_g\geq10$;][]{Kellerman1989}. We determined the minimum and maximum radio-loudness ratios of each source using the extrapolated $S_{5\text{GHz}}=S_{887.5\text{MHz}}(5\text{GHz}/887.5\text{MHz})^{\alpha}$, which are given in Table~\ref{tab:candidates}. We determined ten were radio-quiet, 75 were radio-loud, and 14 were initially radio-quiet ($S_{5\text{GHz}}/S_g<10$) but transitioned to radio-loud. The other two sources had no optical reference. Eight radio-quiet sources had peaked SEDs ($\nu_p=1367.5$~MHz) and two had steep SEDs ($\alpha_{\text{steep}}=-1.0$--$-0.7$). The sources which had transitioned from radio-quiet to radio-loud included eight sources with peaked SEDs ($\nu_p=887.5$--$1655.5$~MHz), two with flat SEDs, and three with steep SEDs ($\alpha_{\text{steep}}=-0.9$--$-0.5$). Thirty-four radio-loud sources had peaked SEDs ($\nu_p=887.5$--$1655.5$~MHz), 12 had flat SEDs, and 20 had steep SEDs ($\alpha_{\text{steep}}=-1.9$--$-0.5$). All ten sources with inverted SEDs were radio-loud. These SEDs were non-contemporaneous, so our spectral classifications were estimates, and would be improved with dedicated radio follow-up observations.
\subsubsection{Spectral Index Evolution}
\begin{sidewaysfigure*}[htbp!]
    \caption{Spectral evolution of four sources in our sample across the RACS-low/VAST light curves. In clockwise order starting from the top left: an inverted spectrum which flattened, a peaked spectrum which flattened below the turnover, a peaked spectrum which did not change over time, and a peaked spectrum which transitioned into a steep spectrum. Orange stars indicate spectral indices for inverted SEDs, purple circles indicate spectral indices for steep SEDs, solid black hexagons indicate inverted spectral indices for peaked spectrum sources, and unfilled black hexagons indicate steep spectral indices for peaked spectrum sources. Orange dashed lines and purple dashed lines indicate the inverted and steep spectral indices for the median VAST epoch. Dark gray regions indicate steep indices ($\alpha\leq-0.5$), medium gray regions indicate flat indices ($|\alpha|<0.5$), and light gray regions indicate inverted indices ($\alpha\geq0.5$). \label{fig:spec_idx}}
    \gridline{\fig{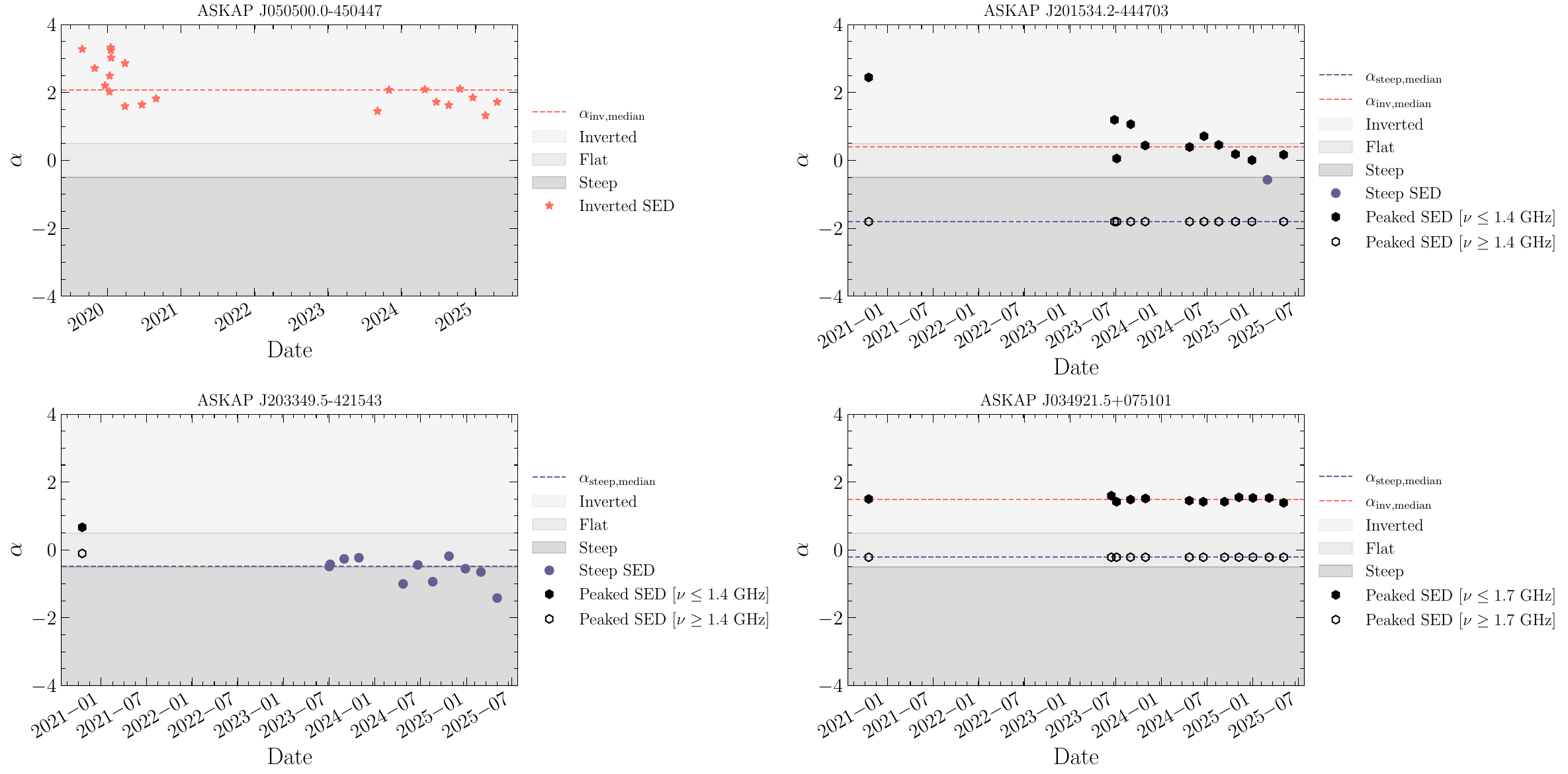}{0.98\textwidth}{}}
\end{sidewaysfigure*}

We repeated our SED fits for each VAST/RACS-low epoch to investigate the spectral evolution with time. We kept the $3$~GHz flux density fixed ($S_{3\text{GHz}}=S_{\text{VLASS, median}}$) for each epoch. We determined the spectral type (inverted, steep, or flat) and spectral indices for each epoch. Our sources exhibited one of two behaviors: in 79 source, the spectral indices oscillated around a constant value, and in 22, they decreased over time. Six inverted sources flattened over time, five peaked spectrum sources flattened below the turnover, and 11 peaked spectrum sources transitioned into steep spectrum sources. The spectral evolution of a constant spectrum source, a flattening inverted spectrum source, a peaked spectrum source flattening below the turnover, and a peaked spectrum source transitioning to a steep spectrum source are shown in Figure~\ref{fig:spec_idx}.

\begin{longdeluxetable}{llllllll}
    \tablecaption{Properties of the radio-brightening in our sample. References for archival $g$-band magnitudes used to calculate $S_{5\text{GHz}}/S_g$ are denoted as follows: SDSS\textsuperscript{s}, DESI Legacy surveys\textsuperscript{l}, SMSS\textsuperscript{m}, and Gaia\textsuperscript{g}. Non-detections in the DESI Legacy Surveys and ZTF are denoted with *\textsuperscript{l} and *\textsuperscript{z}*. References for archival spectroscopic redshift measurements are denoted by: SDSS\textsuperscript{s}, DESI\textsuperscript{d}, the Gaia extragalactic catalog\textsuperscript{g}, and the Calán-Tololo survey\textsuperscript{c}. Spectroscopic redshifts we obtained with dedicated HET and SOAR follow-up are denoted by z\textsuperscript{h} and z\textsuperscript{so}, respectively. Where spectroscopic redshifts were not available, photometric redshifts are indicated by z* with references given as: SDSS\textsuperscript{s*}, DESI\textsuperscript{d*}, MORX\textsuperscript{mx*}, and KiDS\textsuperscript{k*}. Maximum radio luminosity is given as an upper limit, assuming non-beamed, isotropic emission.\label{tab:candidates}}
    \tablehead{
    \colhead{Source} & \colhead{RA} & \colhead{Dec}  & \colhead{$\Delta S_{\nu}$} & \colhead{$S_{5\text{GHz}}/S_g$}& \colhead{$S_{5\text{GHz}}/S_g$} &  \colhead{z} & \colhead{$L_{\nu,\text{max}}$} \\
    \colhead{} & \colhead{(deg)} & \colhead{(deg)} & \colhead{($\%$)} & \colhead{(min)} & \colhead{(max)} & \colhead{} & \colhead{(erg~s\textsuperscript{$-1$}Hz\textsuperscript{$-1$})}} 
\startdata
	ASKAP J004232.9$-$063312 & $\phantom{0}$10.6373 & $\phantom{0}$$-$6.5535 & $\phantom{0}$$\phantom{0}$99 & $\phantom{0}$10000\textsuperscript{l}$\phantom{0}$$\phantom{0}$ & 19000$\phantom{0}$$\phantom{0}$ & 1.8*\textsuperscript{d} & 1.5E+33 \\ 
	ASKAP J004408.0$-$271820 & $\phantom{0}$11.0336 & $-$27.3058 & $\phantom{0}$$\phantom{0}$83 & $\phantom{0}$579\textsuperscript{l}$\phantom{0}$$\phantom{0}$ & $\phantom{0}$1000$\phantom{0}$$\phantom{0}$ & 1.6*\textsuperscript{d} & 4.4E+32 \\ 
	ASKAP J005051.1$-$392305 & $\phantom{0}$12.7133 & $-$39.3848 & $\phantom{0}$$\phantom{0}$85 & $\phantom{0}$252\textsuperscript{l}$\phantom{0}$$\phantom{0}$ & $\phantom{0}$468$\phantom{0}$$\phantom{0}$ & 1.4*\textsuperscript{d} & 2.8E+32 \\ 
	ASKAP J010459.9$-$515129 & $\phantom{0}$16.2497 & $-$51.8583 & $\phantom{0}$$\phantom{0}$81 & $\phantom{0}$$\phantom{0}$4.0\textsuperscript{l}$\phantom{0}$ & $\phantom{0}$$\phantom{0}$7.3$\phantom{0}$ & 0.3*\textsuperscript{d} & 7.3E+30 \\ 
	ASKAP J011948.3$-$382254 & $\phantom{0}$19.9514 & $-$38.3817 & $\phantom{0}$155 & $\phantom{0}$588\textsuperscript{l}$\phantom{0}$$\phantom{0}$ & $\phantom{0}$1000$\phantom{0}$$\phantom{0}$ & 1.0*\textsuperscript{d} & 1.2E+32 \\ 
	ASKAP J012404.4$-$350117 & $\phantom{0}$21.0186 & $-$35.0215 & $\phantom{0}$104 & $\phantom{0}$233\textsuperscript{l}$\phantom{0}$$\phantom{0}$ & $\phantom{0}$475$\phantom{0}$$\phantom{0}$ & 3.3\textsuperscript{g} & 2.0E+34 \\ 
	ASKAP J013753.7$-$524813 & $\phantom{0}$24.4740 & $-$52.8038 & $\phantom{0}$154 & $\phantom{0}$13000\textsuperscript{l}$\phantom{0}$$\phantom{0}$ & 34000$\phantom{0}$$\phantom{0}$ & ... & ... \\ 
	ASKAP J014215.4$-$330202 & $\phantom{0}$25.5645 & $-$33.0339 & $\phantom{0}$141 & $\phantom{0}$$\phantom{0}$7.2\textsuperscript{l}$\phantom{0}$ & $\phantom{0}$$\phantom{0}$17.3$\phantom{0}$$\phantom{0}$ & 1.5*\textsuperscript{d} & 3.4E+32 \\ 
	ASKAP J014517.5$-$354555 & $\phantom{0}$26.3230 & $-$35.7655 & $\phantom{0}$$\phantom{0}$88 & $\phantom{0}$$\phantom{0}$6.6\textsuperscript{l}$\phantom{0}$ & $\phantom{0}$$\phantom{0}$12.5$\phantom{0}$$\phantom{0}$ & 0.5\textsuperscript{g} & 6.6E+31 \\ 
	ASKAP J014742.5$-$554924 & $\phantom{0}$26.9274 & $-$55.8235 & $\phantom{0}$121 & $\phantom{0}$$\phantom{0}$7.6\textsuperscript{l}$\phantom{0}$ & $\phantom{0}$$\phantom{0}$16.7$\phantom{0}$$\phantom{0}$ & 2.8\textsuperscript{g} & 1.8E+33 \\ 
	ASKAP J014921.2$-$020651 & $\phantom{0}$27.3388 & $\phantom{0}$$-$2.1144 & $\phantom{0}$100 & $\phantom{0}$$\phantom{0}$22.3\textsuperscript{s}$\phantom{0}$$\phantom{0}$ & $\phantom{0}$$\phantom{0}$44.6$\phantom{0}$$\phantom{0}$ & 3.1\textsuperscript{s} & 4.0E+33 \\ 
	ASKAP J015150.5$-$011219 & $\phantom{0}$27.9611 & $\phantom{0}$$-$1.2053 & $\phantom{0}$$\phantom{0}$85 & $\phantom{0}$$\phantom{0}$15.1\textsuperscript{s}$\phantom{0}$$\phantom{0}$ & $\phantom{0}$$\phantom{0}$28.0$\phantom{0}$$\phantom{0}$ & 0.8\textsuperscript{s} & 1.9E+32 \\ 
	ASKAP J021149.1$+$060015 & $\phantom{0}$32.9549 & $\phantom{0}$$+$6.0043 & $\phantom{0}$102 & $\phantom{0}$819\textsuperscript{l}$\phantom{0}$$\phantom{0}$ & $\phantom{0}$2000$\phantom{0}$$\phantom{0}$ & 0.9*\textsuperscript{d} & 2.6E+32 \\ 
	ASKAP J022234.4$+$020004 & $\phantom{0}$35.6437 & $\phantom{0}$$+$2.0014 & $\phantom{0}$143 & $\phantom{0}$$\phantom{0}$39.1\textsuperscript{s}$\phantom{0}$$\phantom{0}$ & $\phantom{0}$$\phantom{0}$95.0$\phantom{0}$$\phantom{0}$ & 2.9\textsuperscript{s} & 7.2E+33 \\ 
	ASKAP J023309.6$-$482402 & $\phantom{0}$38.2901 & $-$48.4006 & $\phantom{0}$$\phantom{0}$98 & $\phantom{0}$916\textsuperscript{l}$\phantom{0}$$\phantom{0}$ & $\phantom{0}$2000$\phantom{0}$$\phantom{0}$ & 1.0*\textsuperscript{d} & 1.7E+32 \\ 
	ASKAP J030231.6$-$481818 & $\phantom{0}$45.6317 & $-$48.3051 & $\phantom{0}$125 & $\phantom{0}$408\textsuperscript{l}$\phantom{0}$$\phantom{0}$ & $\phantom{0}$916$\phantom{0}$$\phantom{0}$ & 0.9*\textsuperscript{d} & 3.9E+32 \\ 
	ASKAP J031044.0$+$043610 & $\phantom{0}$47.6834 & $\phantom{0}$$+$4.6029 & $\phantom{0}$$\phantom{0}$93 & $\phantom{0}$101\textsuperscript{s}$\phantom{0}$$\phantom{0}$ & $\phantom{0}$194$\phantom{0}$$\phantom{0}$ & 1.5*\textsuperscript{mx} & 1.2E+33 \\ 
	ASKAP J031738.6$+$033454 & $\phantom{0}$49.4111 & $\phantom{0}$$+$3.5819 & $\phantom{0}$223 & $\phantom{0}$$\phantom{0}$0.9\textsuperscript{s}$\phantom{0}$ & $\phantom{0}$$\phantom{0}$2.9$\phantom{0}$ & 0.3\textsuperscript{g} & 1.1E+31 \\ 
	ASKAP J033602.9$+$051128 & $\phantom{0}$54.0124 & $\phantom{0}$$+$5.1913 & $\phantom{0}$195 & $\phantom{0}$124\textsuperscript{s}$\phantom{0}$$\phantom{0}$ & $\phantom{0}$366$\phantom{0}$$\phantom{0}$ & 0.2*\textsuperscript{s} & 6.7E+30 \\ 
	ASKAP J033834.3$-$542501 & $\phantom{0}$54.6431 & $-$54.4170 & $\phantom{0}$146 & $\phantom{0}$$\phantom{0}$5.7\textsuperscript{l}$\phantom{0}$ & $\phantom{0}$$\phantom{0}$14.1$\phantom{0}$$\phantom{0}$ & 0.9\textsuperscript{g} & 4.0E+32 \\ 
	ASKAP J034921.5$+$075101 & $\phantom{0}$57.3398 & $\phantom{0}$$+$7.8503 & $\phantom{0}$112 & $\phantom{0}$$\phantom{0}$21.2\textsuperscript{l}$\phantom{0}$$\phantom{0}$ & $\phantom{0}$$\phantom{0}$45.0$\phantom{0}$$\phantom{0}$ & 2.1\textsuperscript{g} & 9.5E+32 \\ 
	ASKAP J035501.0$+$091248 & $\phantom{0}$58.7543 & $\phantom{0}$$+$9.2134 & $\phantom{0}$$\phantom{0}$92 & $\phantom{0}$$\phantom{0}$68.8\textsuperscript{l}$\phantom{0}$$\phantom{0}$ & $\phantom{0}$132$\phantom{0}$$\phantom{0}$ & 0.6*\textsuperscript{mx} & 6.8E+31 \\ 
	ASKAP J035553.9$-$461757 & $\phantom{0}$58.9748 & $-$46.2992 & $\phantom{0}$$\phantom{0}$83 & $\phantom{0}$11000\textsuperscript{l}$\phantom{0}$$\phantom{0}$ & 21000$\phantom{0}$$\phantom{0}$ & 1.0*\textsuperscript{d} & 2.3E+32 \\ 
	ASKAP J040202.8$-$321845 & $\phantom{0}$60.5120 & $-$32.3127 & 1831 & $\phantom{0}$$\phantom{0}$1.5\textsuperscript{l}$\phantom{0}$ & $\phantom{0}$$\phantom{0}$28.4$\phantom{0}$$\phantom{0}$ & 1.5\textsuperscript{g} & 3.7E+32 \\ 
	ASKAP J040913.7$-$060838 & $\phantom{0}$62.3074 & $\phantom{0}$$-$6.1442 & $\phantom{0}$$\phantom{0}$98 & $\phantom{0}$$\phantom{0}$14.4\textsuperscript{s}$\phantom{0}$$\phantom{0}$ & $\phantom{0}$$\phantom{0}$28.5$\phantom{0}$$\phantom{0}$ & 2.7\textsuperscript{s} & 4.3E+33 \\ 
	ASKAP J041916.2$-$312707 & $\phantom{0}$64.8176 & $-$31.4522 & $\phantom{0}$$\phantom{0}$87 & $\phantom{0}$$\phantom{0}$1.7\textsuperscript{l}$\phantom{0}$ & $\phantom{0}$$\phantom{0}$3.1$\phantom{0}$ & 0.4\textsuperscript{g} & 1.4E+31 \\ 
	ASKAP J042029.8$-$551757 & $\phantom{0}$65.1245 & $-$55.2993 & $\phantom{0}$195 & $\phantom{0}$$\phantom{0}$1.7\textsuperscript{l}$\phantom{0}$ & $\phantom{0}$$\phantom{0}$0.2$\phantom{0}$ & 1.4\textsuperscript{g} & 2.6E+32 \\ 
	ASKAP J042550.4$-$084716 & $\phantom{0}$66.4601 & $\phantom{0}$$-$8.7879 & $\phantom{0}$119 & $\phantom{0}$$\phantom{0}$53.7\textsuperscript{l}$\phantom{0}$$\phantom{0}$ & $\phantom{0}$117$\phantom{0}$$\phantom{0}$ & 1.4\textsuperscript{d} & 3.8E+32 \\ 
	ASKAP J042559.8$-$394025 & $\phantom{0}$66.4996 & $-$39.6737 & $\phantom{0}$129 & $\phantom{0}$$\phantom{0}$4.0\textsuperscript{l}$\phantom{0}$ & $\phantom{0}$$\phantom{0}$9.3$\phantom{0}$ & 2.4\textsuperscript{g} & 1.3E+33 \\ 
	ASKAP J045256.1$-$564841 & $\phantom{0}$73.2342 & $-$56.8116 & $\phantom{0}$$\phantom{0}$86 & $\phantom{0}$109\textsuperscript{l}$\phantom{0}$$\phantom{0}$ & $\phantom{0}$203$\phantom{0}$$\phantom{0}$ & 1.6*\textsuperscript{d} & 5.8E+32 \\ 
	ASKAP J045302.0$-$385033 & $\phantom{0}$73.2587 & $-$38.8425 & $\phantom{0}$$\phantom{0}$94 & $\phantom{0}$$\phantom{0}$2.9\textsuperscript{l}$\phantom{0}$ & $\phantom{0}$$\phantom{0}$5.7$\phantom{0}$ & 0.3*\textsuperscript{d} & 5.2E+30 \\ 
	ASKAP J050500.0$-$450447 & $\phantom{0}$76.2504 & $-$45.0799 & $\phantom{0}$210 & $\phantom{0}$392\textsuperscript{l}$\phantom{0}$$\phantom{0}$ & $\phantom{0}$1000$\phantom{0}$$\phantom{0}$ & 1.4\textsuperscript{g} & 5.4E+32 \\ 
	ASKAP J052822.8$-$322358 & $\phantom{0}$82.0950 & $-$32.3995 & $\phantom{0}$$\phantom{0}$92 & $\phantom{0}$$\phantom{0}$35.4\textsuperscript{l}$\phantom{0}$$\phantom{0}$ & $\phantom{0}$$\phantom{0}$68.1$\phantom{0}$$\phantom{0}$ & 1.4\textsuperscript{g} & 5.0E+32 \\ 
	ASKAP J053009.0$-$383734 & $\phantom{0}$82.5377 & $-$38.6264 & $\phantom{0}$$\phantom{0}$96 & $\phantom{0}$$\phantom{0}$14.7\textsuperscript{l}$\phantom{0}$$\phantom{0}$ & $\phantom{0}$$\phantom{0}$28.7$\phantom{0}$$\phantom{0}$ & 0.1*\textsuperscript{d} & 3.1E+30 \\ 
	ASKAP J061401.5$-$551306 & $\phantom{0}$93.5063 & $-$55.2186 & $\phantom{0}$100 & $\phantom{0}$217\textsuperscript{l}$\phantom{0}$$\phantom{0}$ & $\phantom{0}$433$\phantom{0}$$\phantom{0}$ & 1.9\textsuperscript{c} & 2.0E+33 \\ 
	ASKAP J075939.8$+$054615 & 119.9162 & $\phantom{0}$$+$5.7708 & $\phantom{0}$133 & $\phantom{0}$448\textsuperscript{l}$\phantom{0}$$\phantom{0}$ & $\phantom{0}$1000$\phantom{0}$$\phantom{0}$ & 2.8*\textsuperscript{d} & 1.6E+33 \\ 
	ASKAP J080329.2$+$045904 & 120.8718 & $\phantom{0}$$+$4.9845 & $\phantom{0}$$\phantom{0}$98 & $\phantom{0}$115\textsuperscript{s}$\phantom{0}$$\phantom{0}$ & $\phantom{0}$228$\phantom{0}$$\phantom{0}$ & 0.6*\textsuperscript{s} & 2.9E+31 \\ 
	ASKAP J082555.9$-$050932 & 126.4830 & $\phantom{0}$$-$5.1589 & $\phantom{0}$109 & $\phantom{0}$$\phantom{0}$4.0\textsuperscript{s}$\phantom{0}$ & $\phantom{0}$$\phantom{0}$8.4$\phantom{0}$ & 2.7\textsuperscript{g} & 1.5E+33 \\ 
	ASKAP J084831.8$-$084845 & 132.1328 & $\phantom{0}$$-$8.8127 & $\phantom{0}$$\phantom{0}$97 & $\phantom{0}$393\textsuperscript{l}$\phantom{0}$$\phantom{0}$ & $\phantom{0}$773$\phantom{0}$$\phantom{0}$ & 2.1\textsuperscript{g} & 1.2E+34 \\ 
	ASKAP J085302.8$+$075154 & 133.2620 & $\phantom{0}$$+$7.8650 & $\phantom{0}$$\phantom{0}$83 & $\phantom{0}$$\phantom{0}$12.7\textsuperscript{l}$\phantom{0}$$\phantom{0}$ & $\phantom{0}$$\phantom{0}$23.2$\phantom{0}$$\phantom{0}$ & 1.2\textsuperscript{d} & 2.0E+32 \\ 
	ASKAP J090140.0$+$075347 & 135.4167 & $\phantom{0}$$+$7.8965 & $\phantom{0}$114 & $\phantom{0}$947\textsuperscript{l}$\phantom{0}$$\phantom{0}$ & $\phantom{0}$2000$\phantom{0}$$\phantom{0}$ & 0.9*\textsuperscript{d} & 8.8E+32 \\ 
	ASKAP J091307.4$-$020936 & 138.2811 & $\phantom{0}$$-$2.1601 & $\phantom{0}$146 & $\phantom{0}$$\phantom{0}$30.7\textsuperscript{s}$\phantom{0}$$\phantom{0}$ & $\phantom{0}$$\phantom{0}$75.4$\phantom{0}$$\phantom{0}$ & 2.2\textsuperscript{s} & 2.0E+33 \\ 
	ASKAP J094055.3$-$043235 & 145.2308 & $\phantom{0}$$-$4.5433 & $\phantom{0}$115 & $\phantom{0}$24000\textsuperscript{l}$\phantom{0}$$\phantom{0}$ & 51000$\phantom{0}$$\phantom{0}$ & ... & ... \\ 
	ASKAP J100154.9$-$370127 & 150.4788 & $-$37.0243 & $\phantom{0}$109 & $\phantom{0}$$\phantom{0}$23.8\textsuperscript{m}$\phantom{0}$$\phantom{0}$ & $\phantom{0}$$\phantom{0}$49.8$\phantom{0}$$\phantom{0}$ & 1.8\textsuperscript{so} & 2.4E+33 \\ 
	ASKAP J101432.5$-$401536 & 153.6358 & $-$40.2603 & $\phantom{0}$122 & ... & ... & ... & ... \\ 
	ASKAP J102644.6$-$082612 & 156.6861 & $\phantom{0}$$-$8.4368 & $\phantom{0}$$\phantom{0}$83 & $\phantom{0}$$\phantom{0}$16.6\textsuperscript{l}$\phantom{0}$$\phantom{0}$ & $\phantom{0}$$\phantom{0}$30.4$\phantom{0}$$\phantom{0}$ & 0.9\textsuperscript{d} & 2.2E+32 \\ 
	ASKAP J103737.9$-$292240 & 159.4082 & $-$29.3779 & $\phantom{0}$110 & $\phantom{0}$967\textsuperscript{l}$\phantom{0}$$\phantom{0}$ & $\phantom{0}$2000$\phantom{0}$$\phantom{0}$ & ... & ... \\ 
	ASKAP J104215.9$-$070919 & 160.5665 & $\phantom{0}$$-$7.1555 & $\phantom{0}$111 & $\phantom{0}$132\textsuperscript{l}$\phantom{0}$$\phantom{0}$ & $\phantom{0}$278$\phantom{0}$$\phantom{0}$ & 1.6\textsuperscript{d} & 1.7E+33 \\ 
	ASKAP J104514.5$-$394654 & 161.3105 & $-$39.7819 & $\phantom{0}$$\phantom{0}$88 & $\phantom{0}$133\textsuperscript{l}$\phantom{0}$$\phantom{0}$ & $\phantom{0}$250$\phantom{0}$$\phantom{0}$ & ... & ... \\ 
	ASKAP J104855.1$-$340145 & 162.2299 & $-$34.0293 & $\phantom{0}$102 & $\phantom{0}$$\phantom{0}$56.0\textsuperscript{l}$\phantom{0}$$\phantom{0}$ & $\phantom{0}$113$\phantom{0}$$\phantom{0}$ & ... & ... \\ 
	ASKAP J104948.5$+$081801 & 162.4526 & $\phantom{0}$$+$8.3002 & $\phantom{0}$$\phantom{0}$95 & $\phantom{0}$200\textsuperscript{s}$\phantom{0}$$\phantom{0}$ & $\phantom{0}$391$\phantom{0}$$\phantom{0}$ & 2.3\textsuperscript{s} & 2.7E+33 \\ 
	ASKAP J105012.6$-$402353 & 162.5526 & $-$40.3982 & $\phantom{0}$104 & $\phantom{0}$466\textsuperscript{l}$\phantom{0}$$\phantom{0}$ & $\phantom{0}$953$\phantom{0}$$\phantom{0}$ & ... & ... \\ 
	ASKAP J105501.7$-$304732 & 163.7572 & $-$30.7924 & $\phantom{0}$122 & $\phantom{0}$$\phantom{0}$1.9\textsuperscript{l}$\phantom{0}$ & $\phantom{0}$$\phantom{0}$4.2$\phantom{0}$ & 0.2\textsuperscript{so} & 4.5E+30 \\ 
	ASKAP J105532.4$-$305141 & 163.8850 & $-$30.8615 & $\phantom{0}$$\phantom{0}$83 & $\phantom{0}$1000\textsuperscript{l}$\phantom{0}$$\phantom{0}$ & $\phantom{0}$3000$\phantom{0}$$\phantom{0}$ & ... & ... \\ 
	ASKAP J105726.7$+$073745 & 164.3614 & $\phantom{0}$$+$7.6293 & $\phantom{0}$146 & $\phantom{0}$$\phantom{0}$4.3\textsuperscript{l}$\phantom{0}$ & $\phantom{0}$$\phantom{0}$10.7$\phantom{0}$$\phantom{0}$ & 0.2\textsuperscript{d} & 3.4E+30 \\ 
	ASKAP J105807.4$+$080923 & 164.5309 & $\phantom{0}$$+$8.1565 & $\phantom{0}$$\phantom{0}$88 & $\phantom{0}$855\textsuperscript{l}$\phantom{0}$$\phantom{0}$ & $\phantom{0}$2000$\phantom{0}$$\phantom{0}$ & 1.3*\textsuperscript{d} & 3.1E+32 \\ 
	ASKAP J110137.5$+$021509 & 165.4063 & $\phantom{0}$$+$2.2527 & $\phantom{0}$$\phantom{0}$96 & $\phantom{0}$7000\textsuperscript{l}$\phantom{0}$$\phantom{0}$ & 13000$\phantom{0}$$\phantom{0}$ & 2.2\textsuperscript{d} & 6.6E+34 \\ 
	ASKAP J112810.8$-$305103 & 172.0451 & $-$30.8510 & $\phantom{0}$286 & $\phantom{0}$422*\textsuperscript{l}$\phantom{0}$$\phantom{0}$ & $\phantom{0}$2000$\phantom{0}$$\phantom{0}$ & ... & ... \\ 
	ASKAP J113340.6$-$374359 & 173.4195 & $-$37.7332 & $\phantom{0}$144 & $\phantom{0}$$\phantom{0}$31.6\textsuperscript{l}$\phantom{0}$$\phantom{0}$ & $\phantom{0}$$\phantom{0}$77.3$\phantom{0}$$\phantom{0}$ & 1.9\textsuperscript{so} & 1.4E+33 \\ 
	ASKAP J113825.8$-$330537 & 174.6079 & $-$33.0938 & $\phantom{0}$115 & $\phantom{0}$4000\textsuperscript{l}$\phantom{0}$$\phantom{0}$ & $\phantom{0}$9000$\phantom{0}$$\phantom{0}$ & ... & ... \\ 
	ASKAP J114402.9$-$422622 & 176.0123 & $-$42.4395 & $\phantom{0}$111 & $\phantom{0}$$\phantom{0}$0.01\textsuperscript{m} & $\phantom{0}$$\phantom{0}$0.03 & 3.1\textsuperscript{g} & 1.6E+33 \\ 
	ASKAP J115600.0$-$315504 & 179.0004 & $-$31.9178 & $\phantom{0}$140 & $\phantom{0}$1000*\textsuperscript{l}$\phantom{0}$$\phantom{0}$ & $\phantom{0}$3000$\phantom{0}$$\phantom{0}$ & ... & ... \\ 
	ASKAP J124514.4$-$285831 & 191.3101 & $-$28.9754 & $\phantom{0}$111 & $\phantom{0}$376\textsuperscript{l}$\phantom{0}$$\phantom{0}$ & $\phantom{0}$794$\phantom{0}$$\phantom{0}$ & ... & ... \\ 
	ASKAP J124957.8$-$372356 & 192.4912 & $-$37.3991 & $\phantom{0}$$\phantom{0}$95 & $\phantom{0}$$\phantom{0}$9.9\textsuperscript{l}$\phantom{0}$ & $\phantom{0}$$\phantom{0}$19.3$\phantom{0}$$\phantom{0}$ & 1.4\textsuperscript{g} & 8.0E+32 \\ 
	ASKAP J125010.2$-$332909 & 192.5425 & $-$33.4861 & $\phantom{0}$462 & $\phantom{0}$$\phantom{0}$4.2\textsuperscript{l}$\phantom{0}$ & $\phantom{0}$$\phantom{0}$23.5$\phantom{0}$$\phantom{0}$ & 2.2\textsuperscript{g} & 1.1E+33 \\ 
	ASKAP J125050.8$-$445757 & 192.7117 & $-$44.9660 & $\phantom{0}$105 & ... & ... & ... & ... \\ 
	ASKAP J125140.3$-$072350 & 192.9182 & $\phantom{0}$$-$7.3973 & $\phantom{0}$257 & $\phantom{0}$2000\textsuperscript{l}$\phantom{0}$$\phantom{0}$ & $\phantom{0}$7000$\phantom{0}$$\phantom{0}$ & 2.0*\textsuperscript{d} & 3.6E+33 \\ 
	ASKAP J125148.9$-$071228 & 192.9541 & $\phantom{0}$$-$7.2079 & $\phantom{0}$$\phantom{0}$87 & $\phantom{0}$$\phantom{0}$43.0\textsuperscript{l}$\phantom{0}$$\phantom{0}$ & $\phantom{0}$$\phantom{0}$80.3$\phantom{0}$$\phantom{0}$ & 0.8\textsuperscript{h} & 8.0E+31 \\ 
	ASKAP J130117.7$+$071211 & 195.3238 & $\phantom{0}$$+$7.2032 & $\phantom{0}$113 & $\phantom{0}$$\phantom{0}$42.6\textsuperscript{l}$\phantom{0}$$\phantom{0}$ & $\phantom{0}$$\phantom{0}$90.8$\phantom{0}$$\phantom{0}$ & 1.0\textsuperscript{h} & 1.1E+32 \\ 
	ASKAP J131510.3$-$371415 & 198.7931 & $-$37.2377 & $\phantom{0}$$\phantom{0}$93 & $\phantom{0}$1000\textsuperscript{l}$\phantom{0}$$\phantom{0}$ & $\phantom{0}$2000$\phantom{0}$$\phantom{0}$ & ... & ... \\ 
	ASKAP J141806.4$-$031153 & 214.5268 & $\phantom{0}$$-$3.1983 & $\phantom{0}$$\phantom{0}$87 & $\phantom{0}$3000\textsuperscript{l}$\phantom{0}$$\phantom{0}$ & $\phantom{0}$5000$\phantom{0}$$\phantom{0}$ & 1.2\textsuperscript{d} & 9.2E+32 \\ 
	ASKAP J150801.4$-$041936 & 227.0060 & $\phantom{0}$$-$4.3267 & $\phantom{0}$100 & $\phantom{0}$510\textsuperscript{l}$\phantom{0}$$\phantom{0}$ & $\phantom{0}$1000$\phantom{0}$$\phantom{0}$ & 1.7\textsuperscript{d} & 1.3E+33 \\ 
	ASKAP J152545.9$+$044126 & 231.4414 & $\phantom{0}$$+$4.6909 & $\phantom{0}$133 & $\phantom{0}$1000\textsuperscript{s}$\phantom{0}$$\phantom{0}$ & $\phantom{0}$3000$\phantom{0}$$\phantom{0}$ & 1.0\textsuperscript{s} & 3.0E+32 \\ 
	ASKAP J153553.6$+$083641 & 233.9736 & $\phantom{0}$$+$8.6116 & $\phantom{0}$258 & $\phantom{0}$466\textsuperscript{l}$\phantom{0}$$\phantom{0}$ & $\phantom{0}$2000$\phantom{0}$$\phantom{0}$ & 1.2*\textsuperscript{d} & 2.6E+32 \\ 
	ASKAP J154903.1$+$060252 & 237.2631 & $\phantom{0}$$+$6.0479 & $\phantom{0}$199 & $\phantom{0}$$\phantom{0}$48.7\textsuperscript{s}$\phantom{0}$$\phantom{0}$ & $\phantom{0}$146$\phantom{0}$$\phantom{0}$ & 0.6\textsuperscript{s} & 5.3E+31 \\ 
	ASKAP J163528.6$+$070506 & 248.8696 & $\phantom{0}$$+$7.0853 & $\phantom{0}$$\phantom{0}$91 & $\phantom{0}$554\textsuperscript{l}$\phantom{0}$$\phantom{0}$ & $\phantom{0}$1000$\phantom{0}$$\phantom{0}$ & 0.9\textsuperscript{d} & 3.0E+32 \\ 
	ASKAP J164431.6$+$033701 & 251.1318 & $\phantom{0}$$+$3.6170 & $\phantom{0}$165 & $\phantom{0}$183\textsuperscript{l}$\phantom{0}$$\phantom{0}$ & $\phantom{0}$485$\phantom{0}$$\phantom{0}$ & 2.1\textsuperscript{d} & 5.0E+33 \\ 
	ASKAP J165107.0$+$012406 & 252.7794 & $\phantom{0}$$+$1.4018 & $\phantom{0}$152 & $\phantom{0}$229\textsuperscript{l}$\phantom{0}$$\phantom{0}$ & $\phantom{0}$578$\phantom{0}$$\phantom{0}$ & 0.3*\textsuperscript{d} & 6.9E+31 \\ 
	ASKAP J165913.7$+$083944 & 254.8074 & $\phantom{0}$$+$8.6623 & $\phantom{0}$119 & $\phantom{0}$$\phantom{0}$8.8\textsuperscript{l}$\phantom{0}$ & $\phantom{0}$$\phantom{0}$19.3$\phantom{0}$$\phantom{0}$ & 0.8\textsuperscript{d} & 1.2E+32 \\ 
	ASKAP J170252.1$+$014024 & 255.7172 & $\phantom{0}$$+$1.6735 & $\phantom{0}$$\phantom{0}$82 & $\phantom{0}$$\phantom{0}$6.4\textsuperscript{l}$\phantom{0}$ & $\phantom{0}$$\phantom{0}$11.7$\phantom{0}$$\phantom{0}$ & 1.6\textsuperscript{d} & 1.8E+33 \\ 
	ASKAP J171659.5$+$033419 & 259.2481 & $\phantom{0}$$+$3.5721 & $\phantom{0}$132 & $\phantom{0}$142*\textsuperscript{z}$\phantom{0}$$\phantom{0}$ & $\phantom{0}$330$\phantom{0}$$\phantom{0}$ & ... & ... \\ 
	ASKAP J174307.9$+$075614 & 265.7830 & $\phantom{0}$$+$7.9372 & $\phantom{0}$197 & $\phantom{0}$$\phantom{0}$45.3*\textsuperscript{z}$\phantom{0}$$\phantom{0}$ & $\phantom{0}$135$\phantom{0}$$\phantom{0}$ & ... & ... \\ 
	ASKAP J175638.9$+$083440 & 269.1622 & $\phantom{0}$$+$8.5780 & $\phantom{0}$114 & $\phantom{0}$$\phantom{0}$9.2\textsuperscript{s}$\phantom{0}$ & $\phantom{0}$$\phantom{0}$19.7$\phantom{0}$$\phantom{0}$ & 0.1*\textsuperscript{s} & 1.6E+30 \\ 
	ASKAP J194551.5$-$012544 & 296.4648 & $\phantom{0}$$-$1.4290 & $\phantom{0}$199 & $\phantom{0}$371*\textsuperscript{z}$\phantom{0}$$\phantom{0}$ & $\phantom{0}$1000$\phantom{0}$$\phantom{0}$ & ... & ... \\ 
	ASKAP J201534.2$-$444703 & 303.8927 & $-$44.7843 & $\phantom{0}$$\phantom{0}$90 & $\phantom{0}$$\phantom{0}$1.0\textsuperscript{l}$\phantom{0}$ & $\phantom{0}$$\phantom{0}$2.0$\phantom{0}$ & 0.8\textsuperscript{g} & 6.2E+31 \\ 
	ASKAP J201813.9$+$025307 & 304.5582 & $\phantom{0}$$+$2.8855 & $\phantom{0}$112 & $\phantom{0}$$\phantom{0}$22.3\textsuperscript{g}$\phantom{0}$$\phantom{0}$ & $\phantom{0}$$\phantom{0}$47.2$\phantom{0}$$\phantom{0}$ & 0.7\textsuperscript{so} & 1.3E+32 \\ 
	ASKAP J203349.5$-$421543 & 308.4565 & $-$42.2620 & $\phantom{0}$$\phantom{0}$96 & $\phantom{0}$$\phantom{0}$20.1\textsuperscript{l}$\phantom{0}$$\phantom{0}$ & $\phantom{0}$$\phantom{0}$39.4$\phantom{0}$$\phantom{0}$ & 0.6\textsuperscript{g} & 4.1E+31 \\ 
	ASKAP J210007.1$-$305617 & 315.0297 & $-$30.9381 & $\phantom{0}$124 & $\phantom{0}$4000*\textsuperscript{l}$\phantom{0}$$\phantom{0}$ & $\phantom{0}$9000$\phantom{0}$$\phantom{0}$ & ... & ... \\ 
	ASKAP J211405.4$+$092525 & 318.5225 & $\phantom{0}$$+$9.4238 & $\phantom{0}$125 & $\phantom{0}$174\textsuperscript{l}$\phantom{0}$$\phantom{0}$ & $\phantom{0}$390$\phantom{0}$$\phantom{0}$ & 0.4*\textsuperscript{d} & 2.1E+31 \\ 
	ASKAP J211616.0$+$025451 & 319.0667 & $\phantom{0}$$+$2.9149 & $\phantom{0}$104 & $\phantom{0}$238\textsuperscript{s}$\phantom{0}$$\phantom{0}$ & $\phantom{0}$485$\phantom{0}$$\phantom{0}$ & 3.7\textsuperscript{s} & 5.1E+33 \\ 
	ASKAP J212956.7$-$385231 & 322.4863 & $-$38.8755 & $\phantom{0}$$\phantom{0}$81 & $\phantom{0}$$\phantom{0}$89.0\textsuperscript{l}$\phantom{0}$$\phantom{0}$ & $\phantom{0}$161$\phantom{0}$$\phantom{0}$ & 1.2\textsuperscript{g} & 1.5E+32 \\ 
	ASKAP J213904.6$-$331126 & 324.7692 & $-$33.1907 & $\phantom{0}$107 & $\phantom{0}$$\phantom{0}$15.3\textsuperscript{m}$\phantom{0}$$\phantom{0}$ & $\phantom{0}$$\phantom{0}$31.8$\phantom{0}$$\phantom{0}$ & 1.2\textsuperscript{g} & 2.3E+32 \\ 
	ASKAP J214147.3$+$061116 & 325.4472 & $\phantom{0}$$+$6.1884 & $\phantom{0}$290 & $\phantom{0}$$\phantom{0}$18.3\textsuperscript{s}$\phantom{0}$$\phantom{0}$ & $\phantom{0}$$\phantom{0}$71.2$\phantom{0}$$\phantom{0}$ & 3.2\textsuperscript{s} & 1.1E+34 \\ 
	ASKAP J215954.4$-$002149 & 329.9769 & $\phantom{0}$$-$0.3639 & $\phantom{0}$$\phantom{0}$92 & $\phantom{0}$$\phantom{0}$7.6\textsuperscript{s}$\phantom{0}$ & $\phantom{0}$$\phantom{0}$14.6$\phantom{0}$$\phantom{0}$ & 2.0\textsuperscript{s} & 1.4E+33 \\ 
	ASKAP J222325.2$-$322403 & 335.8551 & $-$32.4009 & $\phantom{0}$139 & $\phantom{0}$4000*\textsuperscript{l}$\phantom{0}$$\phantom{0}$ & $\phantom{0}$9000$\phantom{0}$$\phantom{0}$ & ... & ... \\ 
	ASKAP J222637.6$-$313933 & 336.6569 & $-$31.6593 & $\phantom{0}$$\phantom{0}$86 & $\phantom{0}$850\textsuperscript{l}$\phantom{0}$$\phantom{0}$ & $\phantom{0}$2000$\phantom{0}$$\phantom{0}$ & 0.9*\textsuperscript{k} & 9.7E+31 \\ 
	ASKAP J224239.7$-$303517 & 340.6657 & $-$30.5881 & $\phantom{0}$131 & $\phantom{0}$$\phantom{0}$10.2\textsuperscript{l}$\phantom{0}$$\phantom{0}$ & $\phantom{0}$$\phantom{0}$23.6$\phantom{0}$$\phantom{0}$ & 0.5\textsuperscript{g} & 2.9E+31 \\ 
	ASKAP J230757.4$+$034139 & 346.9892 & $\phantom{0}$$+$3.6944 & $\phantom{0}$198 & $\phantom{0}$329\textsuperscript{l}$\phantom{0}$$\phantom{0}$ & $\phantom{0}$982$\phantom{0}$$\phantom{0}$ & 0.8*\textsuperscript{d} & 1.1E+32 \\ 
	ASKAP J230856.2$-$360521 & 347.2345 & $-$36.0892 & $\phantom{0}$$\phantom{0}$81 & $\phantom{0}$353\textsuperscript{l}$\phantom{0}$$\phantom{0}$ & $\phantom{0}$640$\phantom{0}$$\phantom{0}$ & ... & ... \\ 
	ASKAP J231817.7$-$344402 & 349.5738 & $-$34.7339 & $\phantom{0}$163 & $\phantom{0}$$\phantom{0}$5.6\textsuperscript{l}$\phantom{0}$ & $\phantom{0}$$\phantom{0}$14.6$\phantom{0}$$\phantom{0}$ & 1.5\textsuperscript{g} & 5.4E+32 \\ 
	ASKAP J233557.8$-$495601 & 353.9910 & $-$49.9338 & $\phantom{0}$102 & $\phantom{0}$$\phantom{0}$23.9\textsuperscript{l}$\phantom{0}$$\phantom{0}$ & $\phantom{0}$$\phantom{0}$48.1$\phantom{0}$$\phantom{0}$ & 1.1*\textsuperscript{d} & 1.4E+32 \\ 
    \enddata
\end{longdeluxetable}
\subsection{Follow-up Observations}
\subsubsection{Spectroscopic Observations with the Hobby-Eberly Telescope}
We observed five sources with unknown or photometric redshifts with the Low-Resolution Spectrograph 2 \citep[LRS2;][]{Chonis2016} on the Hobby-Eberly Telescope \citep[HET;][]{Ramsey1998,Hill2021}, which is a $10$~m telescope with a fixed observing elevation of $55^\circ$. We observed two sources with $g<20$, ASKAP J034921.5$+$075101 and ASKAP J170252.1$+$014024, in the orange channel on LRS2-B ($4635$--$6950$\AA) and three with $r<20.5$,  ASKAP J084831.8$-$084845,  ASKAP J125148.9$-$071228,  ASKAP J130117.7$+$071211, in the red channel on LRS2-R ($6450$--$8400$\AA). We used the {\fontfamily{qcr}\selectfont directACAM} setup for the LRS2-B observations and the {\fontfamily{qcr}\selectfont ACAMblind} setup for the LRS2-R observations. Our observations were carried out during dark sky conditions (no moon), $2\arcsec$ seeing, and spectroscopic sky conditions. The data were reduced, including flux density calibration and sky subtraction, using the {\sc Panacea} pipeline \citep{Zeimann2026}. We determined the redshifts by identifying spectral lines. Our HET spectra and spectral line fits are shown in Figure~\ref{fig:spectra} in Appendix~\ref{appx:spectra}.
\subsubsection{Spectroscopic Observations with the Southern Astrophysical Research Telescope}
We observed four sources,  ASKAP J100154.9$-$370127,  ASKAP J105501.7$-$304732,  ASKAP J113340.6$-$374359, and  ASKAP J201813.9$+$025307, with the Goodman High Throughput Spectrograph \citep[Goodman HTS;][]{Clemens2004} on the Southern Astrophysical Research Telescope (SOAR). Observations were taken using the red camera and {\fontfamily{qcr}\selectfont 400m1} grating covering the wavelength range between $3500$--$7200$\AA\ with a resolution of 1~\AA. We used a standard $1\arcsec$ slit to observe the target. Data were reduced using {\sc PypeIt} \citep{pypeit,pypeit_zenodo}. We determined the redshifts by identifying spectral lines in our SOAR spectra. The SOAR spectral and spectral line fits are shown in Figure~\ref{fig:spectra} in Appendix~\ref{appx:spectra}. 

\subsubsection{Radio SEDs with the Australia Telescope Compact Array}
\begin{figure*}[htbp!]
    \centering
    \gridline{\fig{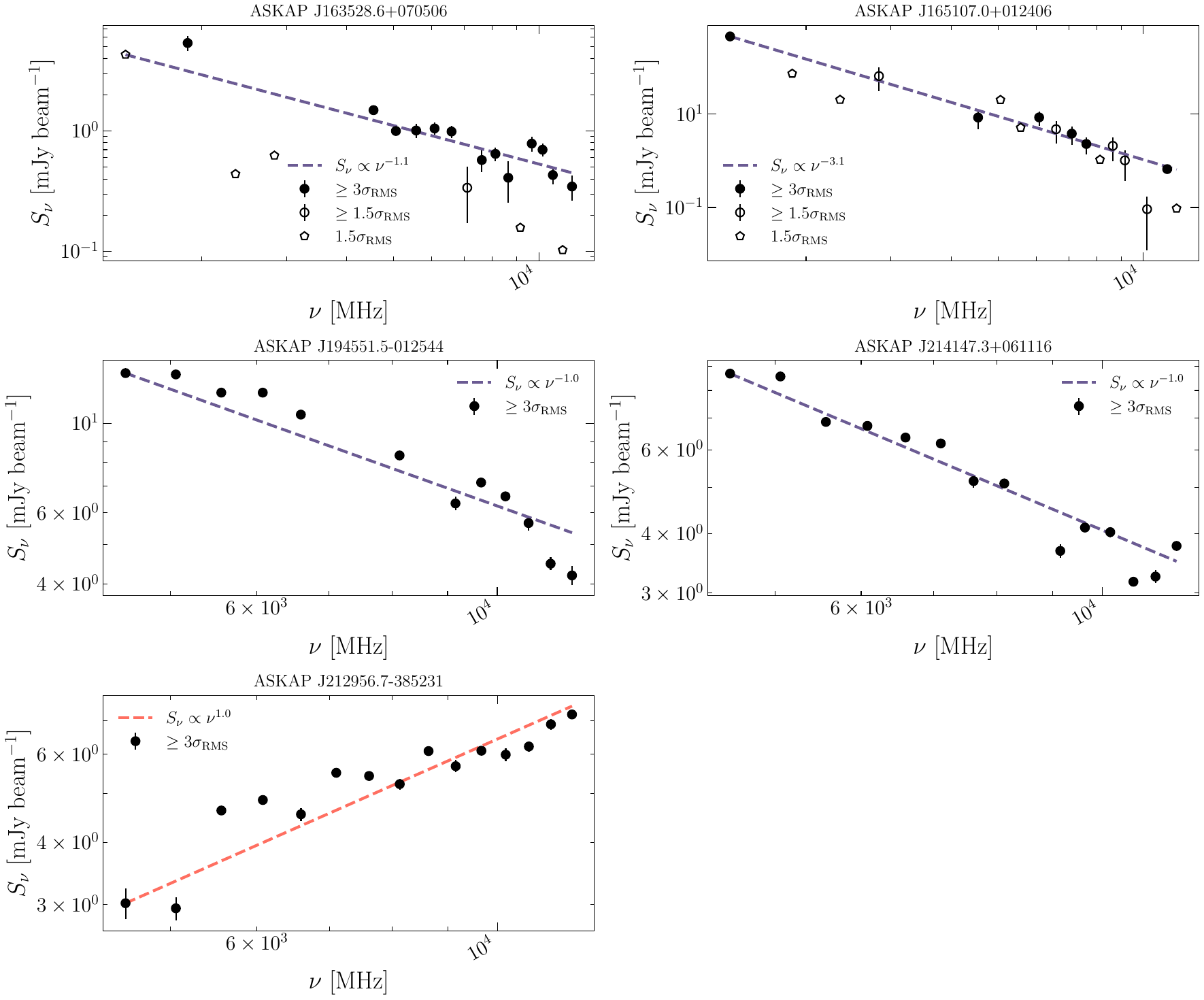}{0.98\textwidth}{}}
    \caption{Radio SEDs from ATCA observations. Solid black circles indicate flux density measurements with SNR$\geq 3\sigma_{\text{RMS}}$, unfilled circles indicate flux density measurements with $3\sigma_{\text{RMS}}>$SNR$\geq 1.5\sigma_{\text{RMS}}$, and unfilled hexagonal markers indicate $1.5\sigma_{\text{RMS}}$ upper limits on non-detections. Dashed purple lines indicate steep power law fits and dashed orange lines indicate inverted power law fits. \label{fig:atca_seds}}
\end{figure*}

We observed five sources, ASKAP J163528.6$+$070506, ASKAP J165107.0$+$012406, ASKAP J194551.5$-$012544, ASKAP J212956.7$-$385231, and ASKAP J214147.3$+$061116 in the L, S, C, and X bands using the Australia Telescope Compact Array \citep[ATCA;][]{Frater1992} under project code C3713. These observations were carried out in the H214 configuration with the new Broadband Integrated GPU Correlator for ATCA (BIGCAT\footnote{\url{https://www.atnf.csiro.au/projects/instrumentation/bigcat/}}). We observed ASKAP J163528.6$+$070506 and ASKAP J165107.0$+$012406 with the $4$~cm and $16$~cm BIGCAT receivers, which have an instantaneous bandwidth of $8$~GHz and $2$~GHz, respectively, and we observed ASKAP J194551.5$-$012544, ASKAP J212956.7$-$385231, and ASKAP J214147.3$+$061116 with the $4$~cm receiver. We observed each source twice per receiver, split into $300$~s scans separated by several hours to maximize UV coverage. Antenna 4 was unavailable during these observations. Since Antenna 6 is located $6$~km away from the remaining five antennas (which have a maximum baseline of $246.8$~m) in the H214 configuration, we excluded data from Antenna 6 during the data reduction process to avoid uneven UV coverage. We used PKS 1934$-$638 as our primary calibrator for all sources in both bands. The data were converted from the raw BIGCAT output (ASDM) to Measurement Sets, which we converted to FITS format. We reduced the observations with {\fontfamily{qcr}\selectfont
miriad} using standard ATCA point source data reduction techniques. We imaged each band in $0.512$~GHz sub-bands and constructed quasi-simultaneous SEDs. For some sources, we were not able to image a few sub-bands due to too much radio frequency interference (RFI). We fit a power law to each SED. The SEDs and best fits are shown in Figure~\ref{fig:atca_seds}.
We found one source, ASKAP J212956.7$-$385231 had an inverted SED with $\alpha=1.0$, suggesting that it is a young radio AGN. The remaining four SEDs revealed steep SEDs, with $\alpha$ ranging from $-3.1$ to $-1.0$. We compared the quasi-contemporaneous ATCA SEDs power law fits with the our prior spectral characterizations to assess the accuracy of the non-contemporaneous SEDs. We had classified ASKAP J163528.6$+$070506 as a peaked spectrum source with $\nu_p=887.5$~MHz and $\nu_{\text{steep}}=-0.9$ between $887.5$~MHz--$3$~GHz, and our ATCA observations revealed a steep ($\alpha=-1.1$) SED between $1.4$--$11.7$~GHz. We had classified ASKAP J165107.0$+$012406 as a steep ($\alpha=-0.6$) spectrum source, and our ATCA observations revealed a steep spectrum with a much steeper index ($\alpha=-3.1$) between $1.4$--$7.6$~GHz. We had classified ASKAP J194551.5$-$012544 as a peaked spectrum source with $\nu_p=1367.5$~MHz and $\alpha_{\text{steep}}=-0.5$ between $1367.5$~MHz--$3$~GHz, and our ATCA observations revealed a steep ($\alpha=-1.0$) SED between $4.5$--$11.7$~GHz. We had classified ASKAP J214147.3$+$061116 as a peaked spectrum source with $\nu_p=1655.5$~MHz and $\alpha_{\text{steep}}=-0.6$ between $1655.5$~MHz--$3$~GHz, and our ATCA observations revealed a steep ($\alpha=-1.0$) SED between $4.5$--$11.7$~GHz. We had classified ASKAP J212956.7$-$385231 as an inverted spectrum source with $\alpha=0.6$ between $887.5$~MHz--$3$~GHz based on its non-contemporaneous SED, and our ATCA observations confirmed the inverted SED, with $\alpha=1.0$ between $4.5$--$11.7$~GHz. While the ATCA observations revealed steeper spectral indices than the non-contemporaneous SEDs for all five sources, we determined the non-contemporaneous SEDs were in agreement with the quasi-contemporaneous SEDs. Our spectral classifications based on archival observations between $887.5$~MHz--$3$~GHz were consistent with the higher frequency ATCA SEDs. The ATCA SEDs had a broader bandwidth and better spectral resolution than the non-contemporaneous SEDs -- especially the two gigahertz peaked spectrum sources, for which we only had $1$--$2$ archival measurements with $\nu>\nu_p$ -- which likely was responsible for some of the discrepancy in spectral indices. 
\subsection{Comparison with AGN Samples Transitioning to Radio-loud at 3~GHz}
\begin{sidewaysfigure*}[htpb!]
    \includegraphics[width=23cm]{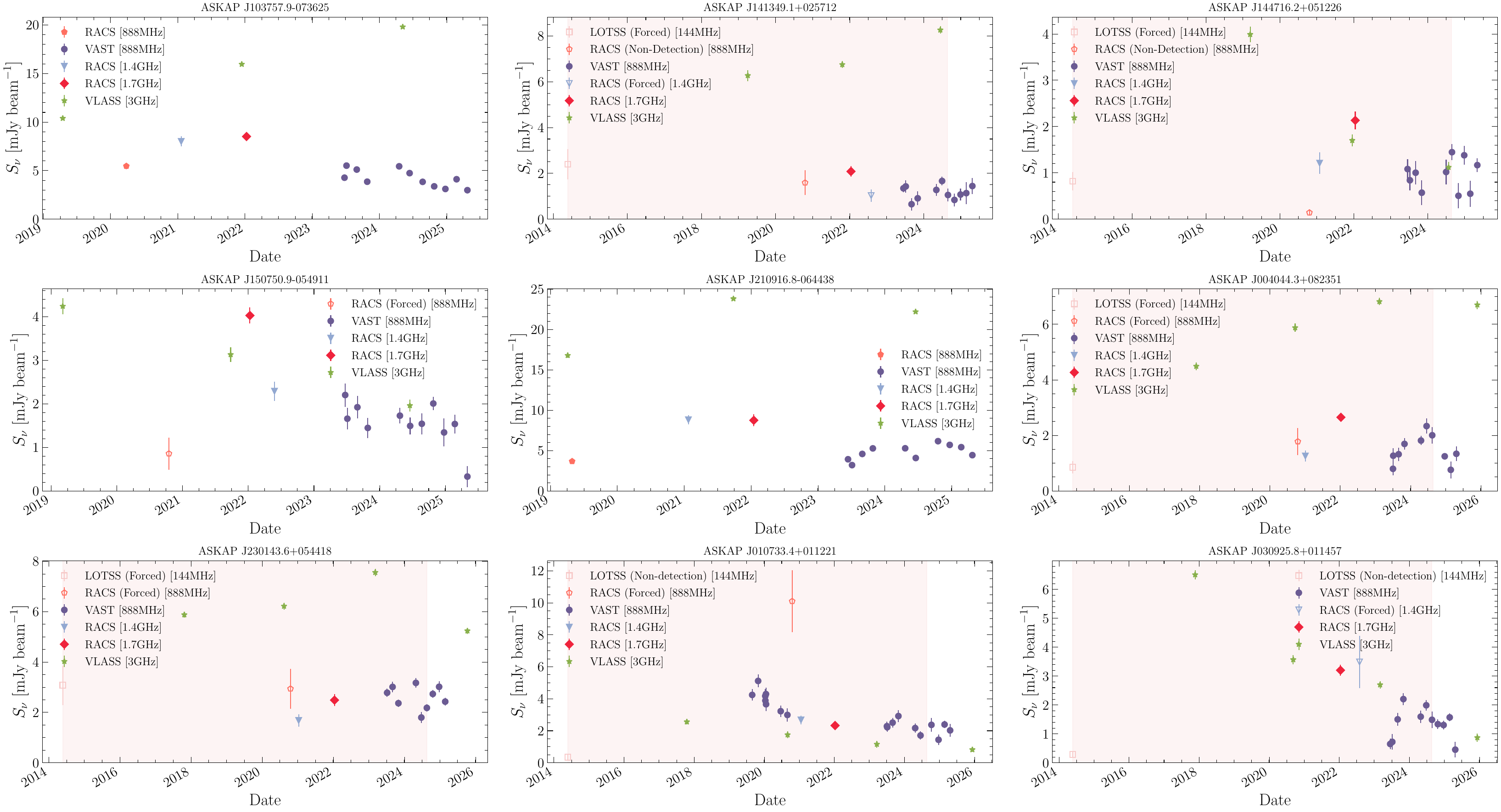}
    \caption{Multi-frequency light curves for the previously published newly radio-loud AGN interpreted to have launched young radio jets \citep{Nyland2020, wolowska2021, Zhang2022}. Purple circles represent VAST measurements, orange hexagons represent RACS-low, blue triangles represent RACS-mid, red diamonds represent RACS-high, green stars represent VLASS, and pink squares represent LoTSS. Epochs with forced photometry or non-detections are denoted with unfilled markers. LoTSS observation dates were unknown, so the survey span is highlighted in pink. ASKAP J103757.9$-$073625, ASKAP J141349.1$+$025712, ASKAP J144716.2$+$051226, ASKAP J150750.9$-$054911, ASKAP J210916.8$-$064438 were identified as gigahertz-peaked spectrum sources in \citet{Nyland2020}. ASKAP J144716.2$+$051226 and ASKAP J210916.8$-$064438 were optically identified broad-line quasars, and the remaining three were infrared-color selected AGN candidates. ASKAP J010733.4$+$011221 and ASKAP J030925.8$+$011457 were radio galaxies with gigahertz-peaked SEDs over multiple epochs in \citet{wolowska2021}. ASKAP J004044.3$+$082351 and ASKAP J230143.6$+$054418 were found to have an inverted spectral index ($\alpha\sim0.85$) between RACS-low and VLASS in \citet{Zhang2022}. ASKAP J004044.3$+$082351 was a LINER and ASKAP J230143.6$+$054418 was a Seyfert galaxy.}
    \label{fig:lit_agn}
\end{sidewaysfigure*}

We compared our radio variable AGN sample to the AGN hosting young radio jets reported in \citet{Nyland2020}, \citet{wolowska2021}, and \citet{Zhang2022}. The sources in \citet{Nyland2020} were either optically identified broad-line quasars with $0.2<z<3.2$ or infrared color selected AGN, and all exhibited peaked quasi-simultaneous SEDs with $\nu_p$ ranging from $2.5$--$22.7$~GHz. The sample presented in \citet{wolowska2021} consisted of nine radio galaxies and three quasars. They found initially that all $12$ sources had peaked quasi-simultaneous SEDs with $\nu_p$ ranging from $2$--$12$~GHz, but at later times, the quasar SEDs flattened and the radio galaxy SEDs remained peaked \citep{wolowska2021}. Very Long Baseline Array follow-up of the \citet{wolowska2021} sources revealed a small jet in some sources, but most were unresolved, and they classified the sources as core-jet or point-like objects. The sample in \citet{Zhang2022} was comprised of low redshift ($z<0.3$) LINER, Seyfert, star-forming, and passive galaxies. While they did not obtain contemporaneous radio SEDs, a comparison with RACS-low indicated these sources had either flat or inverted spectra between $887.5$~MHz and $3$~GHz. 
Nine of the recently discovered AGN were located in the VAST footprint: two broad-line quasars, three infrared color-selected AGN \citep{Nyland2020}, two radio galaxies \citep{wolowska2021}, one LINER, and one Seyfert galaxy \citep{Zhang2022} were located in the VAST field of view. We cross-matched these sources with the VAST, RACS, LoTSS, and VLASS source catalogs. All nine had a counterpart in the VAST catalog. None met our selection criteria for radio variable AGN; one failed the $S_{\nu,\text{max}}\geq1.8S_{\nu,\text{min}}$ cut, seven failed the $S_{\nu,\text{final}}>1.5S_{\nu,\text{initial}}$ cut, and one failed the $S_{\nu,\text{final}}\geq0.7S_{\nu,\text{max}}$ cut. This indicated our sources represented a different class of AGN transients Seven sources had no counterpart in the RACS-low catalog, two had no counterpart in the RACS-mid catalog, and six had no counterpart in the LoTSS catalog. To confirm non-detections in RACS and LoTSS, we obtained forced photometry with the {\fontfamily{qcr}\selectfont miriad} {\fontfamily{qcr}\selectfont
cgcurs} task with {\fontfamily{qcr}\selectfont
options$=$imstat}. The multi-frequency light curves are shown in Figure~\ref{fig:lit_agn}.
\section{Discussion}
\subsection{Non-AGN contaminants}
The infrared color selected AGN catalog we used to construct our initial sample was expected to contain up to $10\%$ false positives \citep{Assef2018}. Fifty-two of our sources were spectroscopically confirmed AGN. While the majority of radio sources at these flux density levels are AGN \citep{Murphy2026}, we investigated whether our sample of 49 non-spectroscopically confirmed AGN may have been contaminated with star forming galaxies or radio stars. By selecting for radio variability and compactness, we excluded star forming galaxies. Radio emission in star forming galaxies is much more diffuse than in AGN, as it originates in HII regions and supernova remnants spread throughout \citep{Helou1985}, and therefore unlikely to scintillate or exhibit strong intrinsic variability. Two sources, ASKAP J010459.9$-$515129 and ASKAP J080329.2$+$045904, were identified in the literature as X-ray sources years before they were observed in VAST, which further suggests they were not star forming galaxies. Radio stars, like compact AGN, are point sources and do exhibit variability \citep{Driessen2024}. The 29 sources with photometric redshifts reached radio luminosities $>10^{30}$~erg~s\textsuperscript{$-1$}Hz\textsuperscript{$-1$}, which was $\sim10$ orders of magnitude too high for a star \citep{Driessen2024}. We cross-matched our sample with the Sydney Radio Star catalog \citep{Driessen2024}, a reliability-optimized catalog of radio stars identified in ASKAP observations. The radio stars were selected based on cross-matches between ASKAP point source catalogs and volume limited stellar catalogs, high circular polarization fractions, proper motion searches, or variability \citep{Driessen2024}. None of our sources had counterparts in the Sydney Radio Star catalog. The eight optical light curves we obtained for non-spectroscopically confirmed sources showed characteristic AGN variability. Therefore, we ruled out the possibility of non-AGN contamination in our sample.
\subsection{Radio Supernovae}
During a SN, interactions between ejecta and the circumstellar medium can cause radio flares which last for years \citep{Bietenholz2021}. We investigated whether SNe in the host galaxies of any AGN in our sample may have caused the observed increase in radio emission. The most luminous radio SNe peak at $L_{\nu}\sim10^{29}$~erg~s\textsuperscript{$-1$}Hz\textsuperscript{$-1$}, and the frequency dependence of $L_{\nu}$ is expected to be weak \citep{Bietenholz2021}. The minimum radio luminosity observed in any of the 81 sources in our sample with known redshifts was $L_\nu=7.5\times10^{29}$~erg~s\textsuperscript{$-1$}Hz\textsuperscript{$-1$}, so even the brightest radio SNe would be undetectable. We found no optical SNe signatures \citep{He2026} in the ZTF light curves, which were contemporaneous with VAST. All sources in our sample, including the 20 with unknown redshifts, had some radio emission in archival data, which was consistent with radio-quiet AGN, but not radio SNe. Due to the high radio luminosities, prior radio activity, and lack of optical SNe flares, we ruled out radio SNe.
\subsection{GRBs}
GRBs can be extremely luminous in the radio, either through on-axis beamed jets or quasi-isotropic afterglows \citep{Chandra2012}. There are two types of GRBs: short GRBs, which are thought to be caused by neutron star mergers and can reach luminosities $L_\nu\sim10^{29}$~ergs~s\textsuperscript{$-1$}Hz\textsuperscript{$-1$}, and long GRBs, which are presumed to originate in SNe and can be much more luminous, up to $L_\nu\sim10^{32}$~ergs~s\textsuperscript{$-1$}Hz\textsuperscript{$-1$} \citep{Chandra2012}. Twenty-seven sources in our sample reached luminosities $>10^{33}$~erg~s\textsuperscript{$-1$}Hz\textsuperscript{$-1$}, which was too high to be due to a GRB. While we could not rule out GRBs in the 74 sources with lower (or unknown) luminosities based on their luminosity, the morphology of their light curves were inconsistent with typical radio light curves for GRBs, which decay within a few hundred days \citep{Chandra2012}. GRBs at high redshift could appear to rise and decay over longer timescales due to time dilation, but these are rare \citep{Tanvir2018}, so it is improbable that our sample included high-z GRBs. The lower luminosity sources had at least some radio emission in archival surveys, which is expected for radio-quiet AGN but not GRBs \citep{Chandra2012,Murphy2026}. Therefore, we ruled out GRBs as a cause of the increasing radio emission.
\subsection{Propagation Effects}
Radio emission from a sufficiently compact extragalactic source will scatter as it propagates through the interstellar and interplanetary medium, which causes the source to appear variable, especially at lower frequencies \citep{Walker1998,Rickett1984}. We investigated whether our sources could be variable due to propagation effects. 
\subsubsection{Interplanetary Scintillation}
Interplanetary scintillation \citep[IPS;][]{Rickett1984}, which is caused by plasma in the solar wind distorting the incoming radio waves, is strongest at very low frequencies, and much weaker at at $887.5$~MHz. IPS is correlated with the motion of the Sun, which leads to annual cycles of variability and the strongest scintillation being observed in objects with small solar elongations \citep{Marchili2025}. Our sample was distributed isotropically across the VAST footprint, not concentrated near the ecliptic, which would not be true if IPS was the dominant cause of variability \citep{Marchili2025}. We found no evidence for annual periodicity, which is expected for IPS, in our sample. Given the relatively weak effects of IPS expected at $887.5$~MHz, the lack of any discernible annual periodicity, and the arbitrary location of our sources with respect to the ecliptic, we concluded that IPS was not responsible for the observed radio variability in our sample. 
\subsubsection{Extreme Scattering Events}
Radio light curves of quasars can show variability due to extreme scattering events [ESEs;][]\citep{Fiedler1987}, which are thought to be caused by particularly dense plasma structures within the Milky Way along the line of sight. ESEs were originally discovered in high-cadence $2.3$~GHz and $8.3$~GHz monitoring of radio AGN, and are characterized by U-shaped drops in flux density bracketed by smaller increases, or, more rarely, by strong spikes \citep{Fiedler1987}. Unlike our observed light curves, ESEs show variability over timescales shorter than $\sim1$~year and do not result in sustained increased flux densities \citep{Fiedler1987}. The lensing structures responsible for ESEs are not well understood, since they require large, long-lived over-pressured regions, which cannot be generated via Kolmogorov turbulence in the interstellar medium \citep{Bannister2016}. The temporal ESE behavior has not been well characterized observationally at lower frequencies ($<1$~GHz) \citep{Dong2018, Bannister2016, Murphy2013}, and ESE signatures are strongly frequency-dependent compared to intrinsic AGN variability \citep{Bannister2016}. ESE models assuming a two-dimensional axisymmetric lensing structure (such as in the case of self-gravitating clouds) have predicted four different classes of ESE light curve morphologies, depending on the orientation of the lens with respect to the line of sight \citep{Dong2018}. For some orientations, the models predicted the U-shaped light curves observed in ESEs at $2.3$~GHz and $8.3$~GHz and the three strong spikes observed in some ESEs at $8.3$~GHz \citep{Dong2018}, neither of which resembled the morphologies of our VAST light curves. At lower frequencies, the models predicted ESE light curves with only one or two spikes, and no decreases in flux density \citep{Dong2018}. However, the axisymmetric plasma lens is only one proposed ESE model \citep{Bannister2016, Dong2018}, and the predicted ESE timescales depend on the size and distance of the lensing structure, which is unknown \citep{Dong2018}. While a single or double-peaked ESE signature predicted by theoretical simulations may be consistent with some of our VAST light curves, we note that our light curves were inconsistent with all observed ESEs \citep{Dong2018, Bannister2016, Fiedler1987}. Rapid follow-up observations could be used to rule out ESE-like spectral signatures in future work \citep{Bannister2016}.
\subsubsection{Interstellar Scintillation}
\begin{figure*}[htbp!]
    \gridline{\fig{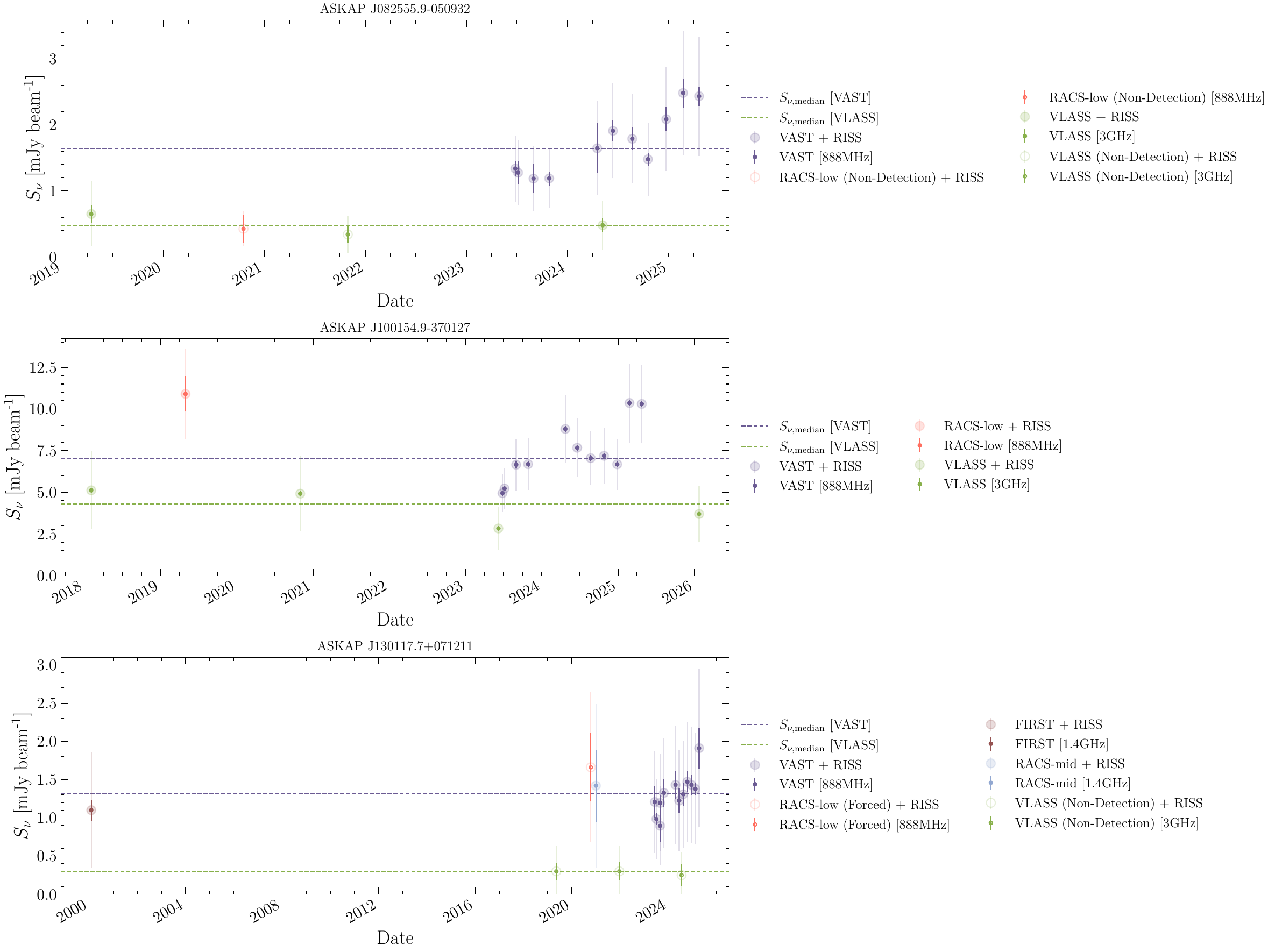}{0.98\textwidth}{}}
    \caption{Our scintillation-corrected light curves, which incorporated an additional uncertainty in the flux density measurements, revealed three different behaviors at $887.5$~MHz, which are shown here. From top to bottom: brightening light curves which could not be explained by RISS (26); variable light curves which could not be explained by RISS, but the addition of the RACS-low epoch indicates a persistent radio source (15); light curves which were consistent with RISS-modulated non-variable emission (60). At higher frequencies ($3$~GHz and $1.4$~GHz), all light curves were consistent with RISS as the only source of variability. Purple markers indicate VAST epochs, orange markers indicate RACS-low epochs, green markers indicate VLASS epochs, blue markers indicate RACS-mid epochs, and brown markers indicate FIRST epochs. Solid markers indicate original errorbars for catalog detections, unfilled markers indicate original errorbars for forced detections and non-detections, and translucent markers indicate the RISS-corrected uncertainties. Purple and green dashed lines indicate the median flux density across VAST and VLASS, respectively.\label{fig:scintillation}}
\end{figure*}

Interstellar scintillation is caused by turbulence in the interstellar medium along our line of sight. At $887.5$~MHz, our observations were firmly in the strong scattering regime \citep{Walker1998,Rickett1984}, and two different types of distortions may have occurred: diffractive interstellar scintillation \citep[DISS;][]{Lang1969}, rapid, large intensity fluctuations caused by the superposition of scattering through individual patches of plasma; and refractive interstellar scintillation \citep[RISS;][]{Rickett1984}, slower and smaller fluctuations introduced by refocusing of the light as it propagates. Only extremely compact objects -- pulsars and GRBs in their earliest phases -- are subject to DISS \citep{Lang1969}, and AGN are too large to be affected. The compactness requirements for RISS are less stringent \citep{Rickett1984,Hancock2019}, so RISS was likely responsible for some variability in our AGN. 
We introduced an additional error component to account for variability due to RISS in our sources. The flux modulation due to RISS is dependent on the observing frequency and the density of the intervening interstellar medium. We used {\sc
RISS19}\citep{Hancock2019} to compute the modulation index $m$, which is a measure of scatter due to scintillation, at the coordinate location of each source based on H$\alpha$ maps of the sky. Following the strategy used by \citet{Gulati2026}, we introduced an additional error to our flux density measurements, $\sigma_{S_{\nu,\text{RISS}}}=mS_{\nu}$, which we added in quadrature to determine the total error. We implemented this correction for the VAST, RACS-low, RACS-mid, FIRST, and VLASS epochs. If the difference between the maximum and minimum flux density was outside the RISS-corrected error bars, we classified the source as intrinsically variable. The scintillation-corrected light curves revealed that 60 sources were consistent with RISS-modulated non-variable emission at $887.5$~MHz. The 41 sources with $887.5$~MHz variability which could not be explained by RISS alone fell into two categories: 15 had faded between RACS-low and the first VAST epoch, indicating they were persistent radio emitters with some intrinsic, but periodic variability; 26 had brightened across the RACS-low and VAST epochs. Figure~\ref{fig:scintillation} shows scintillation-corrected light curves for each type of source: variable due to RISS only, periodically variable due to some intrinsic cause, and increasing in brightness due to intrinsic cause. At higher frequencies, the scintillation-corrected light curves revealed that any variability in the 83 VLASS light curves and the 30 two-epoch $1.4$~GHz light curves could be attributed entirely to RISS. Eighteen sources had brightened at $887.5$~MHz but not at $3$~GHz over the same time period, which suggested the emitting region had expanded and become less optically thick at lower frequencies. 

\begin{figure*}[htbp!]
    \caption{Time-varying spectral behavior of our sources. Orange markers indicate the sources which had brightened due to intrinsic causes and purple markers indicate persistent sources. Marker shapes denote the original spectral classifications we determined using the median VAST flux density: steep (triangle markers), inverted (star markers), peaked (hexagonal markers), or flat (circle markers). The left figure shows the change in spectral index across the light curve ($\alpha_{\text{final}}-\alpha_{\text{initial}}$) versus the initial spectral index. The right figure shows the change in spectral index between the first and last epochs versus the maximum difference in $\alpha$ for each source. The intrinsically brightening AGN tended to have inverted SEDs initially, and decreased more dramatically, while the persistent sources had flatter SEDs which did not change over time. \label{fig:spec_idx_pop}}
    \gridline{\fig{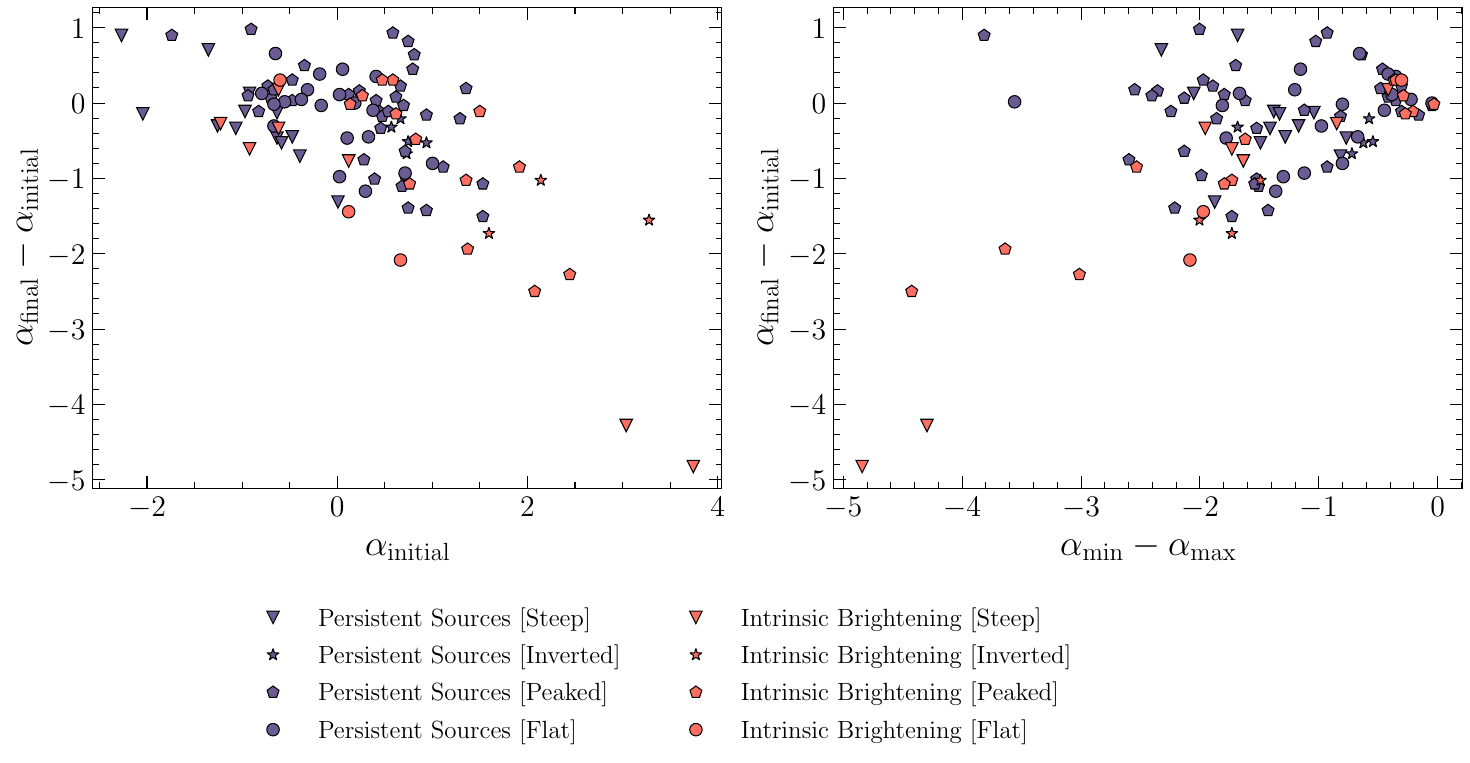}{0.98\textwidth}{}}
\end{figure*}

The time-varying spectral behavior of the persistent radio sources -- those with variability we attributed to RISS or periodic intrinsic causes -- and the sources brightening due to intrinsic causes are shown in Figure~\ref{fig:spec_idx_pop}. Thirteen ($50$\%) of the intrinsically brightening AGN had spectral indices which decreased over time, but only 9 ($12$\%) of the persistent radio sources showed decreasing spectral indices. The sources we found to be brightening due to intrinsic causes exhibited the most dramatic changes in spectral index.

Seven sources (Table~\ref{tab:youngjets}) we determined were variable due to scintillation or other periodic causes had inverted SEDs ($\alpha>0.5$), and all were radio-loud. This indicates these are jetted AGN in their early phases of evolution. Our ATCA observations confirmed ASKAP J212956.7$-$385231 had an inverted SED between $4.5$--$11.7$~GHz, which suggests this is an especially compact radio AGN. High-frequency peaked sources ($\nu_p\geq5$~GHz in the observer frame) are the most compact class of peaked spectrum sources, with linear sizes $<500$~pc \citep{Ballieux2024}. ASKAP J212956.7$-$385231, which had a spectroscopic redshift $z=1.2$, had an observer frame $\nu_p\geq12$~GHz, which corresponded to a rest frame turnover $\nu_{0,p}\geq26.3$~GHz. The low-frequency absorption in compact radio sources is thought to be due primarily to SSA \citep{Odea2021}, which has $\alpha_{\text{inv}}=2.5$ regardless of the underlying electron energy distribution \citep{Rybicki1979}. The spectral indices observed in the seven inverted SEDs, including in our quasi-simultaneous ATCA observations of ASKAP J212956.7$-$385231, were too shallow to be explained by SSA alone. This suggested the absorption mechanism was more complex, and included multiple SSA components or FFA. The light curves and SEDs for these sources are included in Figure~\ref{fig:nv_babies} in Appendix~\ref{appx:nv_babies}.
\begin{deluxetable}{lllr}[htbp!]
    \tablecaption{Radio-loud AGN with persistent radio emission and inverted SEDs, which suggest they host jets in the early phases of their evolution. We included the spectral slope determined using the ATCA SED for ASKAP J212956.7$-$385231.}
    \label{tab:youngjets}
    \tablewidth{0.45\textwidth}
    \setlength{\tabcolsep}{2.1pt}
    \tablehead{
    \colhead{ASKAP Source} & \colhead{$\alpha_{\text{inv}}$} & \colhead{z} & \colhead{Variability Type}}
    \startdata
    ASKAP J013753.7$-$524813 & 0.7 & ... & RISS \\
    ASKAP J035553.9$-$461757 & 0.5 & 1.0 & RISS \\
    ASKAP J094055.3$-$043235 & 0.5 & ... & Periodic \\
    ASKAP J150801.4$-$041936 & 0.5 & 1.7 & RISS \\
    ASKAP J152545.9$+$044126 & 0.6 & 1.0 & RISS \\
    ASKAP J154903.1$+$060252 & 0.8 & 0.6 & RISS \\
    ASKAP J212956.7$-$385231 & 1.0 & 1.2 & RISS \\
    \enddata
\end{deluxetable}
\subsection{Transient AGN Activity}
We determined 26 sources (Table~\ref{tab:favorites}) had brightened at $887.5$~MHz due to intrinsic AGN activity.

This sample consisted of three AGN with inverted SEDs, 13 with peaked SEDs (three with $\nu_p=1655.5$~MHz and ten with $\nu_p=1367.5$~MHz), nine with steep SEDs, and one with a flat SED ($\alpha=-0.4$). Eighteen AGN were radio-loud ($S_{5\text{GHz}}/S_g\ge10$), four were radio-quiet ($S_{5\text{GHz}}/S_g<10$), two had transitioned from radio-quiet to radio-loud, and one had no optical counterpart. The radio-quiet AGN included three sources with peaked SEDs and one with a steep SED. The two AGN which had transitioned from radio-quiet to radio-loud, ASKAP J040202.8$-$321845 and ASKAP J105726.7$+$073745, were especially promising candidates for hosting newly forming radio jets. Thirteen sources had spectral indices which decreased over time, and the other 13 had spectral indices which remained roughly constant.

\begin{deluxetable*}{llllllll}[htbp!]
    \tablecaption{Sources brightening due to intrinsic AGN activity. Indicated spectral types, spectral indices, and turnover frequencies were determined using the median VAST flux density. Radio-loudness (RL) classes are indicated as: radio-loud (RL), radio-quiet (RQ), or transitioning from RQ to RL (T). Photometric redshifts are denoted $z$*.}
    \label{tab:favorites}
    \tablewidth{0.45\textwidth}
    \setlength{\tabcolsep}{2.1pt}
    \tablehead{
    \colhead{ASKAP Source} & \colhead{Spectral Type} & \colhead{$\alpha_{\text{inv}}$} & \colhead{$\alpha_{\text{steep}}$} & \colhead{$\nu_p$ [GHz]} & \colhead{RL Class} &  \colhead{z} & \colhead{SED Evolution}}
    \startdata
    ASKAP J031738.6$+$033454 & Peaked & 0.5 & -1.8 & 1.4 & RQ & 0.3 & Constant \\
    ASKAP J033602.9$+$051128 & Peaked & 0.5 & -0.6 & 1.4 & RL &0.2* & Constant \\
    ASKAP J034921.5$+$075101 & Peaked & 1.5 & -0.2 & 1.7 & RL & 2.1 & Constant \\
    ASKAP J035501.0$+$091248 & Peaked & 0.2 & -0.6 & 1.7 & RL &0.6* & Constant \\
    ASKAP J040202.8$-$321845 & Peaked & 1.0 & -0.4 & 1.4 & T & 1.5 & Peaked to Steep \\
    ASKAP J050500.0$-$450447 & Inverted & 2.1 & ... & ... & RL &1.4 & Flattened \\
    ASKAP J061401.5$-$551306 & Inverted & 1.6 & ... & ... & RL &1.9 & Flattened \\
    ASKAP J082555.9$-$050932 & Steep & ... & -0.7 & ... & RQ & 2.7 & Peaked to Steep \\
    ASKAP J085302.8$+$075154 & Peaked & 0.1 & -1.2 & 1.4 & RL&1.2 & Constant \\
    ASKAP J101432.5$-$401536 & Flat & ... & -0.4 & ... & ...&...& Peaked to Steep\\
    ASKAP J103737.9$-$292240 & Steep & ... & -0.6 & ... & RL&...& Constant \\
    ASKAP J104514.5$-$394654 & Peaked & 0.4 & -0.5 & 1.4 & RL&...& Constant \\
    ASKAP J104855.1$-$340145 & Steep & ... & -1.3 & ... & RL&...& Constant \\
    ASKAP J105012.6$-$402353 & Steep & ... & -0.7 & ... & RL&...& Constant \\
    ASKAP J105726.7$+$073745 & Steep & ... & -0.5 & ... & T&0.2 & Constant \\
    ASKAP J112810.8$-$305103 & Peaked & 1.3 & -1.2 & 1.4 & RL&...& Peaked to Steep \\
    ASKAP J113825.8$-$330537 & Steep & ... & -0.5 & ... & RL&...& Peaked to Steep \\
    ASKAP J114402.9$-$422622 & Peaked & 0.9 & -4.0 & 1.4 & RQ&3.1& Flattened $\leq\nu_p$ \\
    ASKAP J125050.8$-$445757 & Peaked & 0.6 & -1.1 & 1.4 & ...&... & Peaked to Steep \\
    ASKAP J171659.5$+$033419 & Steep & ... & -0.7 & ... & RL&...& Peaked to Steep \\
    ASKAP J174307.9$+$075614 & Steep & ... & -0.6 & ... & RL&...& Constant \\
    ASKAP J194551.5$-$012544 & Peaked & 2.2 & -0.5 & 1.4 & RL&...& Flattened $\leq\nu_p$\\
    ASKAP J201534.2$-$444703 & Peaked & 0.4 & -1.8 & 1.4 & RQ&0.8 & Flattened $\leq\nu_p$\\
    ASKAP J203349.5$-$421543 & Steep & ... & -0.5 & ... & RL&0.6 & Peaked to Steep \\
    ASKAP J214147.3$+$061116 & Peaked & 0.6 & -0.6 & 1.7 & RL&3.2 & Constant \\
    ASKAP J224239.7$-$303517 & Inverted & 0.6 & ... & ... & RL&0.5& Flattened\\
    \enddata
\end{deluxetable*}
For these sources, we investigated four possible causes of intrinsic variability: beaming effects, non-jetted TDEs, jetted TDEs, and jets launched by changes in the AGN.

\subsubsection{Blazars}
Minor brightness fluctuations caused by shocks or instabilities in the AGN jet will be amplified due to relativistic beaming if the jet is aligned close to our line of sight \citep{degouveia2005}. We had excluded known or suspected blazars from our initial AGN catalog before cross-matching with VAST, but our sample could have been contaminated with previously unidentified blazars.

There are two main types of blazars, BL Lacs and Flat Spectrum Radio Quasars \citep[FSRQs;][]{Massaro2015}. BL Lacs are characterized by featureless optical spectra and FSRQs are characterized by flat ($|\alpha|<0.5$) radio SEDs \citep{Massaro2015}. We obtained optical spectra with clear spectral features for four sources (one from SDSS-IV, two from DESI, and one using the HET). Another nine sources had cataloged spectroscopic redshifts in the Gaia Extragalactic and CT survey catalogs; while we did not directly obtain the original spectra, they must have had spectral features in order for their spectroscopic redshifts to be included in the survey catalogs with no redshift warnings. BL Lac contamination was unlikely for the 13 spectroscopically confirmed AGN. Additional spectroscopic follow-up of the remaining 13 AGN is needed to rule out BL Lac contamination.

Only one source (ASKAP J101432.5$-$401536) in our sample of intrinsically brightening AGN had an SED we classified as flat ($\alpha=-0.4$). However, the time-varying spectral index fits indicated ASKAP J101432.5$-$401536 had an SED which was transitioning from peaked to steep. Given the spectral evolution and the proximity of $\alpha$ to the cutoff for a steep spectrum ($\alpha\leq-0.5$), we concluded this source was unlikely to be a FSRQ. The remaining 25 sources had SEDs which were too steep to be FSRQs, including two sources with quasi-simultaneous ATCA SEDs. 

Given our initial blazar exclusion criteria, the observed optical spectral features, and the steep radio spectral indices, we concluded relativistic beaming in on-axis jets was unlikely to explain the radio-brightening in the 26 AGN. 
\subsubsection{Nuclear Transients and Off-Axis Radio Jets}
An off-axis radio jet is characterized by a broken power-law radio SED \citep{Rybicki1979,Odea2021}. At higher frequencies, the emitting region is optically thin ($\alpha<0$), and we observe a synchrotron spectrum \citep{Odea2021}. The shape of the spectrum from synchrotron radiation depends on the electron energy distribution in the source; in AGN and TDEs, a power-law electron energy distribution is usually assumed, since this results in the typical observed power-law radio SED \citep{Odea2021}. At lower frequencies, the jet is optically thick ($\alpha>0$). The primary absorption mechanism is thought to be SSA, which results in a power-law spectrum with $\alpha=5/2$ regardless of the underlying energy distribution \citep{Rybicki1979,Odea2021}. FFA is suspected to play a role in the low-frequency absorption, especially in denser circumnuclear environments, which complicates the spectral shape \citep{Odea2021}. As the synchrotron emission region expands, the source becomes less optically thick, and the turnover frequency shifts lower \citep{Odea2021, Ballieux2024}. The most compact ($<500$~pc) radio AGN peak above $5$~GHz, and the most extended ($>1$~kpc) either peak below $\sim1$~GHz or lack the low-frequency absorption entirely \citep{Ballieux2024}. This is expected to occur in jets launched by both AGN activity and TDEs \citep{Odea2021, Alexander2020}. 16 of the intrinsically brightening AGN had peaked or inverted SEDs with $\nu_p\geq1367.5$~MHz, indicating these were gigahertz or high-frequency peaked sources. In the most dramatic case, ASKAP J224239.7$-$303517 had an inverted SED between $887.5$~MHz--$3$~GHz ($\nu_p>3$~GHz). This suggested these AGN host young, compact ($<1$~kpc in size) jets. The spectral index flattened in all three sources with inverted SEDs, ASKAP J050500.0$-$450447, ASKAP J061401.5$-$551306, and ASKAP J224239.7$-$303517, as expected for an expanding emission region. Seven sources (six radio-loud and one radio-quiet) had peaked SEDs and constant spectral indices across the RACS-low/VAST epochs, which indicates the optical depth of the emission region had not changed as the source brightened. Three sources (ASKAP J040202.8$-$321845, ASKAP J112810.8$-$305103, and ASKAP J125050.8$-$445757) had peaked SEDs which transitioned to steep SEDs, indicating $\nu_p$ shifted from $\geq1367.5$~MHz to $\leq887.5$~MHz, consistent with a more evolved, expanding radio jet. ASKAP J112810.8$-$305103 was radio-loud, ASKAP J125050.8$-$445757 had no optical counterpart, and ASKAP J040202.8$-$321845 had transitioned from radio-quiet to radio-loud. For the remaining three sources (ASKAP J114402.9$-$422622, ASKAP J194551.5$-$012544, and ASKAP J201534.2$-$444703) with peaked SEDs, the spectral index flattened below $\nu_p$. This indicates that the emission regions in the sources with flattening $\alpha_{\text{inv}}$ have expanded, but remain more compact than the peaked spectrum sources which transitioned to steep spectrum sources. ASKAP J114402.9$-$422622 and ASKAP J201534.2$-$444703 were radio-quiet AGN, and ASKAP J194551.5$-$012544 was radio-loud. All six sources had the same low-frequency limit ($887.5$~MHz) and the same turnover ($\nu_p=1367.5$~MHz) in the SEDs we fit using the median VAST flux density, so their differing spectral evolution suggested they had expanded at different rates.

Five sources with steep SEDS, ASKAP J082555.9$-$050932, ASKAP J101432.5$-$401536, ASKAP J113825.8$-$330537, ASKAP J171659.5$+$033419, and ASKAP J203349.5$-$421543, initially had peaked SEDs which transitioned into steep SEDs, placing them in the same class as the more evolved peaked SED sources. The spectral indices remained constant in the other four sources with steep SEDs, ASKAP J103737.9$-$292240, ASKAP J104855.1$-$340145, ASKAP J105012.6$-$402353, ASKAP J105726.7$+$073745, and ASKAP J174307.9$+$075614. This suggested the jets in these sources were the most evolved, and the brightening was due to the jet becoming more powerful. ASKAP J105726.7$+$073745 and ASKAP J174307.9$+$075614, which had steep indices from $144$~MHz--$3$~GHz, were compact steep spectrum sources, the largest and most evolved class of compact radio AGN \citep{Odea2021}.
\subsubsection{Tidal Disruption Events}
TDEs, which occur when a star is ripped apart by tidal forces from the central supermassive black hole, can be accompanied by radio flares \citep{Dykaar2024,Alexander2020}. Most known radio TDEs were first identified in other observing bands and occurred in non-active galaxies, since TDEs can mimic typical AGN variability in the radio \citep{Dykaar2024,Alexander2020}. As a result, the observational properties of radio-selected TDEs in AGN are poorly constrained. TDE radio emission, which is dominated by synchrotron, can originate in either short-lived jets (jetted TDEs) or from the interaction between stellar debris and the surrounding environment (non-jetted TDEs) \citep{Alexander2020,Dykaar2024}. Non-jetted TDEs and off-axis jetted TDEs are much less luminous, and can be classified as radio-quiet TDEs, while on-axis jetted TDEs are relativistically beamed toward us and appear radio-loud \citep{Alexander2020}. Radio-loud TDEs are much rarer, but can be observed out to much higher redshifts. Since our sample consisted of higher redshift AGN, with $z=0.14$--$3.1$, they were more likely to be on-axis jetted events.

Distinguishing between TDEs and AGN jet activity based on radio variability is challenging. Multi-wavelength, multi-epoch observations of TDEs which launched jets have shown that, similarly to AGN, the radio emission originates primarily inside the jet, rather than via interaction with the interstellar medium \citep{Alexander2020}. The observational characteristics of TDEs on longer timescales ($>1$~year post-discovery) have not been well constrained, since only a handful of radio TDEs have been closely monitored on similar time scales to the light curves in our sample \citep{Rhodes2025}. 

A handful of radio-selected TDEs have been discovered in the VAST pilot survey based on their light curve morphologies \citep{Dykaar2024}. To establish their light curve morphology criteria, \citet{Dykaar2024} first simulated the VAST light curves of on-axis jetted, off-axis jetted, and non-jetted TDEs with different outflow energies and ambient interstellar medium densities. The TDE flares in their simulations and observational sample began to fade after $\sim1$~year \citep{Dykaar2024}, while our sources continued to brighten over a longer period. Compared to the VAST pilot TDEs, our sources had higher redshifts and luminosities. The TDEs in the \citet{Dykaar2024} sample ranged in maximum luminosity (with no K-correction) from $4.1\times10^{29}$~erg~s\textsuperscript{$-1$}Hz\textsuperscript{$-1$}--$2.0\times10^{32}$~erg~s\textsuperscript{$-1$}Hz\textsuperscript{$-1$}, while ours ranged from $3.8\times10^{30}$~erg~s\textsuperscript{$-1$}Hz\textsuperscript{$-1$}--$2.0\times10^{33}$~erg~s\textsuperscript{$-1$}Hz\textsuperscript{$-1$}. The TDE sample ranged in redshift from $0.05$--$0.8$, and our final sample ranged from $z=0.1$--$3.1$.

However, long-term, multi-frequency follow-up observations of jetted TDEs have shown radio flares can persist for years afterward. \citet{Rhodes2025} monitored AT 2022cmc, an optically identified TDE at $z=1.2$ suspected of launching a relativistic jet at six observing frequencies between $1.3$--$102$~GHz from $\sim100$--$\sim1000$ days post-discovery. They found that the radio emission had already began to fade at $102$~GHz and $87$~GHz by $\sim100$~days, but did not peak at frequencies between $3$--$5$~GHz until $\>600$~days \citep{Rhodes2025}. The brightening observed in our sources over a similar time frame could be the delayed low-frequency fading from a TDE. At all times, AT 2022cmc exhibited a peaked SED, but $\nu_p$ shifted from $>15$~GHz at $\sim100$~days to $3$~GHz at $\sim900$~days \citep{Rhodes2025}. In eight of our sources, $\nu_p$ shifted from $\nu_p\geq1367.5$~MHz to below $887.5$~MHz over a similar time span. Below the turnover, the spectral index remained constant across all epochs for AT 2022cmc \citep{Rhodes2025}, in contrast to the six sources in our sample which showed flattening low-frequency SEDs. Long-term monitoring of a larger sample of radio TDEs is needed to determine whether the behavior observed in AT 2022cmc is representative of all jetted radio TDEs. 

Non-jetted TDEs which were not initially detected in the radio have been observed to brighten at $3$--$6$~GHz at late times ($700$--$3200$~days), which has been interpreted as delayed outflows \citep{Cendes2024}. Out of a sample of 24 optically-selected, low redshift ($z<0.16$) TDEs, ten were found to have new or re-brightening radio emission in ten sources on timescales comparable to the brightening observed in our sample \citep{Cendes2024}. Quasi-simultaneous multi-frequency observations revealed gigahertz-peaked SEDs in five of the late-time brightening TDEs \citep{Cendes2024}. Multi-epoch SEDs were obtained for three of the peaked spectrum TDEs, and these showed that the turnover frequency had decreased over a few hundred days \citep{Cendes2024}. While some of our sources showed similar spectral evolution, it is unlikely that the radio emission was due to delayed outflows, rather than a jet. The TDEs observed by \citet{Cendes2024} were detected at similar flux levels to our sources ($\leq2$~mJy~beam\textsuperscript{$-1$}) but were significantly lower redshift ($z\leq0.16$). For the fifteen sources with known redshifts ($z=0.2$--$3.2$) in our sample of intrinsically brightening AGN, the outflow-driven radio emission observed by \citet{Cendes2024} would be undetectable in VAST. The remaining eleven sources were unlikely to be AGN at such low redshifts ($z<0.16$), since they were very faint ($g>22$ or non-detections in surveys with detection limits $g\sim22$). Therefore, we ruled out non-jetted TDEs as the cause of the increasing radio emission in our sample.

The mass of the central supermassive black hole can be one tool for differentiating AGN-launched jets from TDE-launched jets. For a TDE to occur, the tidal disruption radius must lie outside the event horizon, which requires $M_\bullet<10^8 M_\odot$ for a solar-type star disrupted by a non-spinning black hole \citep{Hammerstein2023}. Both theoretical predictions and observational results suggest TDEs are more common in galaxies with lower mass black holes ($M_\bullet \sim 10^5-10^6M_\odot$) \citep{Hammerstein2023}. The highest black hole masses ($M_\bullet\sim10^7 M_\odot$) have been observed in featureless TDEs, which lack the broad Balmer and HeII emission lines typically observed in TDEs and are several orders of magnitude more optically luminous than other TDE classes \citep{Hammerstein2023}. For a spinning black hole or a giant star, the maximum black hole mass increases slightly ($M_\bullet\sim10^8M_\odot$), but the fraction of TDEs caused by spinning black holes is predicted to be marginal, and tidal disruptions of giant stars should occur over longer timescales and with lower optical luminosities \citep{Hammerstein2023}. 

The black hole mass for one source in our sample of intrinsically brightening AGN, ASKAP J214147.3$+$061116, has been estimated to be $\log{M_\bullet/M_\odot}=9.6$ \citep{Kozlowski2017}, which is too high for a TDE. In remaining 25 sources in our sample, the high radio-loudness and lack of flares in contemporary optical observations could be explained by a TDE involving a giant star. Further observations are needed to estimate the black hole masses to rule out this possibility.
\subsubsection{Young Radio Jets}
We concluded that the increasing radio emission observed in the 26 sources with variability due to intrinsic AGN activity was due to changes in radio jets. The jets appeared to be in various stages of their evolution, and only 14 showed changes in their SEDs over time. The sources with inverted or peaked SEDs likely hosted the youngest jets, while the sources with constant steep SEDs were more evolved jets which were becoming more powerful. Aside from ASKAP J214147.3$+$061116, further investigation is needed to determine whether these jets were launched by changes in the AGN or by TDEs. Very Long Baseline Interferometry observations can constrain the spatial extent and orientation of the jets \citep{wolowska2021}, and optical or infrared spectroscopic observations can be used to estimate the virial black hole masses \citep{Kozlowski2017}. Long-term observations of larger samples of optical and X-ray-selected TDEs in the radio are needed to better characterize their temporal evolution after $100$~days \citep{Rhodes2025}. 
\subsection{Behavior of Previously Identified Newly Radio-Loud AGN}
We investigated the recent behaviors the nine previously discovered radio-brightening AGN \citep{Nyland2020,wolowska2021,Zhang2022} in the VAST footprint. Our results are summarized in Table~\ref{tab:old_agn}. None of these sources exhibited VAST light curves meeting our selection criteria, and none had continued to brighten in both VLASS and VAST. Notably, three sources faded in both surveys: ASKAP J150750.9$-$054911, ASKAP J010733.4$+$011221 and ASKAP J030925.8$+$011457. ASKAP J150750.9$-$054911 is a gigahertz-peaked spectrum AGN \citep{Nyland2020}, and ASKAP J010733.4$+$011221 and ASKAP J030925.8$+$011457 are radio galaxies \citep{wolowska2021}. The radio galaxies had lower black hole masses ($\log{M_\bullet/M_\odot}=7.1$ and $\log{M_\bullet/M_\odot}=6.7$, respectively) \citep{wolowska2021}, which, combined with their multi-frequency fading, suggest these may be TDEs. There has been no spectroscopic follow up of ASKAP J150750.9$-$054911, an infrared-color selected AGN, which leaves the cause of radio variability ambiguous. The two quasars \citep{Nyland2020}, the Seyfert galaxy, and the LINER \citep{Zhang2022} were reported with black hole masses ranging from $\log{M_\bullet/M_\odot}=8.4$--$9.1$, which ruled out TDEs in these sources. The diverse behaviors of these sources suggested these samples may include a mix of short term jet activity and long term onsets. 
\begin{deluxetable}{lllllll}
    \tablecaption{AGN previously discovered to have launched young radio jets in the VAST field of view. All redshift values are spectroscopic and presented in the original reference paper, except those denoted with *, which we obtained from the DESI catalog. Black hole masses are presented in the original reference paper. Gigahertz-peaked SEDs are indicated by GPS.}
    \label{tab:old_agn}
    \tablewidth{0.45\textwidth}
    \setlength{\tabcolsep}{2.1pt}
    \tablehead{
    \colhead{ASKAP Source} & \colhead{Reference} & \colhead{AGN Subclass} & \colhead{SED} & \colhead{Fading} & \colhead{z} & \colhead{$\log{M_\bullet}/M_\odot$}}
    \startdata
	J103757.9$-$073625 & \citet{Nyland2020} & ... & GPS & No & 1.1*$\phantom{0}$ & ... \\
    J141349.1$+$025712 & \citet{Nyland2020} & ... & GPS & No & 2.4*$\phantom{0}$ & ... \\
    J144716.2$+$051226 & \citet{Nyland2020} & Quasar & GPS & No & 1.7$\phantom{0}$ & 9.0 \\
    J150750.9$-$054911 & \citet{Nyland2020} & ... & GPS & Yes & ... & ...\\
    J210916.8$-$064438 & \citet{Nyland2020} & Quasar & GPS & No & 1.1$\phantom{0}$ & 9.1 \\
    J010733.4$+$011221 & \citet{wolowska2021} & Radio galaxy & GPS & Yes & 0.12 & 7.1 \\
    J030925.8$+$011457 & \citet{wolowska2021} & Radio galaxy & GPS & Yes & 0.04 &  6.8 \\
    J004044.3$+$082351 & \citet{Zhang2022} & LINER & Inverted & No & 0.21 & 9.1 \\
    J230143.6$+$054418 & \citet{Zhang2022} & Seyfert & Inverted & No & 0.14 & 8.4 \\
    \enddata
\end{deluxetable}

\section{Conclusions}
We identified a sample of 101 compact radio AGN brightening by $80-1800\%$ in the VAST survey out of an original catalog of 64,972 radio-detected AGN. We investigated the causes of the observed radio brightening by obtaining optical spectroscopic observations with the HET and SOAR and radio SEDs using archival data and ATCA observations.

We determined that our sources were too luminous to be explained by SNe in the host galaxy or late-time radio outflows from TDEs. The light curve morphologies were inconsistent with GRB afterglows, IPS, or observed ESEs.

We determined that 60 sources showed variability which was consistent with RISS, and 15 had faded between RACS-low and the first VAST epoch. Seven of these sources had inverted SEDs across all observing frequencies (up to $12$~GHz), indicating the emission was due to jets in their very early stages. 

We found 26 sources had brightened at $887.5$~MHz due to intrinsic AGN activity, but were non-variable at higher frequencies ($1$--$3$~GHz). We ruled out beaming as the cause of variability based on the presence of features in their optical spectra and their steep ($|\alpha|>0.5$) radio spectral indices.

Six of the intrinsically brightening AGN had inverted or peaked SEDs which flattened below $\nu_p$, consistent with an expanding young radio jet. Eight sources had peaked SEDs initially, which transitioned to steep SEDs, including one AGN which transitioned from radio-quiet to radio-loud over the same time frame. Seven sources had peaked SEDs which did not change over time. We concluded that the AGN with peaked SEDs hosted young jets, but were evolving at different rates, which suggests they formed under different physical conditions. Five sources had steep SEDs which did not evolve in time, which indicated they were established jets becoming more powerful.

We concluded that the intrinsic radio brightening was due to changes in very young radio jets. Aside from one source with a reported virial black hole mass $\log{M_\bullet/M_\odot}=9.6$, we could not rule out jetted TDEs in favor of jets launched by changes in the AGN accretion.

We obtained multi-frequency radio light curves of nine previously discovered young jets in AGN observed by VLASS. We found that the emission had faded at $887.5$~MHz and $3$~GHz in several sources, which suggested these samples were a mixture of short-term activity and long-term jet onsets.

Future instruments, such as the Deep Synoptic Array \citep[DSA;][]{Hallinan2019}, will accelerate the study of these novel radio variable AGN. The DSA all-sky survey will cover the entire Northern sky ($\delta>-30^\circ$) to a depth of $2\mu$Jy beam\textsuperscript{-1} every four months \citep{Hallinan2019}, which will enable the discovery of many more, and fainter, radio-state transitioning AGN. The survey products will include a source catalog with 10-point SEDs spanning the $0.7$--$2$GHz band for each source \citep{Hallinan2019}. This spectral data will be extremely valuable, since it will allow us to identify and classify different AGN transients, including ESEs, relativistic beaming, and intrinsic changes in off-axis jets.
\begin{acknowledgments}
DLK was supported by NSF grant AST-2511757. I.A. was supported by the National Science Foundation award AST 2505775, NASA grant 24-ADAP24- 0159, Scialog award SA-LSST-2024-102a and LSST2025-112b.
This scientific work uses data obtained from Inyarrimanha Ilgari Bundara, the CSIRO Murchison Radio-astronomy Observatory. We acknowledge the Wajarri Yamaji People as the Traditional Owners and native title holders of the Observatory site. CSIRO’s ASKAP radio telescope is part of the Australia Telescope National Facility \footnote{\url{https://ror.org/05qajvd42}}. Operation of ASKAP is funded by the Australian Government with support from the National Collaborative Research Infrastructure Strategy. ASKAP uses the resources of the Pawsey Supercomputing Research Centre. Establishment of ASKAP, Inyarrimanha Ilgari Bundara, the CSIRO Murchison Radio-astronomy Observatory and the Pawsey Supercomputing Research Centre are initiatives of the Australian Government, with support from the Government of Western Australia and the Science and Industry Endowment Fund.
The Australia Telescope Compact Array is part of the Australia Telescope National Facility \footnote{\url{https://ror.org/05qajvd42}} which is funded by the Australian Government for operation as a National Facility managed by CSIRO.
This paper includes archived data obtained through the CSIRO ASKAP Science Data Archive, CASDA \footnote{\url{http://data.csiro.au}}.
This research made use of the Nimbus Research Cloud (Pawsey Supercomputing Research Centre, 2023;  Nimbus Research Cloud; Perth, Western Australia\footnote{\url{https://doi.org/10.48569/v0j3-qd51}}). This research was supported by use of the Nectar Research Cloud, a collaborative Australian research platform supported by the NCRIS-funded Australian Research Data Commons (ARDC).
This research used VAST Tools\footnote{\url{https://github.com/askap-vast/vast-tools}} \citep{vast_tools}, a Python module to interact with results from the VAST Pipeline and the VAST Survey data. The overall design of the VAST transient pipeline \footnote{\url{https://github.com/askap-vast/vast-pipeline}} \citep{Stewart2024} is described in detail in \cite{Murphy2021} and \cite{Pintaldi2021}.
This research was supported by the Sydney Informatics Hub, a Core Research Facility of the University of Sydney.
This work uses observations obtained with the Hobby-Eberly Telescope (HET), which is a joint
project of the University of Texas at Austin, the Pennsylvania State University, Ludwig-
Maximilians-Universitaet Muenchen, and Georg-August Universitaet Goettingen. The HET is named in honor of its principal benefactors, William P. Hobby and Robert E. Eberly. The Texas Advanced Computing Center (TACC) at the University of Texas at Austin provided high-performance computing, visualization, and storage resources that have contributed to the results reported within this paper. The Low Resolution Spectrograph 2 (LRS2) was developed and funded by the University of Texas at Austin McDonald Observatory and Department of Astronomy, and by Pennsylvania
State University. We thank the Leibniz-Institut fur Astrophysik Potsdam (AIP) and the Institut fur Astrophysik Goettingen (IAG) for their contributions to the construction of the integral field
units.
The authors acknowledge the Texas Advanced Computing Center (TACC)\footnote{\url{http://www.tacc.utexas.edu}} at The University of Texas at Austin for providing computational resources that have contributed to the research results reported within this paper.
Based in part on observations obtained at the Southern Astrophysical Research (SOAR) telescope, which is a joint project of the Minist\'{e}rio da Ci\^{e}ncia, Tecnologia e Inova\c{c}\~{o}es (MCTI/LNA) do Brasil, the US National Science Foundation’s NOIRLab, the University of North Carolina at Chapel Hill (UNC), and Michigan State University (MSU).
This research has made use of the VizieR catalogue access tool, CDS,
Strasbourg, France. The original description of the VizieR service was published in \citet{vizier}.
This work made use of Astropy: \footnote{\url{https://www.astropy.org}} a community-developed core Python package and an ecosystem of tools and resources for astronomy \citep{astropy:2013, astropy:2018, astropy:2022}.
This work has made use of data from the European Space Agency (ESA)
mission {\it Gaia}\footnote{\url{https://www.cosmos.esa.int/gaia}}, processed by
the {\it Gaia} Data Processing and Analysis Consortium (DPAC)\footnote{
\url{https://www.cosmos.esa.int/web/gaia/dpac/consortium}}. Funding
for the DPAC has been provided by national institutions, in particular
the institutions participating in the {\it Gaia} Multilateral Agreement.
This research used data obtained with the Dark Energy Spectroscopic Instrument (DESI). DESI construction and operations is managed by the Lawrence Berkeley National Laboratory. This material is based upon work supported by the U.S. Department of Energy, Office of Science, Office of High-Energy Physics, under Contract No. DE–AC02–05CH11231, and by the National Energy Research Scientific Computing Center, a DOE Office of Science User Facility under the same contract. Additional support for DESI was provided by the U.S. National Science Foundation (NSF), Division of Astronomical Sciences under Contract No. AST-0950945 to the NSF’s National Optical-Infrared Astronomy Research Laboratory; the Science and Technology Facilities Council of the United Kingdom; the Gordon and Betty Moore Foundation; the Heising-Simons Foundation; the French Alternative Energies and Atomic Energy Commission (CEA); the National Council of Humanities, Science and Technology of Mexico (CONAHCYT); the Ministry of Science and Innovation of Spain (MICINN), and by the DESI Member Institutions\footnote{\url{www.desi.lbl.gov/collaborating-institutions}}. The DESI collaboration is honored to be permitted to conduct scientific research on I’oligam Du’ag (Kitt Peak), a mountain with particular significance to the Tohono O’odham Nation. Any opinions, findings, and conclusions or recommendations expressed in this material are those of the author(s) and do not necessarily reflect the views of the U.S. National Science Foundation, the U.S. Department of Energy, or any of the listed funding agencies.
The ZTF forced-photometry service was funded under the Heising-Simons Foundation grant \#12540303 (PI: Graham).
This work used data from the Sloan Digital Sky Survey \citep{Blanton2017, Gunn2006,Smee2013}. Funding for the Sloan Digital Sky Survey IV has been provided by the Alfred P. Sloan Foundation, the U.S. Department of Energy Office of Science, and the Participating Institutions. SDSS-IV acknowledges support and resources from the Center for High Performance Computing at the University of Utah. SDSS-IV\footnote{\url{www.sdss4.org}} is managed by the Astrophysical Research Consortium for the Participating Institutions of the SDSS Collaboration including the Brazilian Participation Group, the Carnegie Institution for Science, Carnegie Mellon University, Center for Astrophysics | Harvard \& Smithsonian, the Chilean Participation Group, the French Participation Group, Instituto de Astrofisica de Canarias, The Johns Hopkins University, Kavli Institute for the Physics and Mathematics of the Universe (IPMU) /University of Tokyo, the Korean Participation Group, Lawrence Berkeley National Laboratory, Leibniz Institut f\"ur Astrophysik Potsdam (AIP), Max-Planck-Institut f\"ur Astronomie (MPIA Heidelberg), Max-Planck Institut f\"ur Astrophysik (MPA Garching), Max-Planck-Institut f\"ur Extraterrestrische Physik (MPE), National Astronomical Observatories of China, New Mexico State University, New York University, University of Notre Dame, Observat\'ario Nacional / MCTI, The Ohio State University, Pennsylvania State University, Shanghai Astronomical Observatory, United Kingdom Participation Group, Universidad Nacional Aut\'onoma deM\'exico, University of Arizona, University of Colorado Boulder, University of Oxford, University of Portsmouth, University of Utah, University of Virginia, University of Washington, University of Wisconsin, Vanderbilt University, and Yale University.
The Legacy Surveys consist of three individual and complementary projects: the Dark Energy Camera Legacy Survey (DECaLS; Proposal ID \#2014B-0404; PIs: David Schlegel and Arjun Dey), the Beijing-Arizona Sky Survey (BASS; NOAO Prop. ID \#2015A-0801; PIs: Zhou Xu and Xiaohui Fan), and the Mayall z-band Legacy Survey (MzLS; Prop. ID \#2016A-0453; PI: Arjun Dey). DECaLS, BASS and MzLS together include data obtained, respectively, at the Blanco telescope, Cerro Tololo Inter-American Observatory, NSF’s NOIRLab; the Bok telescope, Steward Observatory, University of Arizona; and the Mayall telescope, Kitt Peak National Observatory, NOIRLab. Pipeline processing and analyses of the data were supported by NOIRLab and the Lawrence Berkeley National Laboratory (LBNL). The Legacy Surveys project is honored to be permitted to conduct astronomical research on Iolkam Du’ag (Kitt Peak), a mountain with particular significance to the Tohono O’odham Nation. NOIRLab is operated by the Association of Universities for Research in Astronomy (AURA) under a cooperative agreement with the National Science Foundation. LBNL is managed by the Regents of the University of California under contract to the U.S. Department of Energy.
The Legacy Survey team makes use of data products from the Near-Earth Object Wide-field Infrared Survey Explorer (NEOWISE), which is a project of the Jet Propulsion Laboratory/California Institute of Technology. NEOWISE is funded by the National Aeronautics and Space Administration. The Legacy Surveys imaging of the DESI footprint is supported by the Director, Office of Science, Office of High Energy Physics of the U.S. Department of Energy under Contract No. DE-AC02-05CH1123, by the National Energy Research Scientific Computing Center, a DOE Office of Science User Facility under the same contract; and by the U.S. National Science Foundation, Division of Astronomical Sciences under Contract No. AST-0950945 to NOAO.
LOFAR data products were provided by the LOFAR Surveys Key Science project (LSKSP)\footnote{\url{https://lofar-surveys.org/}} and were derived from observations with the International LOFAR Telescope (ILT). LOFAR \citep{Vanhaarlem2013} is the Low Frequency Array, designed and constructed by ASTRON. It has observing, data processing, and data storage facilities in several countries, which are owned by various parties (each with their own funding sources), and which are collectively operated by the LOFAR ERIC under a joint scientific policy. The efforts of the LSKSP have benefited from funding from the European Research Council, NOVA, NWO, CNRS-INSU, the SURF Co-operative, the UK Science and Technology Funding Council and the Jülich Supercomputing Centre.
\end{acknowledgments}


\begin{contribution}


\end{contribution}

\bibliography{ref}{}
\bibliographystyle{aasjournalv7}
\appendix
\section{VAST Epochs Failing Quality Criteria}
\label{appx:bad_epochs}
\begin{figure}[hbpt!]
    \gridline{\fig{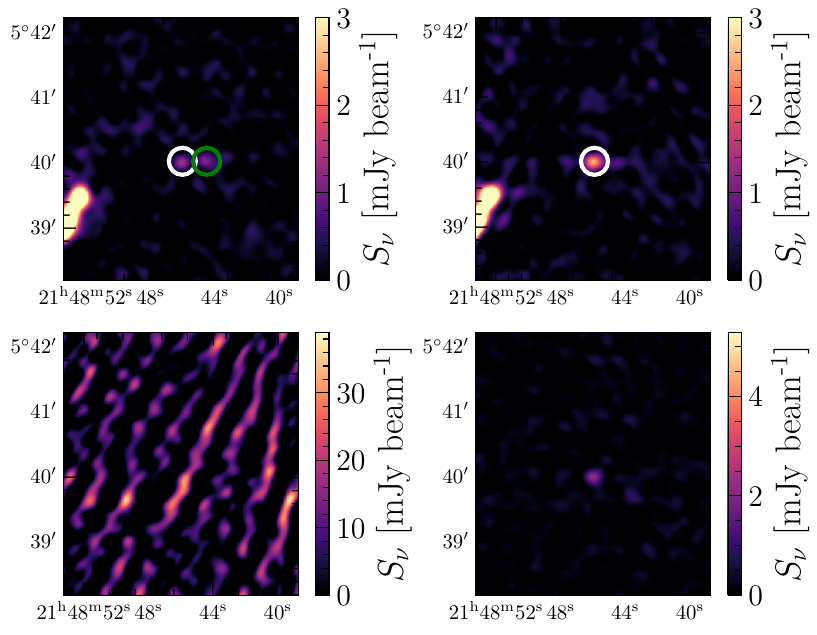}{0.45\textwidth}{}}
    \caption{Two types of unreliable epochs we removed from our final sample. The removed epochs are shown on the left and the kept epochs are shown on the right. Top panel: ASKAP J214845.9$+$054000 (white oval), one of the sources which failed our data quality requirements. In the left epoch above, there is an overlapping nearby source, denoted by the green oval. Any flux density variability between the left and right epochs may have been due to the contamination by the neighboring source's flux density in the left epoch. Bottom panel: Two epochs of ASKAP J050355.2$-$333453, another source failing our data quality requirements. The flux density extracted from the left image is unreliable, since the source was too far from the center of the beam during the observation.\label{fig:bad_epochs}}
\end{figure}
\section{Non-detections in Archival Radio Surveys}
\label{appx:forced_photometry}
\begin{figure}[hbpt!]
    \gridline{\fig{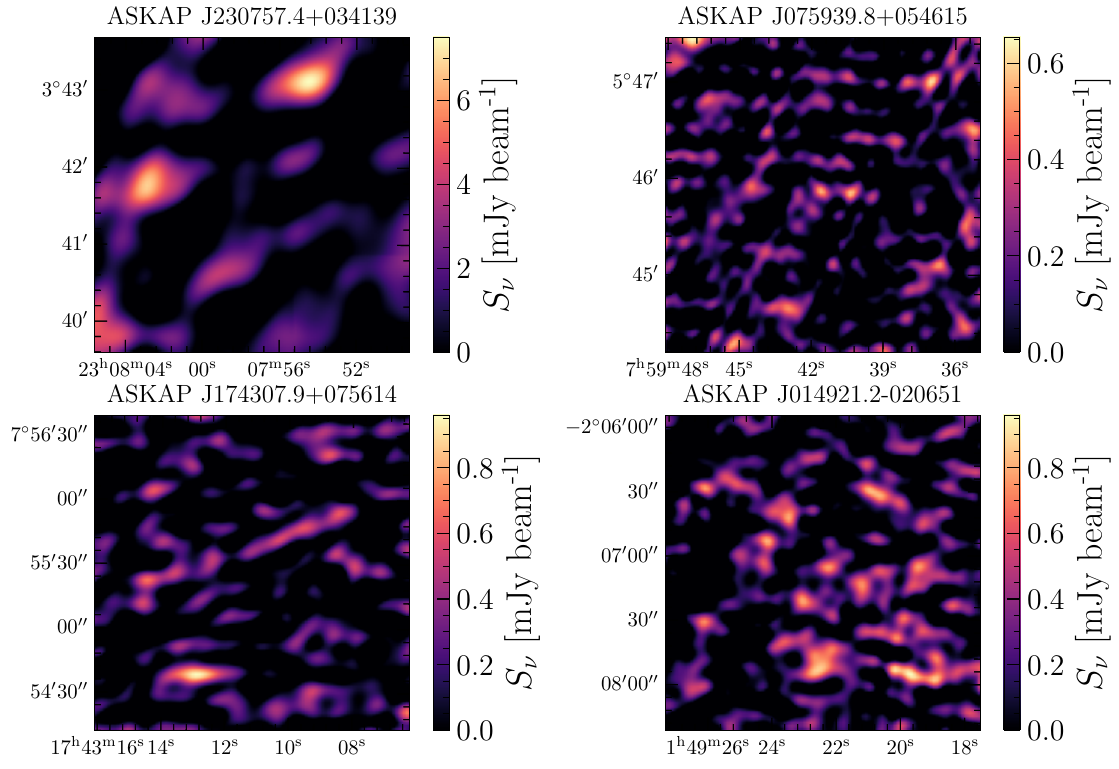}{0.5\textwidth}{}}
    \caption{Images cutouts of four source positions in RACS-low (top left), RACS-mid (top right), RACS-high (bottom left), and LoTSS (bottom right). We confirmed each to be a non-detection using the {\fontfamily{qcr}\selectfont
miriad cgcurs} task. \label{fig:forced_photometry}}
\end{figure}

\section{SOAR Goodman HTS and HET LRS2 Spectra}
\label{appx:spectra}
\begin{sidewaysfigure*}[htpb!]
    \caption{Spectra and redshift determinations for sources we observed with SOAR and the HET. We observed ASKAP J034921.5$+$075101, ASKAP J084831.8$-$084845, ASKAP J125148.9$-$071228, ASKAP J130117.7$+$071211, and ASKAP J170252.1$+$014024 with the HET LRS2. We observed ASKAP J100154.9$-$370127, ASKAP J105501.7$-$304732, ASKAP J113340.6$-$374359, and ASKAP J201813.9$+$025307 with the SOAR Goodman HTS. Wavelength is left in the observing frame. \label{fig:spectra}}
    \gridline{\fig{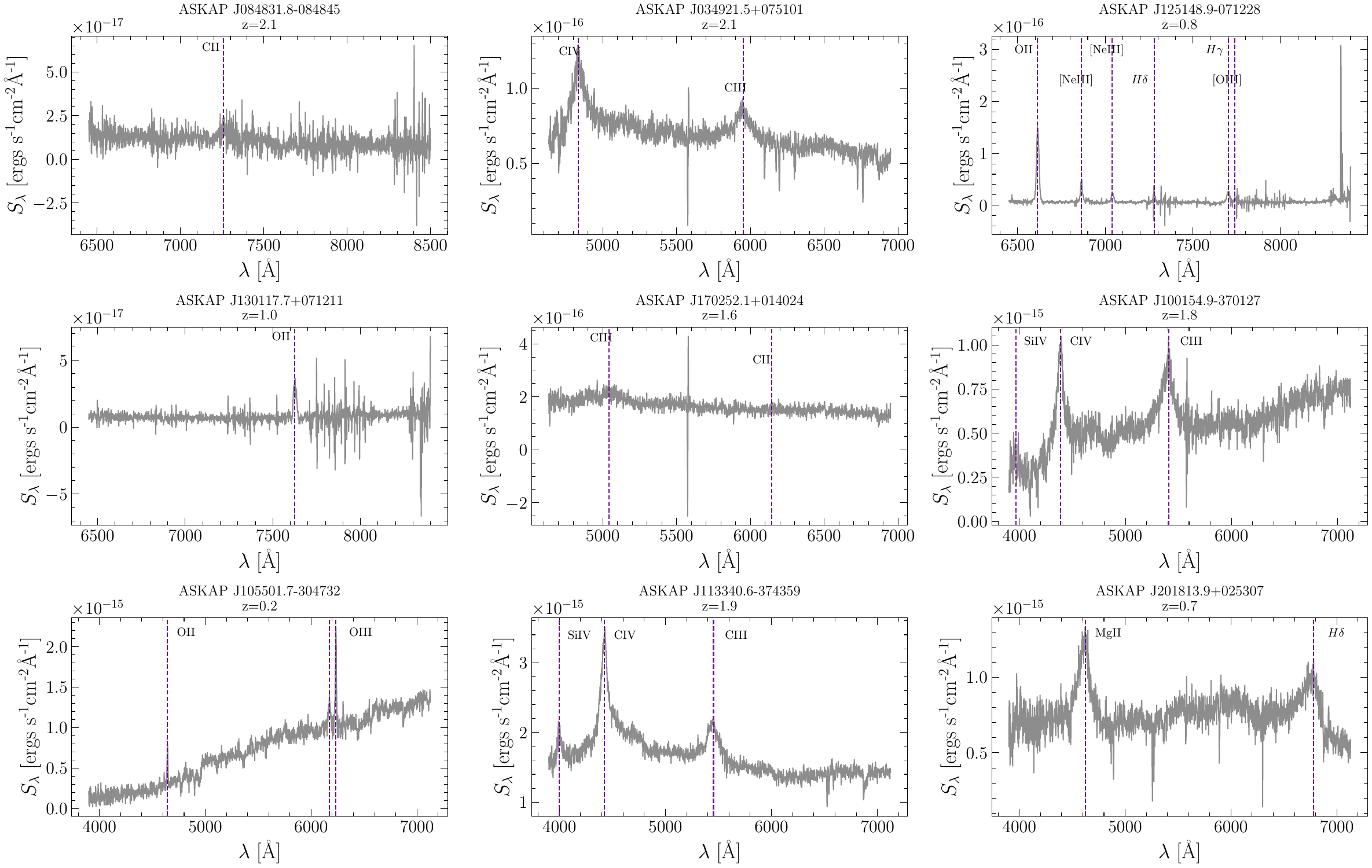}{0.98\textwidth}{}}
\end{sidewaysfigure*}
\clearpage
\section{AGN with Inverted SEDs and Persistent Radio Emission}
\label{appx:nv_babies}
\begin{figure*}[htpb!]
    \caption{Multi-frequency light curves and radio SEDs of the sources we determined had persistent radio emission, but were nonetheless radio jets in the early phases of their evolution. The inverted SEDs were not steep enough for the low-frequency absorption to be explained by SSA alone. \label{fig:nv_babies}}
    \gridline{\fig{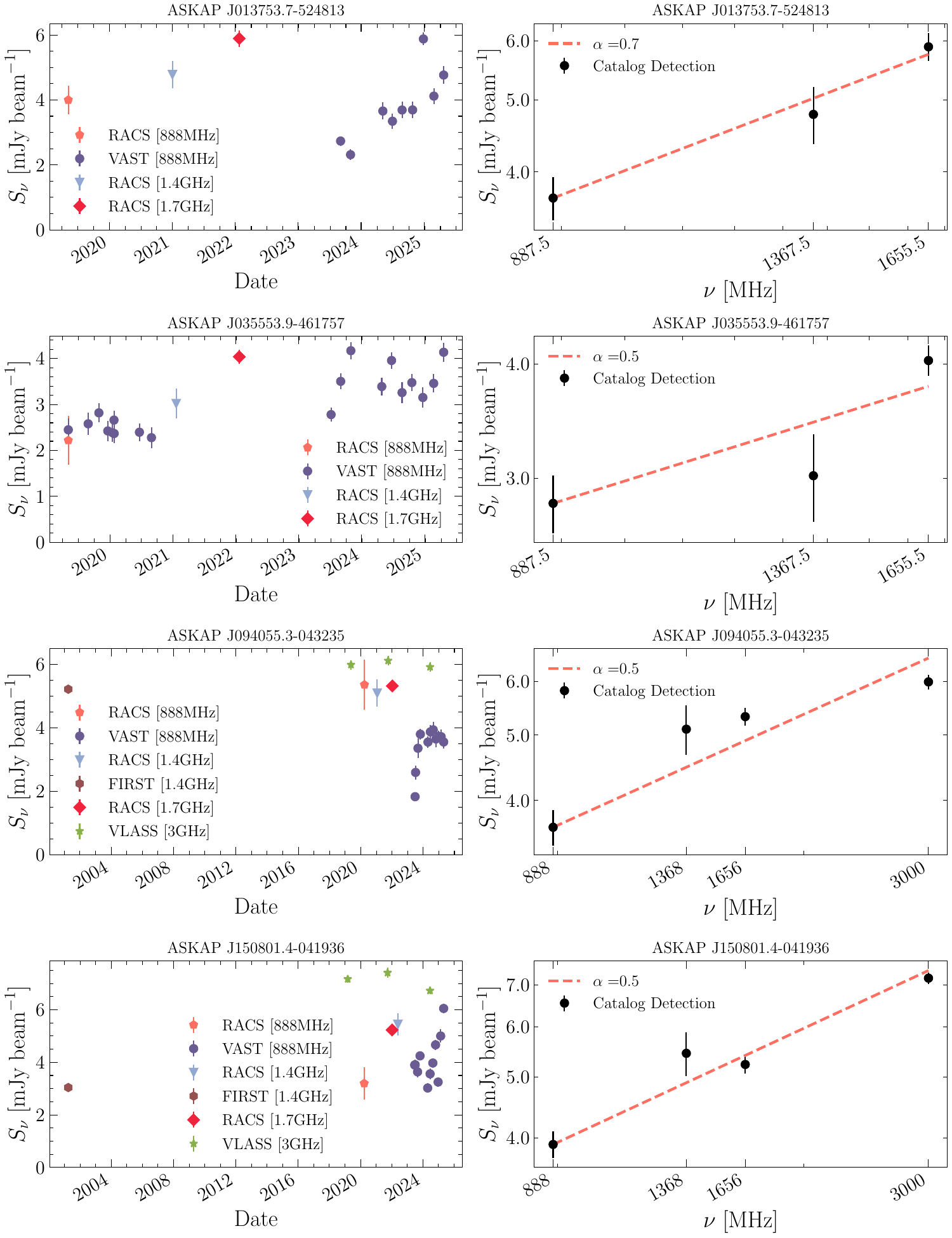}{0.8\textwidth}{}}
\end{figure*}

\begin{figure*}[htpb!]
    \gridline{\fig{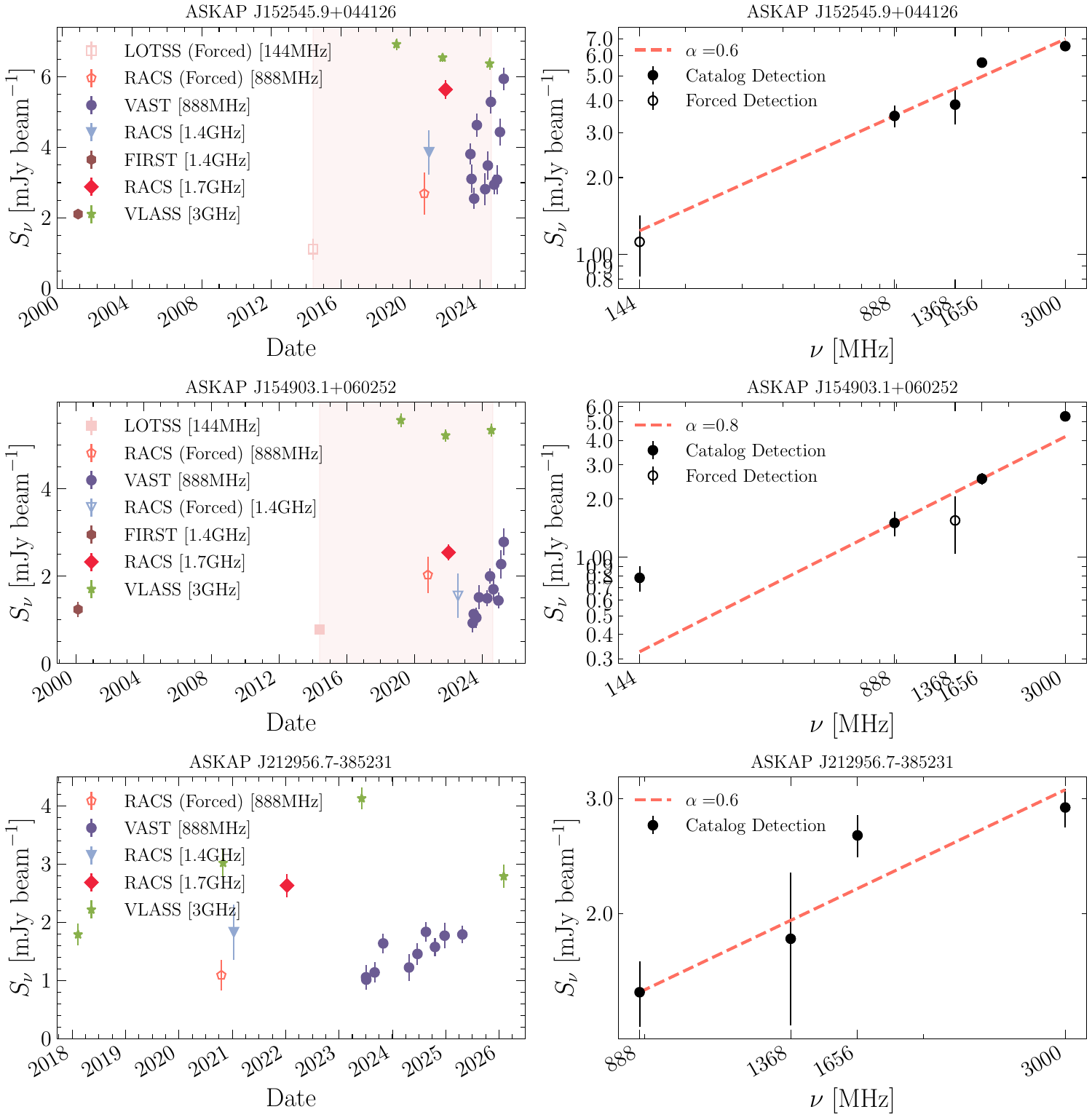}{0.85\textwidth}{}}
\end{figure*}
\clearpage
\section{AGN with Young Jet Activity}
\label{appx:favs}
\begin{figure*}[htpb!]
    \caption{Multi-frequency light curves and radio SEDs for the young radio jet candidates. \label{fig:var_babies}}
    \gridline{\fig{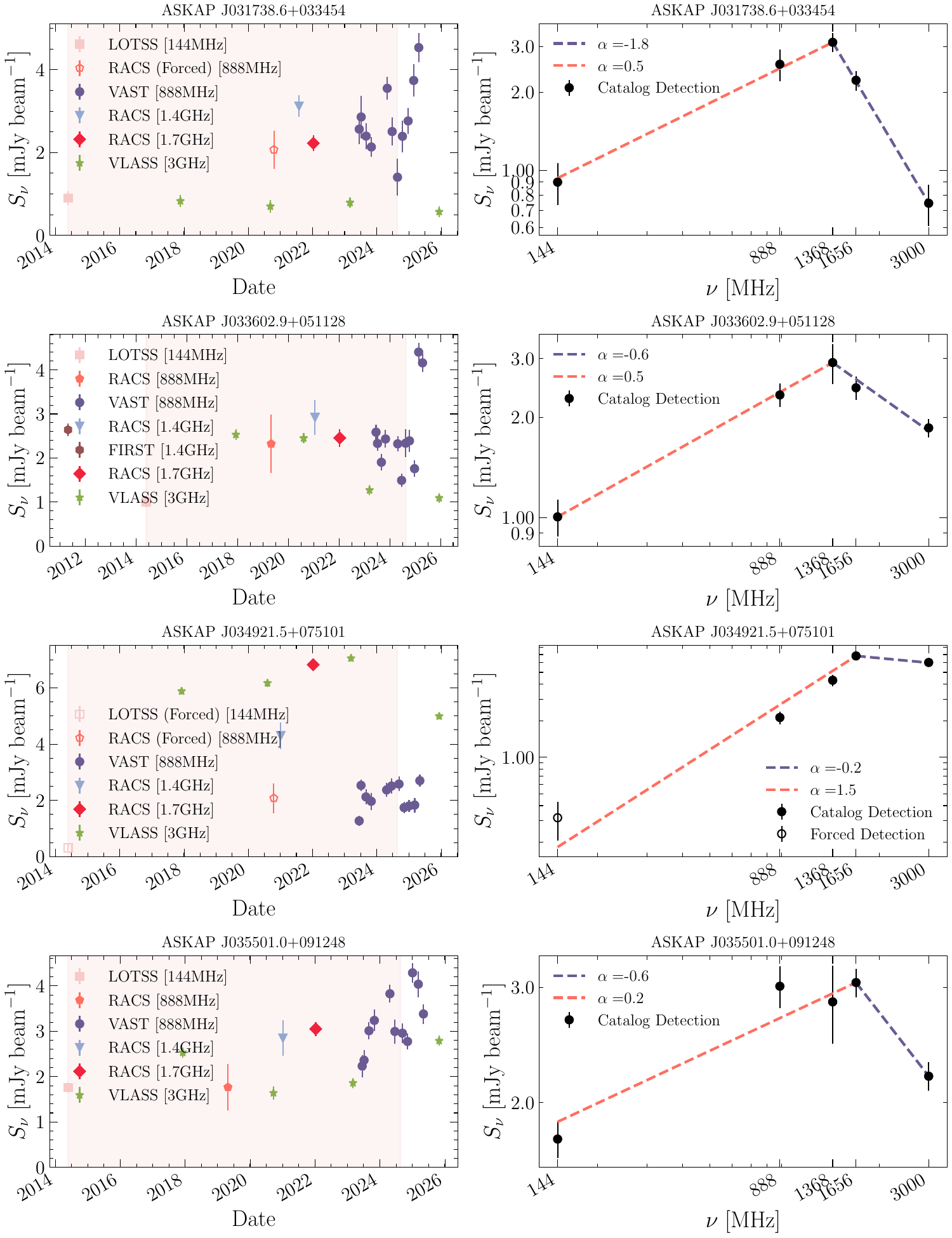}{0.8\textwidth}{}}
\end{figure*}
\clearpage
\begin{figure*}[htpb!]
    \gridline{\fig{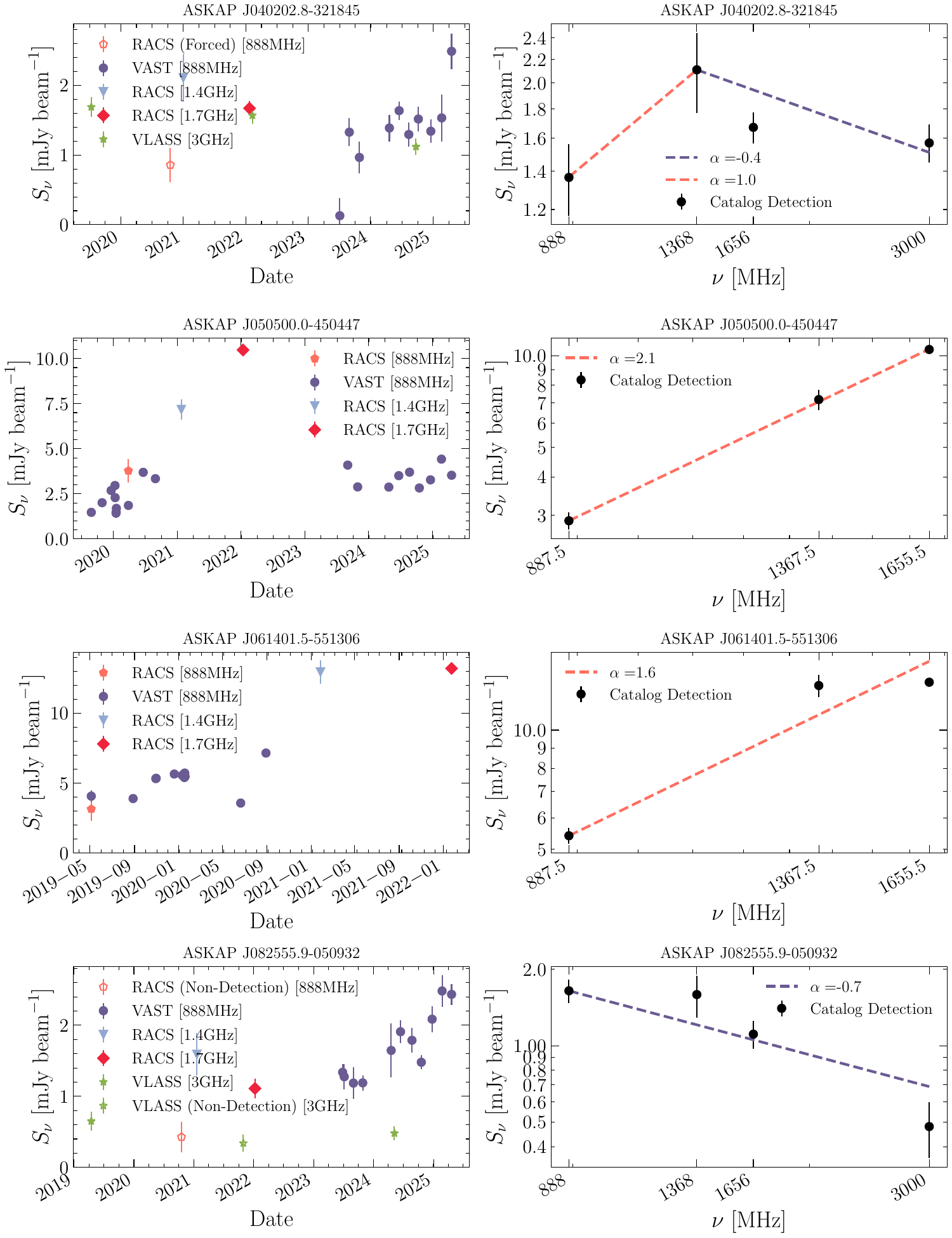}{0.92\textwidth}{}}
\end{figure*}
\clearpage
\begin{figure*}[htpb!]
    \gridline{\fig{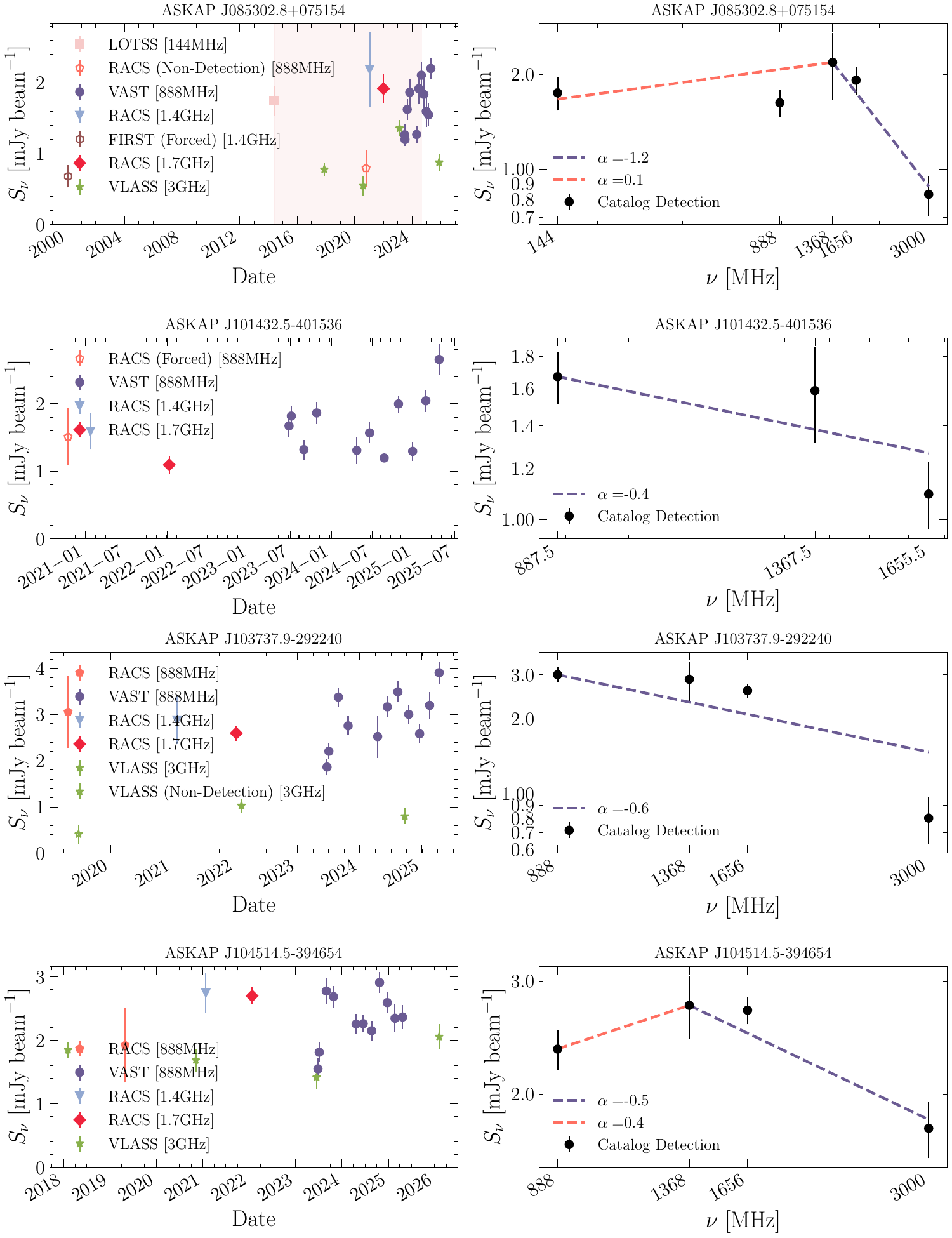}{0.92\textwidth}{}}
\end{figure*}
\clearpage
\begin{figure*}[htpb!]
    \gridline{\fig{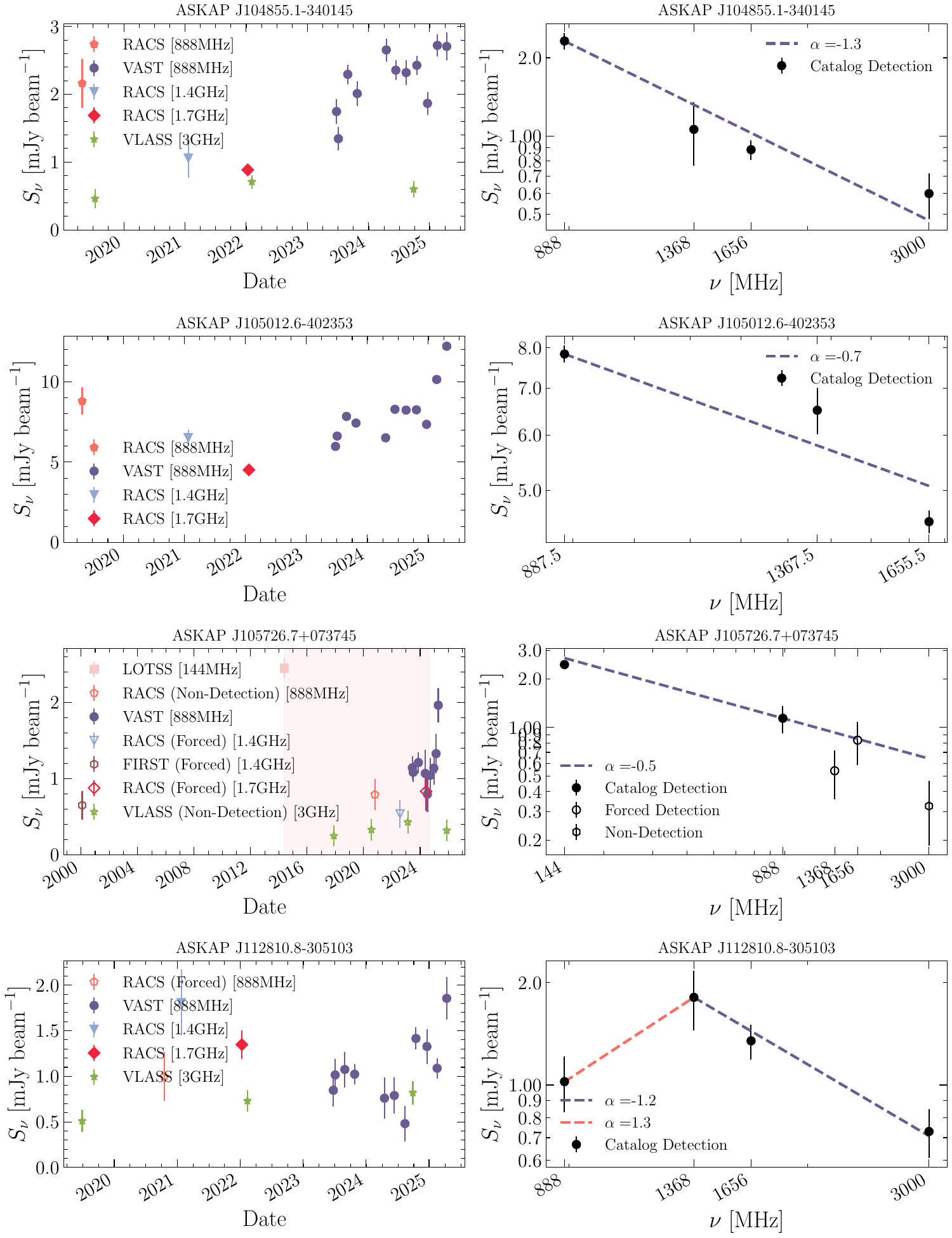}{0.92\textwidth}{}}
\end{figure*}
\clearpage
\begin{figure*}[htpb!]
    \gridline{\fig{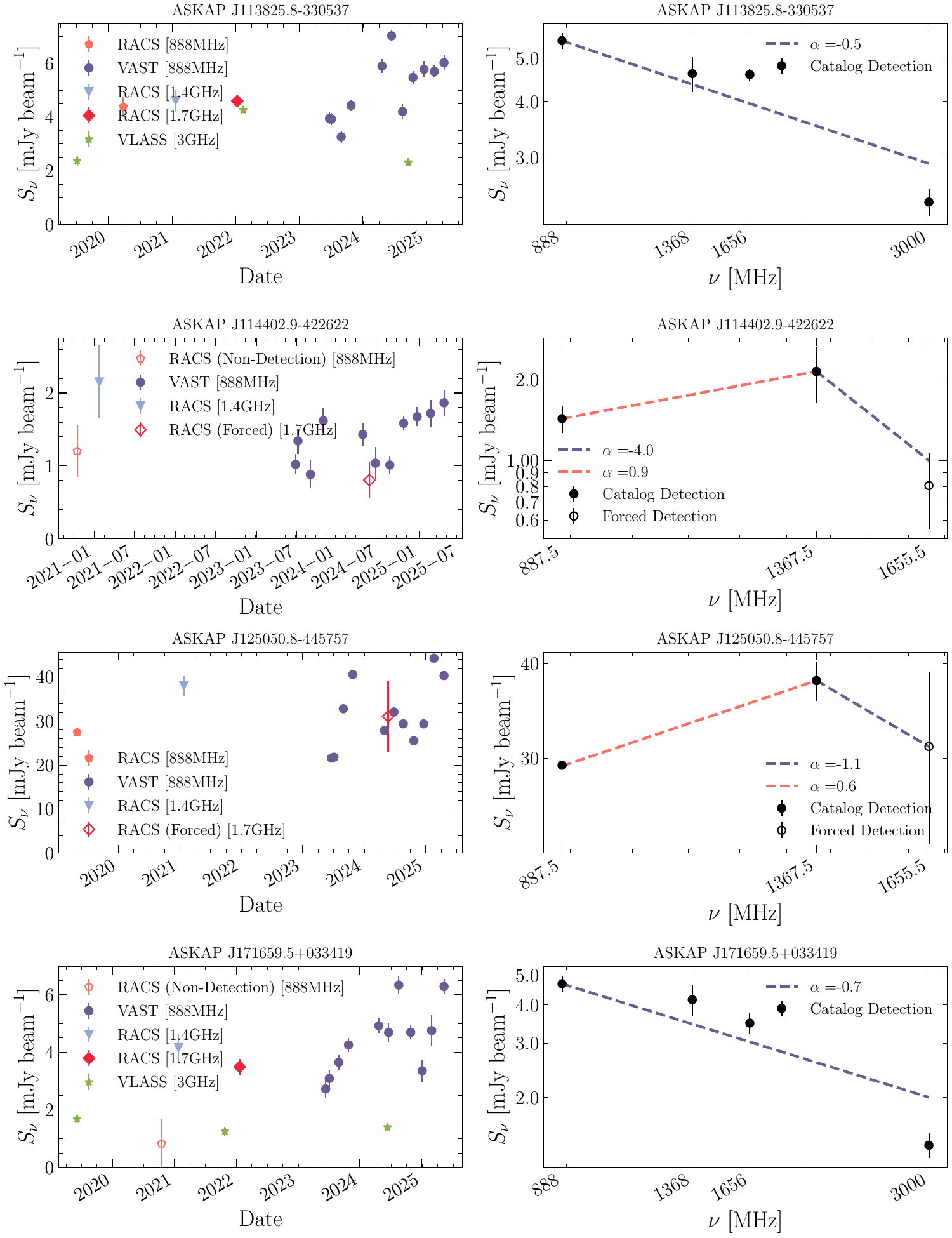}{0.92\textwidth}{}}
\end{figure*}
\clearpage
\begin{figure*}[htpb!]
    \gridline{\fig{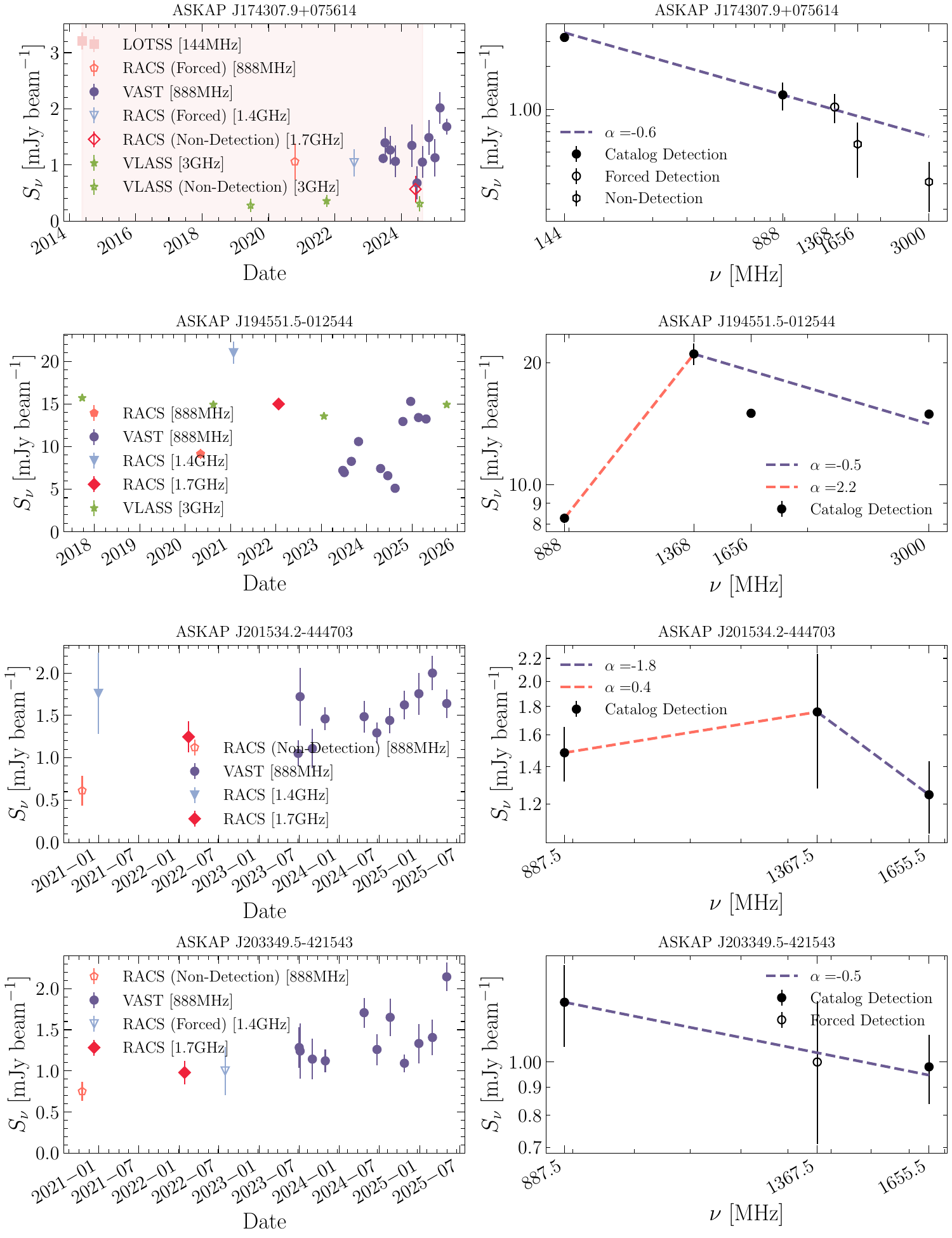}{0.92\textwidth}{}}
\end{figure*}
\clearpage
\begin{figure*}[htpb!]
    \gridline{\fig{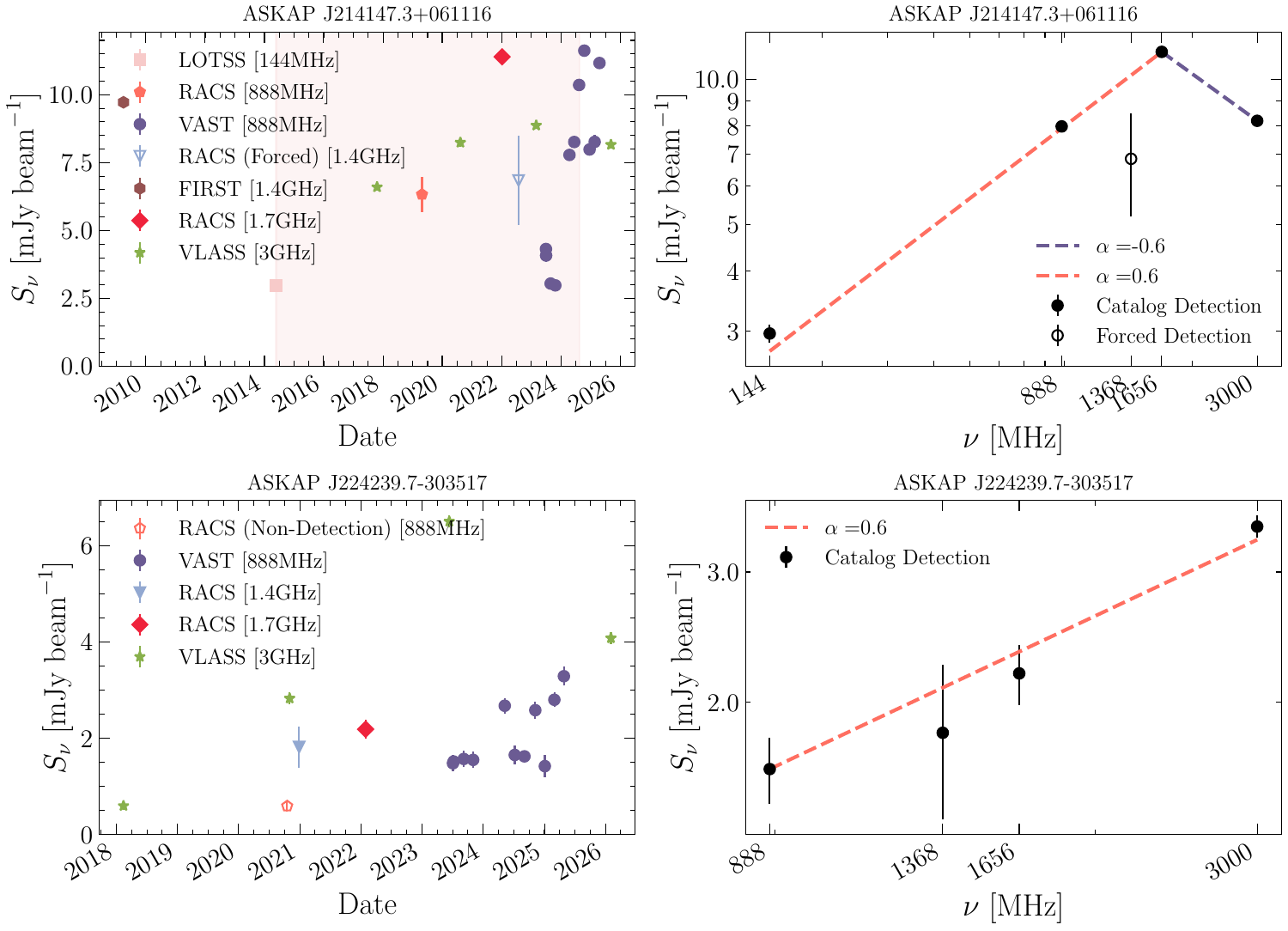}{0.92\textwidth}{}}
\end{figure*}
\clearpage

\section{VAST Light Curves Meeting Our Criteria for Radio Variable AGN}
\label{appx:vast_light_curves}
\begin{figure*}[htpb!] 
\caption{The full set of VAST light curves meeting our criteria for radio variable AGN. \label{fig:all_vast_lcs}}
    \gridline{\fig{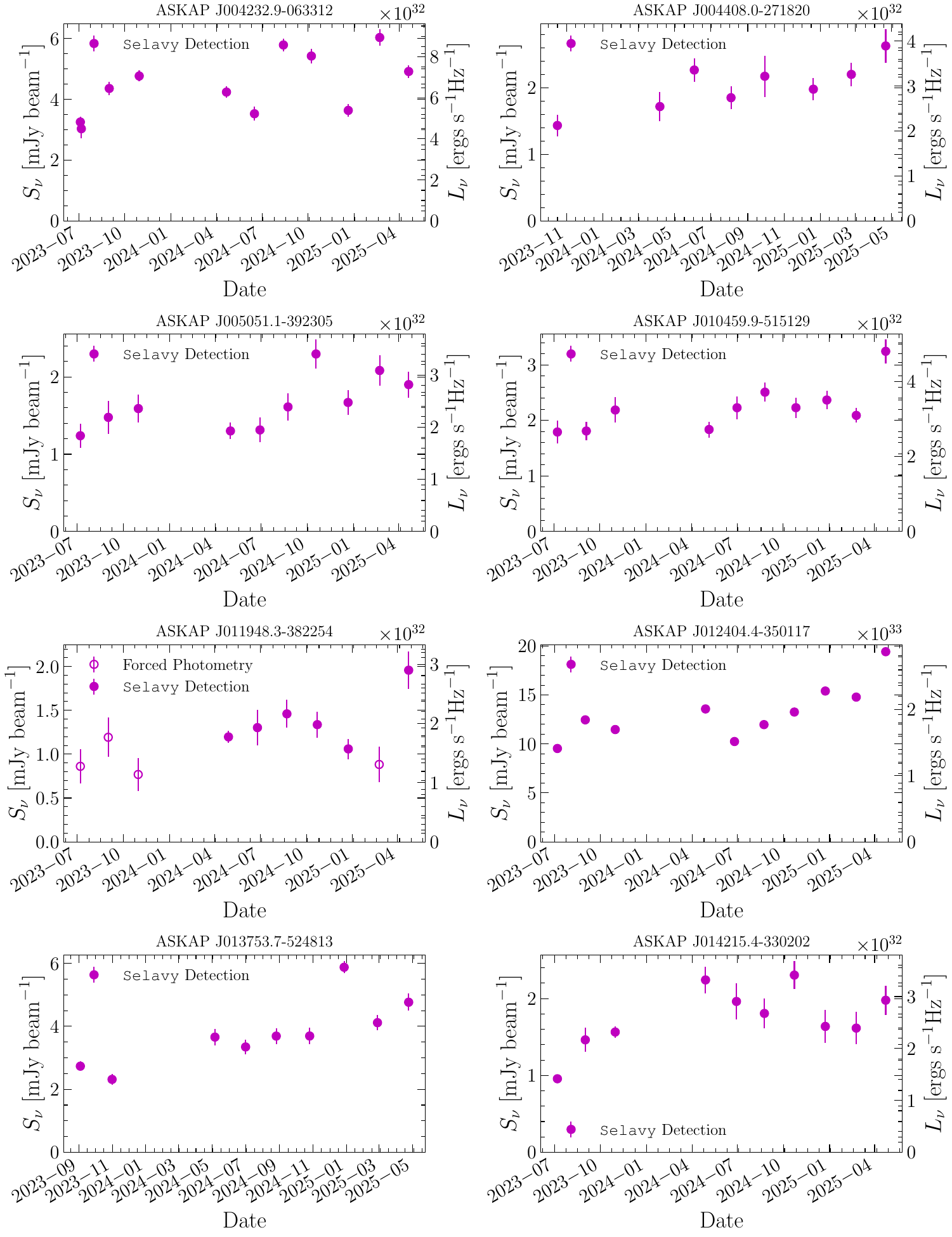}{0.8\textwidth}{}}
 \end{figure*} 
\clearpage
\begin{figure*}[htpb!] 
\gridline{\fig{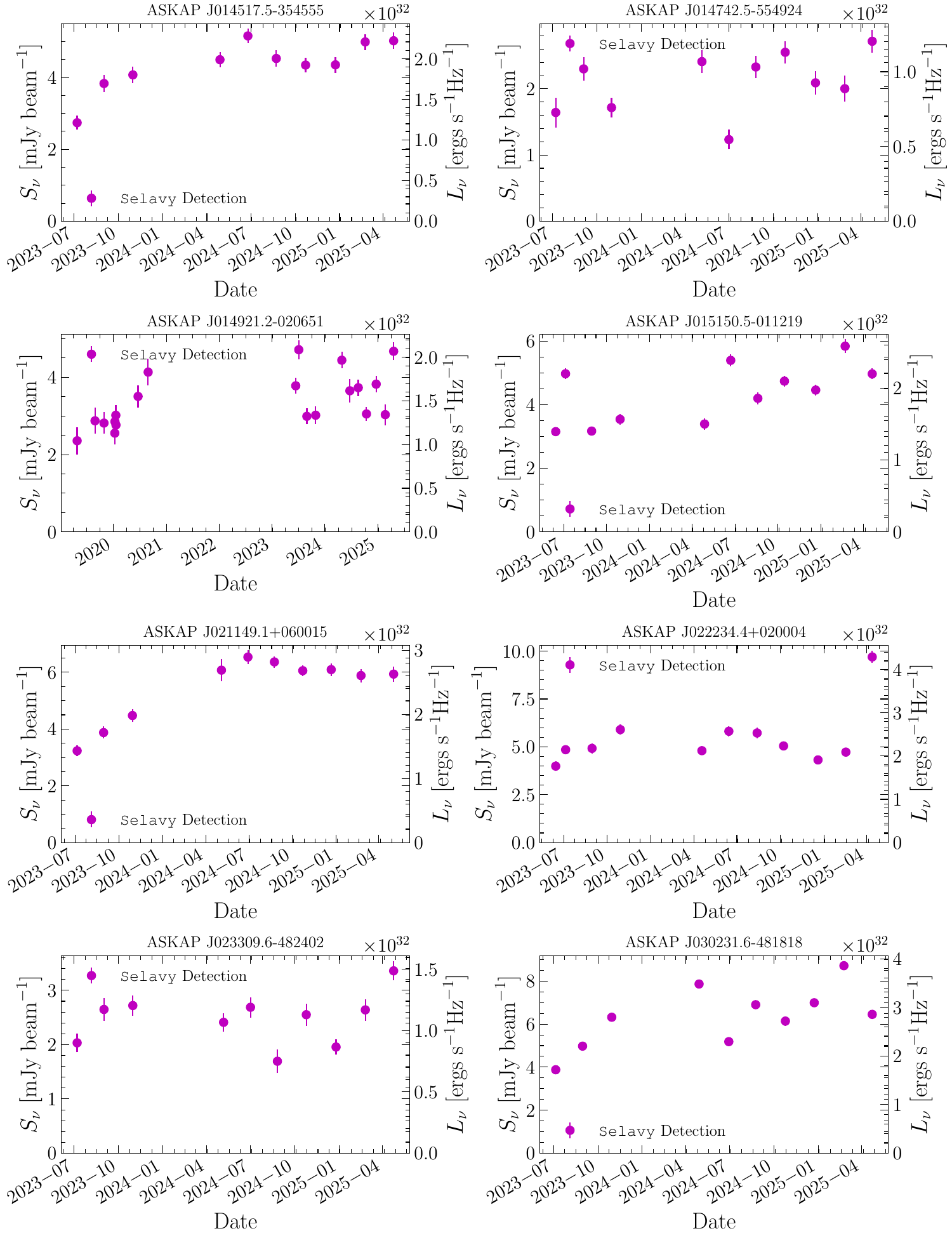}{0.92\textwidth}{}}
 \end{figure*} 
 \clearpage
 \begin{figure*}[htpb!] 
\gridline{\fig{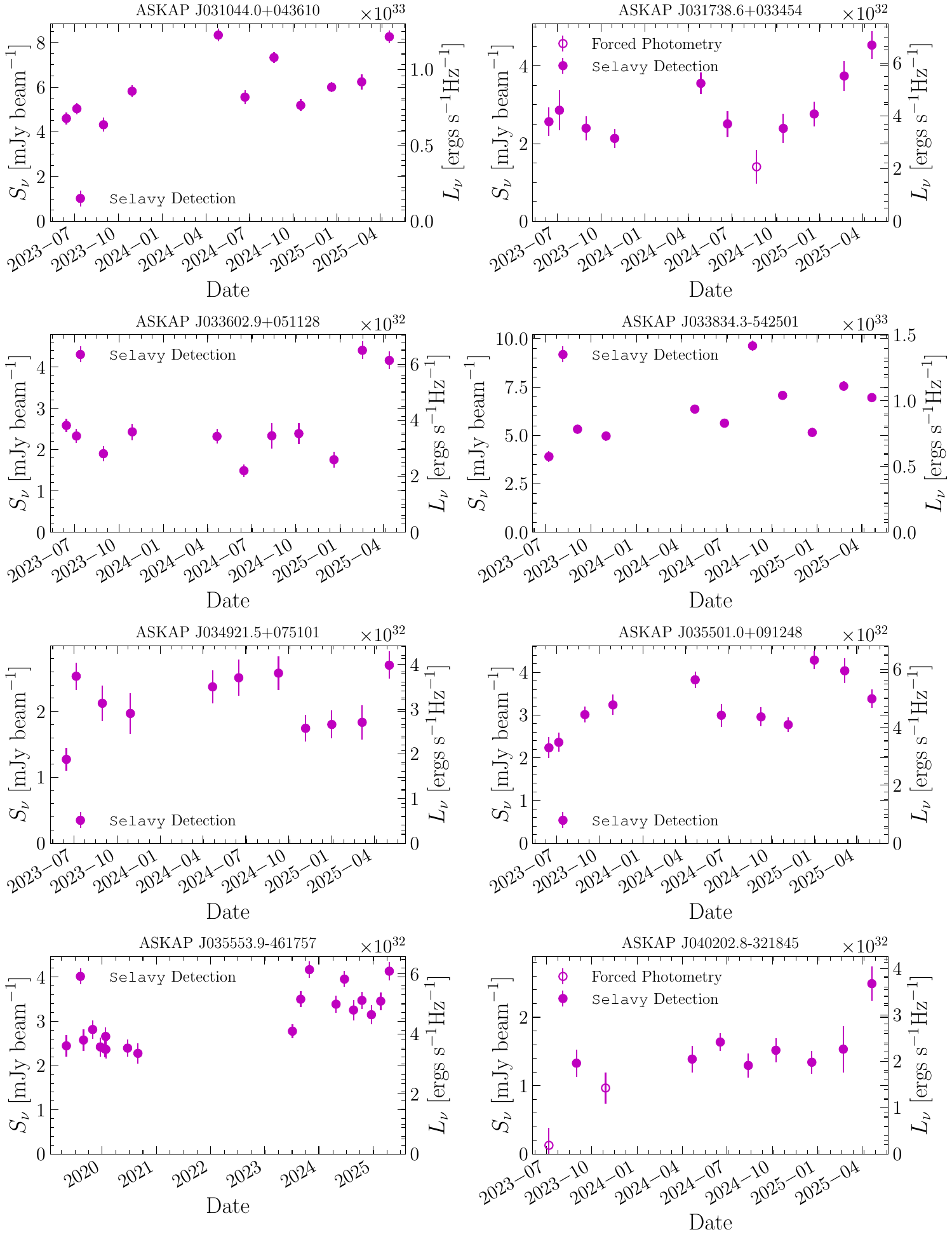}{0.92\textwidth}{}}
 \end{figure*} 
 \clearpage
 \begin{figure*}[htpb!] 
\gridline{\fig{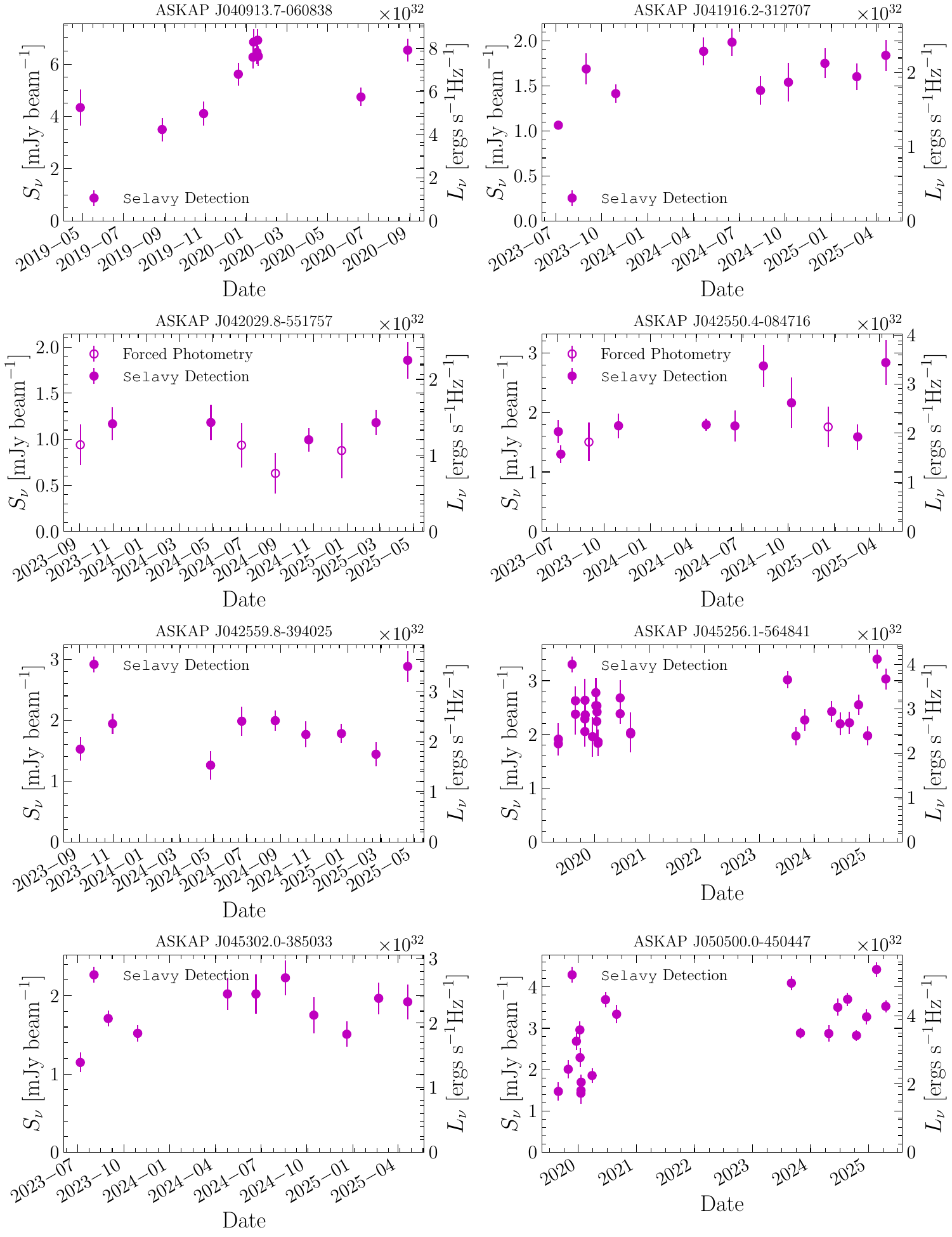}{0.92\textwidth}{}}
 \end{figure*} 
 \clearpage
 \begin{figure*}[htpb!] 
\gridline{\fig{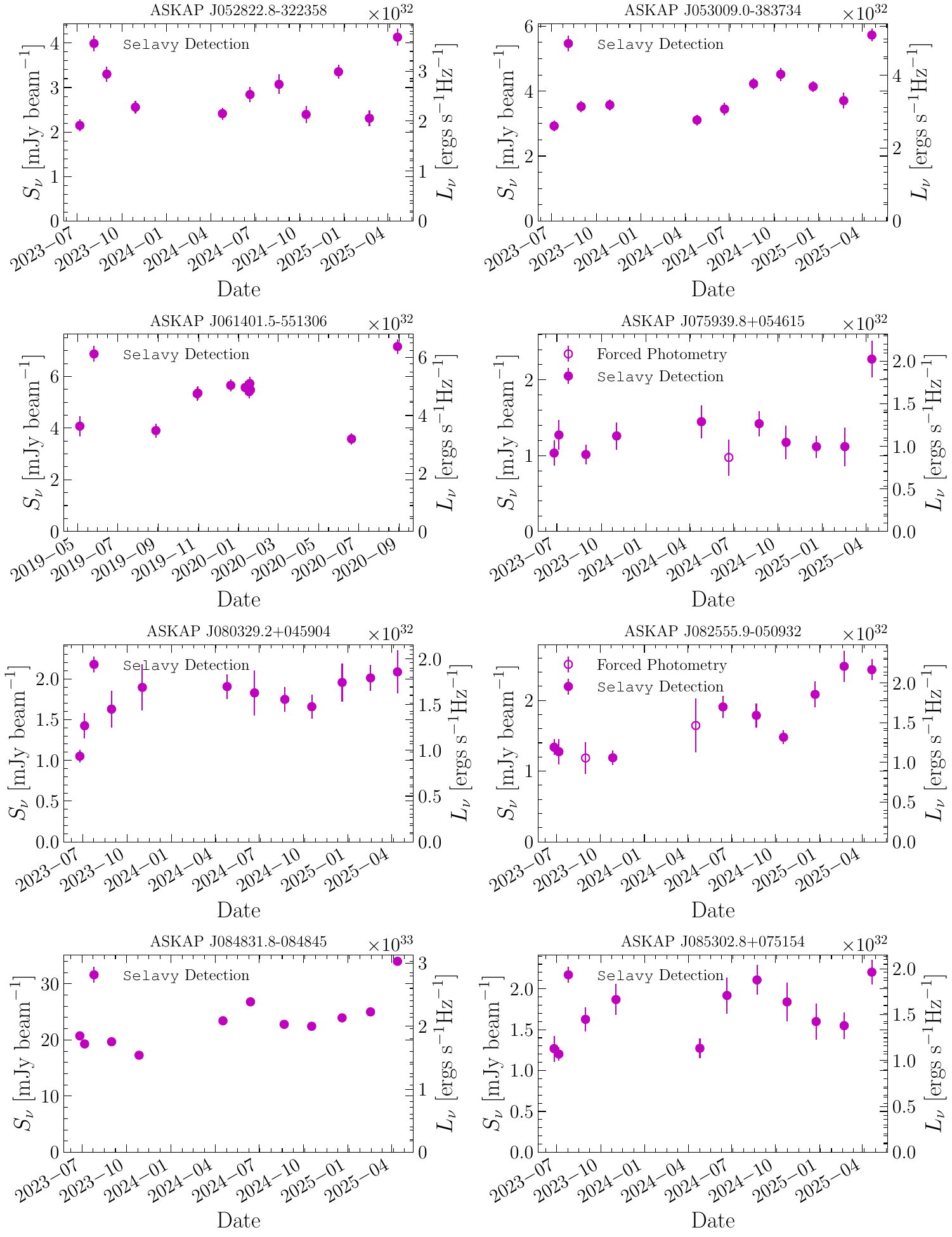}{0.92\textwidth}{}}
 \end{figure*} 
 \clearpage
 \begin{figure*}[htpb!] 
\gridline{\fig{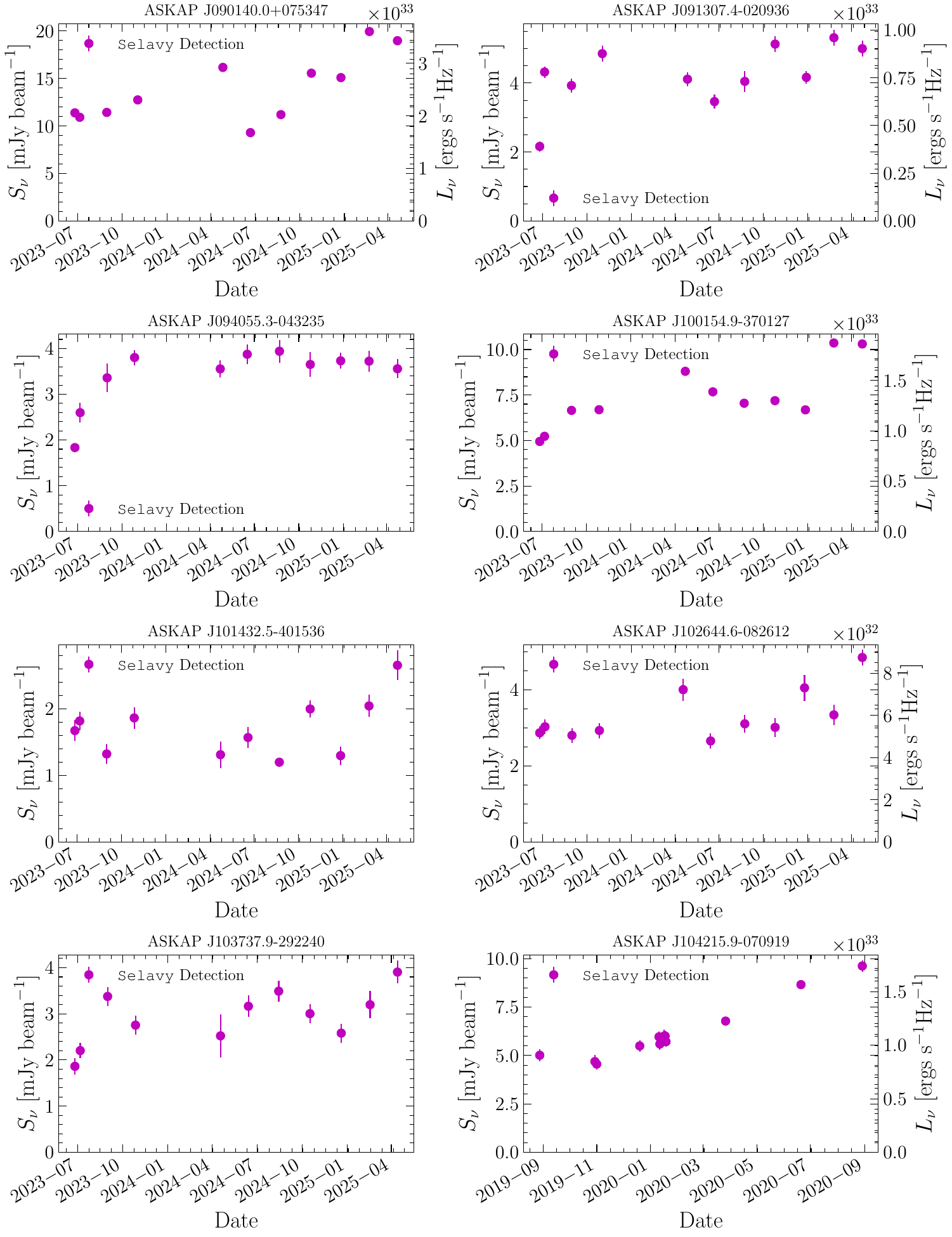}{0.92\textwidth}{}}
 \end{figure*} 
 \clearpage
 \begin{figure*}[htpb!] 
\gridline{\fig{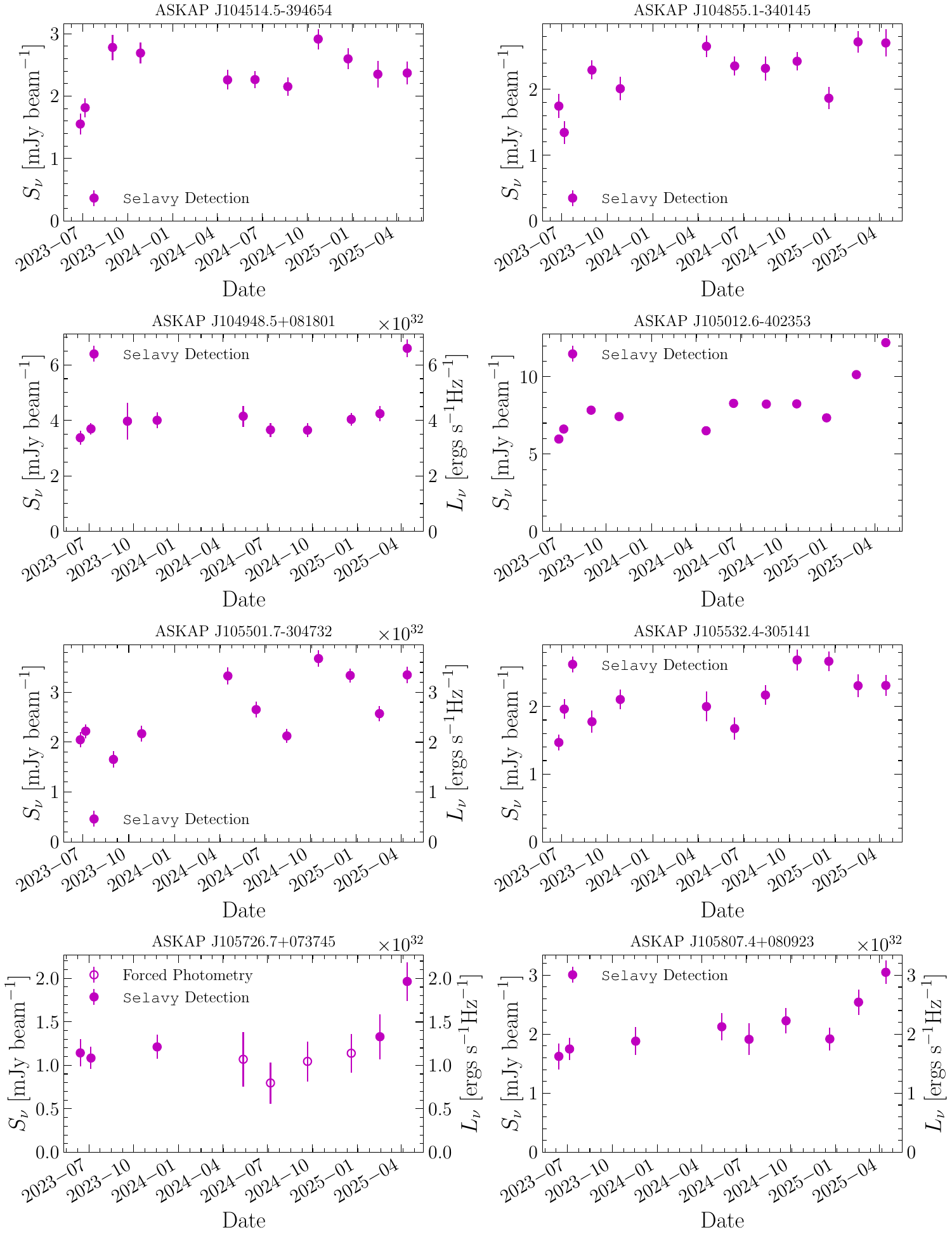}{0.92\textwidth}{}}
 \end{figure*} 
 \clearpage
 \begin{figure*}[htpb!] 
\gridline{\fig{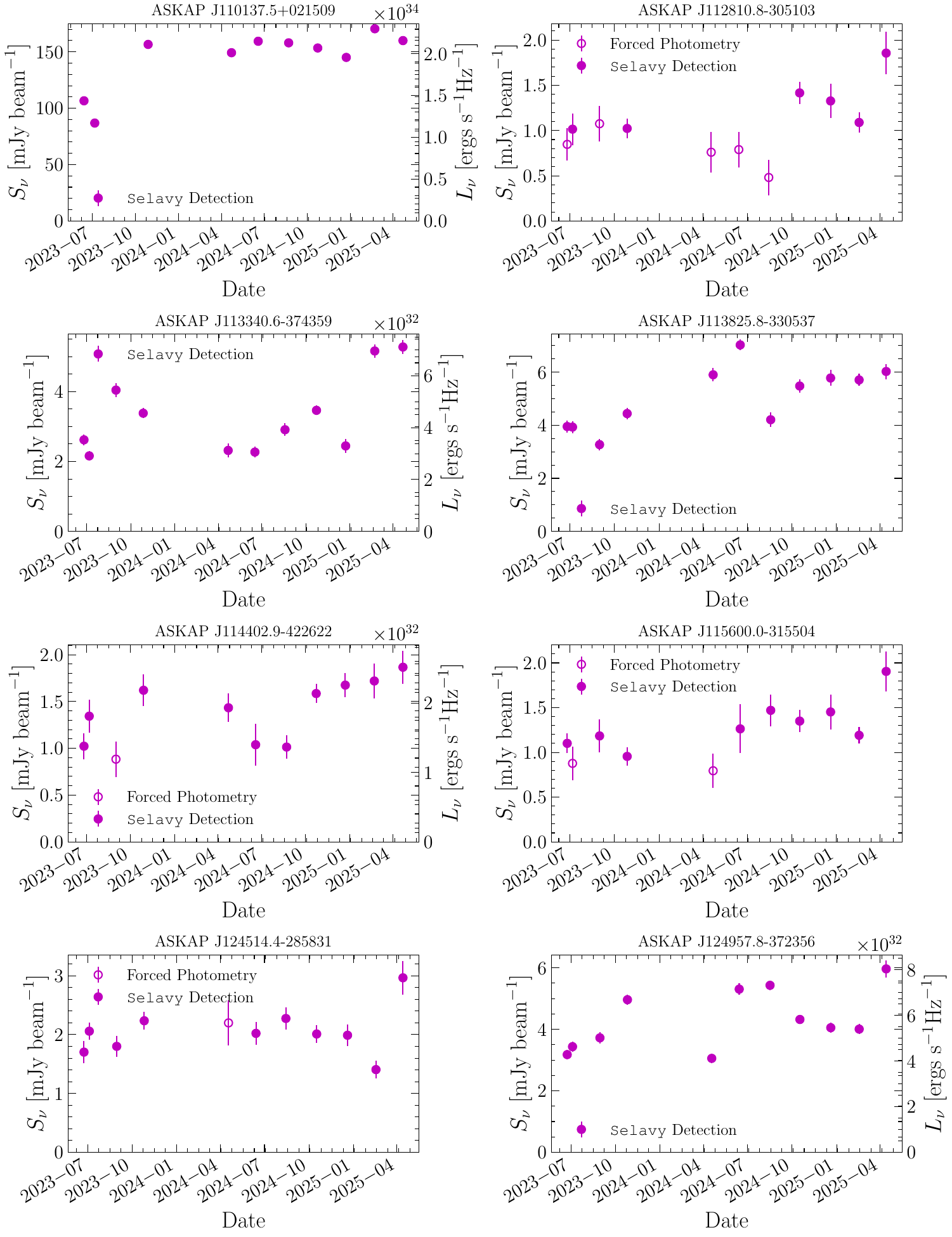}{0.92\textwidth}{}}
 \end{figure*} 
 \clearpage
 \begin{figure*}[htpb!] 
\gridline{\fig{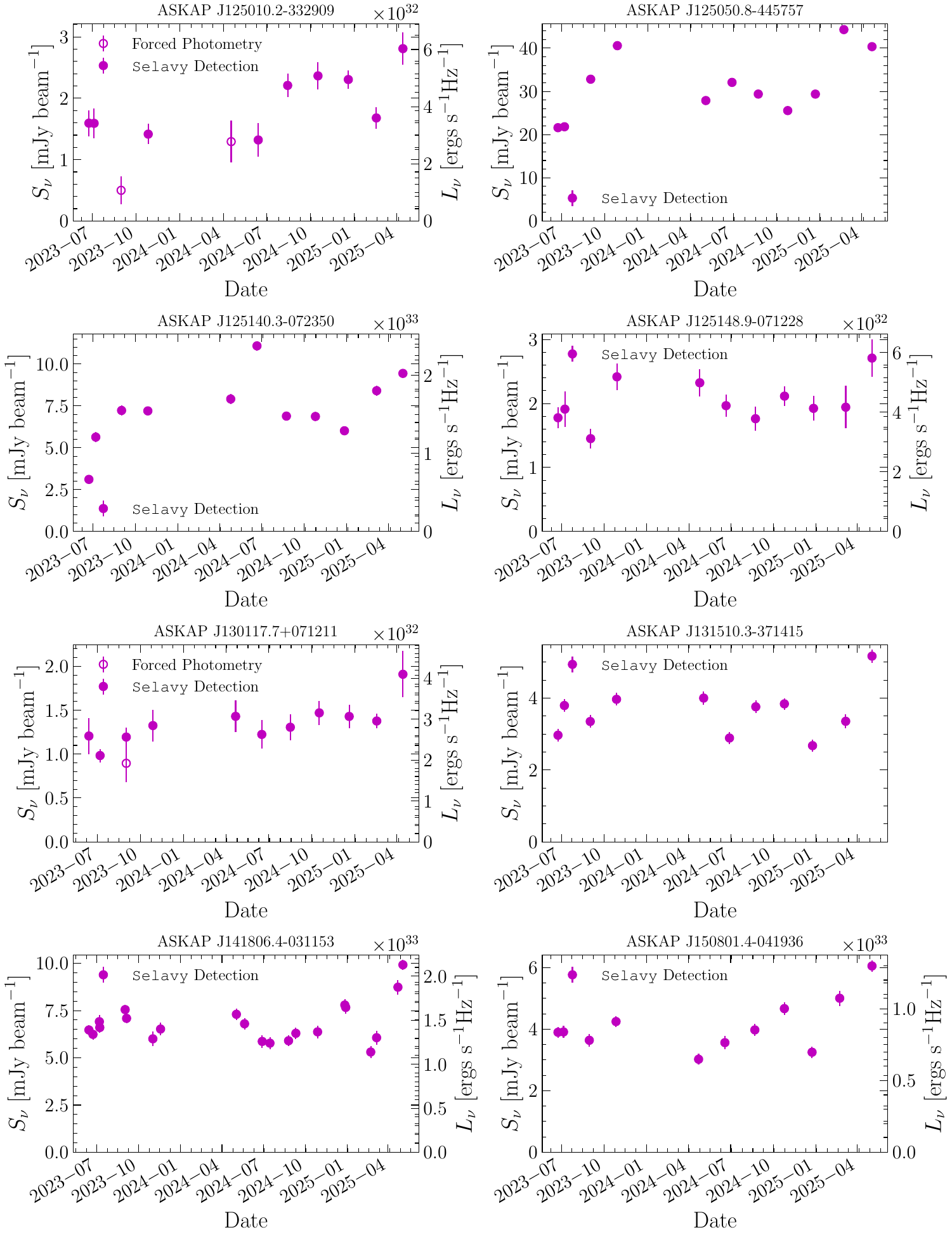}{0.92\textwidth}{}}
 \end{figure*} 
 \clearpage
 \begin{figure*}[htpb!] 
\gridline{\fig{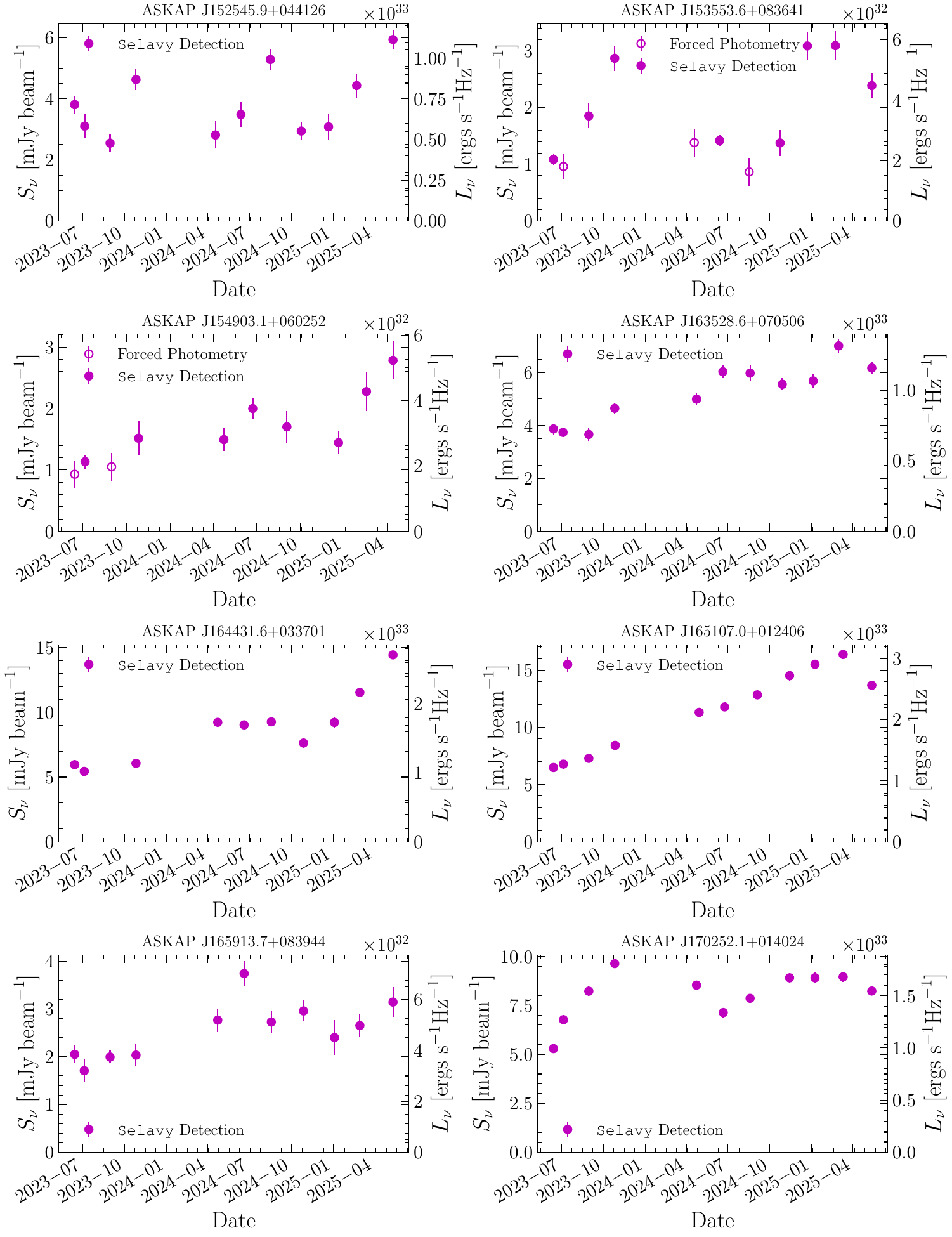}{0.92\textwidth}{}}
 \end{figure*} 
 \clearpage
 \begin{figure*}[htpb!] 
\gridline{\fig{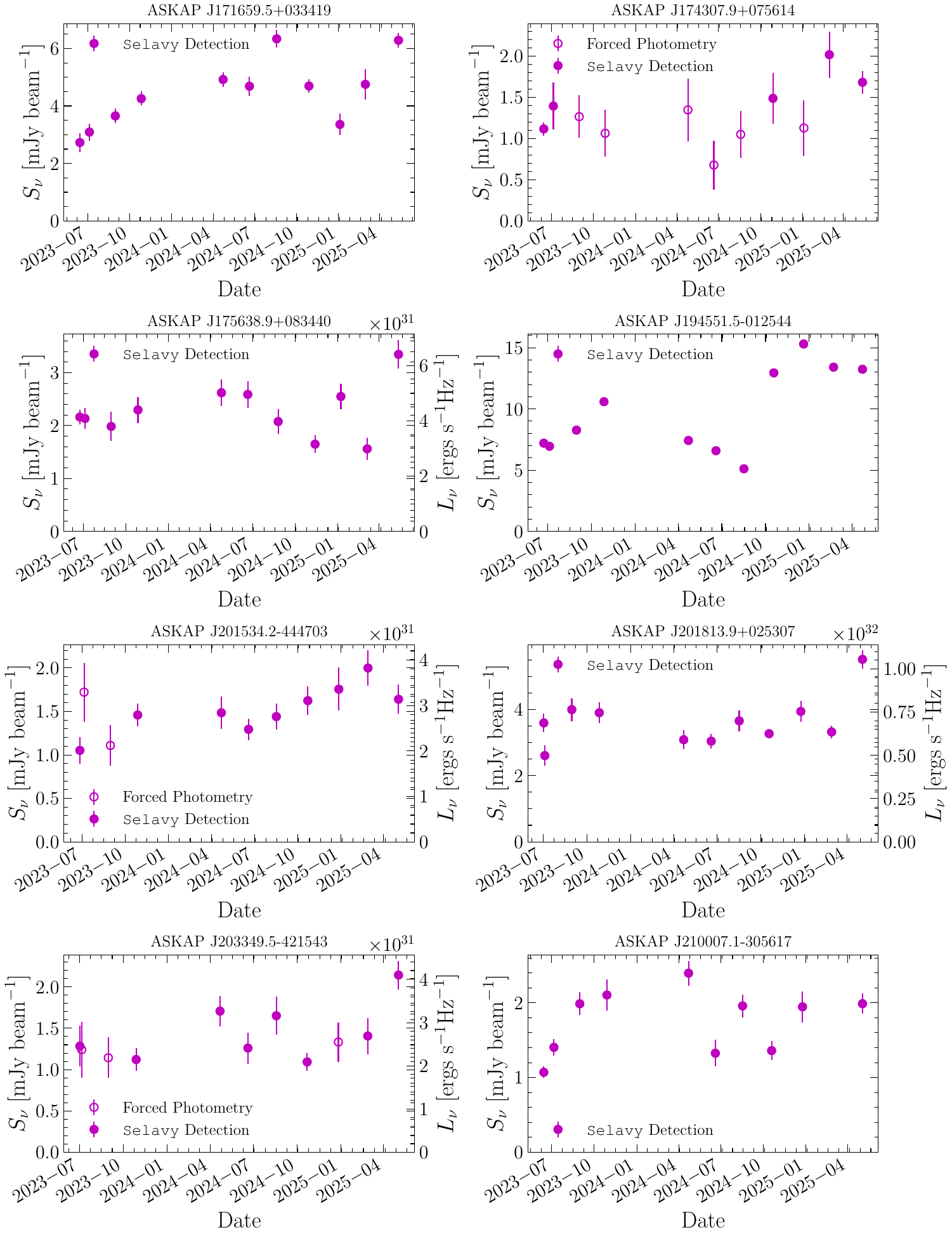}{0.92\textwidth}{}}
 \end{figure*} 
 \clearpage
 \begin{figure*}[htpb!] 
\gridline{\fig{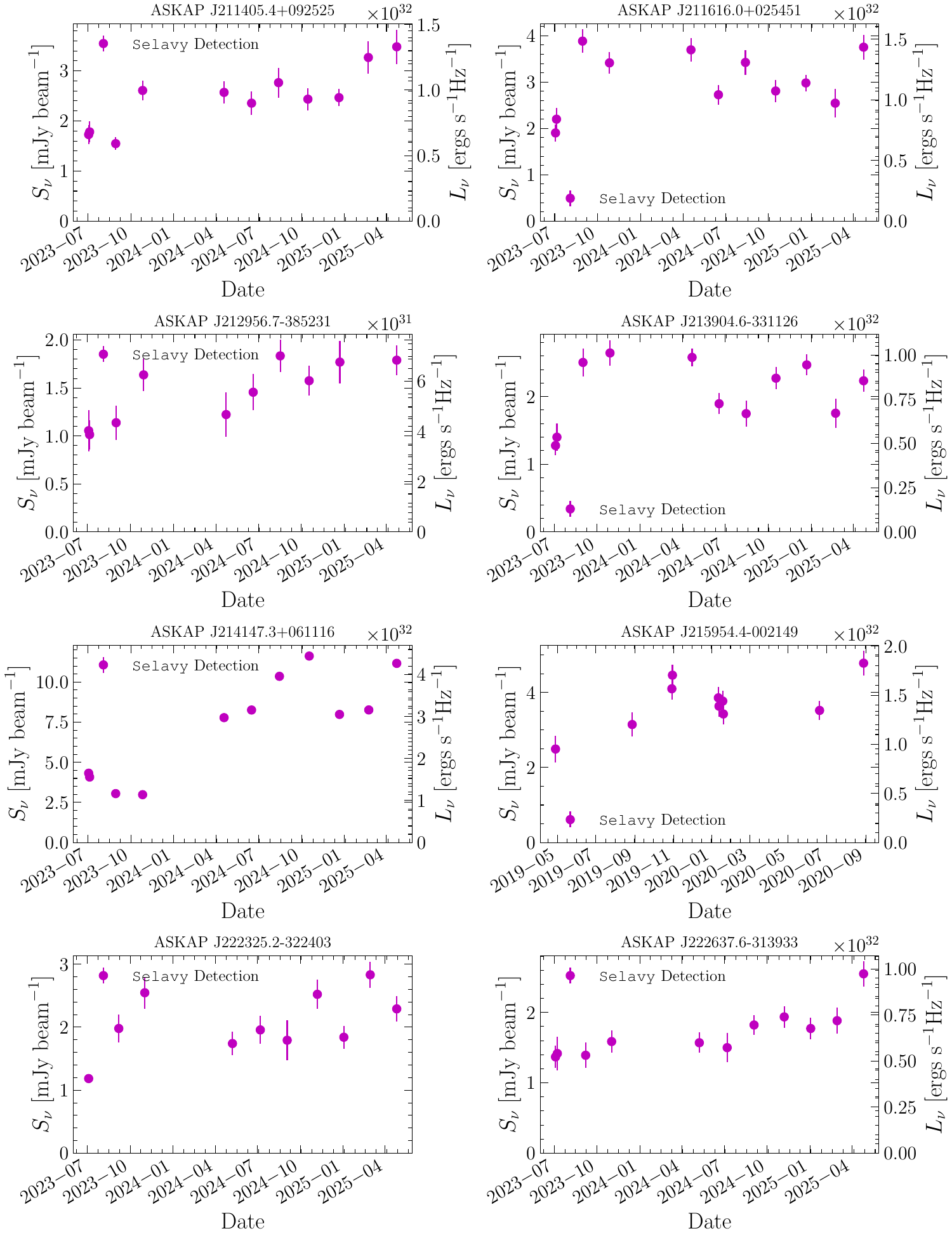}{0.92\textwidth}{}}
 \end{figure*} 


\end{document}